\documentclass[a4paper,11pt]{article}
\usepackage[utf8]{inputenc}
\usepackage{mathtools}
\usepackage{placeins}
\usepackage{graphicx}
\usepackage[font=footnotesize]{caption,subcaption} 
\usepackage{array} 
\usepackage{listings} 
\usepackage{lineno} 
\usepackage{hyperref} 
\usepackage[percent]{overpic} 
\usepackage{float} 
\usepackage{chngcntr} 
\usepackage{color}
\usepackage{authblk}
\usepackage{fullpage}
\usepackage{dcolumn}
\usepackage{bm}
\usepackage{url}
\usepackage{array}
\usepackage{booktabs} 
\usepackage{multicol}

\DeclareMathOperator*{\argmax}{argmax}
\newcolumntype{C}[1]{>{\centering\let\newline\\\arraybackslash\hspace{0pt}}m{#1}}

\title{Best reply structure and equilibrium convergence \\ in generic games}
\author[1,2]{Marco Pangallo$^{\star,}$}
\author[1,2]{Torsten Heinrich}
\author[1,2,3,4]{J. Doyne Farmer}
\affil[1]{\small Institute for New Economic Thinking at the Oxford Martin School, University of Oxford, Oxford OX2 6ED, UK}
\affil[2]{Mathematical Institute, University of Oxford, Oxford OX1 3LP, UK}
\affil[3]{Computer Science Department, University of Oxford, Oxford OX1 3QD, UK}
\affil[4]{Santa Fe Institute, Santa Fe, NM 87501, US}
\date{\today}

\begin{document}

\maketitle

\begin{abstract}
Game theory is widely used as a behavioral model for strategic interactions in biology and social science. It is common practice to assume that players quickly converge to an equilibrium, e.g. a Nash equilibrium.  This can be studied in terms of best reply dynamics, in which each player myopically uses the best response to her opponent's last move.  Existing research shows that convergence can be problematic when there are best reply cycles.   Here we calculate how typical this is by studying the space of all possible two-player normal form games and counting the frequency of best reply cycles.  The two key parameters are the number of moves, which defines how complicated the game is, and the anti-correlation of the payoffs, which determines how competitive it is. We find that as games get more complicated and more competitive, best reply cycles become dominant.  The existence of best reply cycles predicts non-convergence of six different learning algorithms that have support from human experiments.  Our results imply that for complicated and competitive games equilibrium is typically an unrealistic assumption.  Alternatively, if for some reason ``real'' games are special and do not possess cycles, we raise the interesting question of why this should be so. 
\end{abstract}



\renewcommand*{\thefootnote}{\fnsymbol{footnote}}
\footnotetext[1]{\mbox{Corresponding author: marco.pangallo@maths.ox.ac.uk}}
\renewcommand*{\thefootnote}{\arabic{footnote}}

\clearpage

\begin{multicols}{2}

Cycles and feedback loops are common sources of instability in natural and social systems. Here we investigate the relation between cycles and instability in generic settings that can be modeled as two-player games. These include strategic interactions between individual players \cite{myerson2013game}, evolutionary processes \cite{smith1982evolution}, social phenomena such as the emergence of cooperation \cite{axelrod1981evolution} and language formation \cite{Nowak06071999}, congestion on roads and on the internet \cite{rosenthal1973class} and many other applications. We introduce a formalism---that we call \textit{best reply structure}---to characterize instability in terms of an approximated representation of the game, in a similar spirit to the seminal contributions by Kauffman and May on gene regulation \cite{kauffman1969metabolic} and ecosystem stability \cite{may1973qualitative}.

In game theory instability can be understood as the failure of strategies to converge to a fixed point, such as a Nash equilibrium, as a game is played repeatedly \cite{fudenberg1998theory}. It is well-known that convergence is likely to fail in games such as Matching Pennies or Rock Paper Scissors \cite{shapley1964some,gintis2000game,sato2002chaos}, in which the best replies of the game form a cycle (in a sense that will be clarified below).  Very general convergence results have been proven for various types of acyclic games \cite{nachbar1990evolutionary,foster1998nonconvergence,monderer1996fictitious,milgrom1990rationalizability,arieli2016stochastic}.  But  \textit{how typical} are acyclic games?  Do acyclic games span the space of games that are likely to be encountered in realistic settings?  Or are they special?

Here we systematically study this problem for all possible two-player normal form games.  We characterize classes of games in terms of an \textit{ensemble} in which we construct the payoff matrices at random and then hold them fixed as the game is played. Our formalism predicts the typical frequency of convergence as the parameters of the ensemble are varied. We show that best reply cycles become likely and convergence typically fails as games become (i) more complicated, in the sense that the number of moves per player is large, and (ii) more competitive, in the sense that the payoffs to the two players for any given combination of moves are anti-correlated. For example, with 10 moves per player and correlation -0.7, acyclic games make up only 2.7\% of the total. As a consequence, in generic complicated and competitive games equilibrium convergence is typically an unrealistic assumption.  

While studying the generic properties of an ensemble of systems is a common approach in the natural sciences, it is unusual in game theory. Therefore, before describing our contribution and the relation with the literature in more detail, we clarify why we consider this approach useful for game theory.

A natural point of comparison is the work in theoretical ecology by Robert May \cite{may1972will}, who used an ensemble of randomly generated predator-prey interactions as a \textit{null model} of a generic ecosystem, and showed that large ecosystems tend to be unstable. Real ecosystems are not random, rather they are shaped by evolutionary selection and other forces. Many real ecosystems have also existed for long periods of time, suggesting they are in fact stable.  This indicated that real ecosystems are not typical members of the ensemble, and raised the important question of precisely how they are atypical and why they are stable.   Forty five years later, this remains a subject of active research.

Here we apply the same approach to game theory, taking an ensemble of random games as a null model for real-world scenarios that can be represented as games.  Pricing in oligopolistic markets, innovation strategies in competing firms, buying and selling in financial markets, auctions, electoral strategies in competing parties, traffic on roads and sending packets through the internet are all examples of complicated and competitive games. In contrast to ecology, from an empirical point of view it is not clear {\it a priori} whether they are stable:  When is equilibrium a good behavioral model? The rules of these games are designed and not random, but insofar as they can be modeled by normal form games, they are all members of the ensemble we study here.  If complicated and competitive real games are typical members of their ensemble, our results indicate that equilibrium is likely a poor approximation.   

Alternatively, if human-designed games are atypical and cycles are rare, why is this so?  This may vary case by case, but if human-designed games tend to be atypical, our strategic conflicts must have special properties. Whether this is true, and why human design might cause atypical behavior, is far from obvious.  If human-designed games are atypical, then why this is so is an interesting question that deserves further study.  

To better understand our formalism, consider one of the simplest learning algorithms,  \textit{best reply dynamics}. Under this algorithm each player myopically responds with the best reply to her opponent's last move.  The best reply dynamics converges to attractors that can be fixed points, corresponding to pure strategy Nash equilibria, or cycles. We show that a very simple measure---the relative ``size'' of best reply cycles vs fixed points---approximately predicts (R-squared $>$ 0.75) the non-convergence frequency of several well-known and more realistic learning algorithms (reinforcement learning, fictitious play, replicator dynamics, experience-weighted attraction, level-k learning). Some of these learning algorithms have support from human experiments and incorporate forward-looking bounded rationality, suggesting that our results describe the behavior of real players, at least to some extent.

There exists an enormous literature in game theory about the equilibrium convergence properties of learning algorithms; the role of best replies is widely acknowledged even in introductory courses. This literature is often mathematically rigorous and favors exact results in specific classes of games \cite{nachbar1990evolutionary,foster1998nonconvergence,monderer1996fictitious,milgrom1990rationalizability,arieli2016stochastic}. Our work is complementary to this literature, as we provide approximate results for generic games and validate our results with extensive numerical simulations. This makes it possible for us to study some problems that have not been addressed before. For example, we are able to compute the probability of convergence in games that have \textit{both} best reply cycles and fixed points within the same game. 

Once we have established that the best reply structure has predictive value, we determine how it changes with the number of moves and the correlation of the payoffs. We use combinatorial methods to analytically compute the frequency of cycles of different lengths under the microcanonical ensemble. The idea of using methods inspired from statistical mechanics is not new in game theory \cite{blume1993statistical}. However, while existing research has quantified properties of pure strategy Nash equilibria \cite{goldberg1968probability,dresher1970probability,powers1990limiting}, mixed strategy equilibria \cite{berg1998matrix,berg2000statistical} and Pareto equilibria \cite{cohen1998cooperation}, we are the first to quantify the frequency and length of best reply cycles. This gives intuition into why convergence to equilibrium fails in generic complicated and competitive games \cite{galla2013complex} and introduces a formalism that can be extended in many directions and in different fields.  For example, our results are also related to the stability of food webs \cite{may1973qualitative,may1972will} through replicator dynamics, and our formalism can be mapped to Boolean networks, first introduced by Kauffman \cite{kauffman1969metabolic} as a model of gene regulation.  

When convergence to equilibrium fails we often observe chaotic learning dynamics \cite{skyrms1992chaos,galla2013complex}. For the six learning algorithms we analyze here the players do not converge to any sort of intertemporal ``chaotic equilibrium'' \cite{blume1982learning,boldrin1986indeterminacy,hommes1998consistent}, in the sense that their expectations do not match the outcomes of the game even in a statistical sense.  In many cases the resulting attractor is high dimensional, making it difficult for a `rational' player to outperform other players by forecasting their moves using statistical methods.  Once at least one player systematically deviates from equilibrium, learning and heuristics can outperform equilibrium thinking \cite{gigerenzer1999simple} and can be a better description for the behavior of players. Chain recurrent sets \cite{papadimitriou2016nash} and sink equilibria \cite{goemans2005sink} are solution concepts that may apply in this case. 

\section*{Results}

\subsection*{Best reply structure}

Assume a two player normal form game in which the players are Row and Column, each playing moves $i,j = 1, \ldots , N$. A {\it best reply} is the move that gives the best payoff in response to a given move by an opponent. We call \textit{best reply structure} the arrangement of the best replies in the payoff matrix. 

\begin{table}[H]
\footnotesize
\centering
\caption{Terminology}
\begin{tabular}{C{1.6cm}C{5cm}}
\toprule
Best reply & Move that gives the best payoff in response to a given move by an opponent.\\
\hline
Best reply structure & Arrangement of the best replies in the payoff matrix.   \\
\hline
Best reply dynamics & Simple learning algorithm in which the players myopically choose the best reply to the last move of their opponent  \\
\hline
Best reply $k$-cycle &  Closed loop of best replies of length $k$ (each player moves $k$ times) \\
\hline
Best reply fixed point & Combination of moves that is a best reply by both players to a specific move of their opponent (pure Nash Eq.)  \\
\hline 
Best reply vector $\boldsymbol{v}$ & Set of attractors of best reply dynamics, ordered from the longest cycles to the fixed points \\
\hline
Best reply configuration & Unique set of best replies by both players to all moves of their opponent  \\
\hline
Free move / free best reply & Move that is neither part of a cycle or fixed point \\
\bottomrule
\end{tabular}
\end{table}

To illustrate this concept we use a simple learning algorithm, {\it best reply dynamics}, in which each player myopically responds with the best reply to the opponent's last move.  We consider a particular version of best reply dynamics in which the two players alternate moves, each making her best response to her opponent's last move.

\begin{figure}[H]
\centering
\includegraphics[width=.45\textwidth]{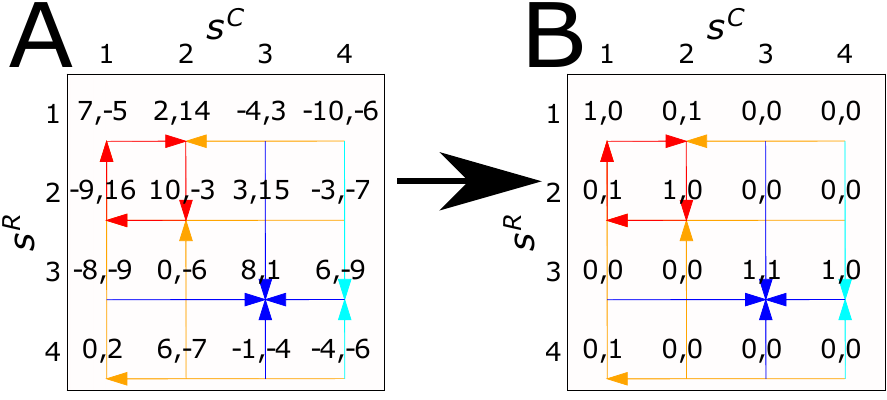}
\caption{Illustration of the best reply structure. $S^R = \{1,2,3,4\}$ and $S^C = \{1,2,3,4\}$ are the possible moves of players Row and Column and each cell in the matrix represents their payoffs (Row is given first).  The best response arrows point to the cell corresponding to the best reply.  The vertical arrows correspond to player Row and the horizontal arrows to player Column.  
The arrows are colored red if they are part of a cycle, orange if they are not part of a cycle but lead to one, blue if they lead directly to a fixed point, and cyan if they lead to a fixed point in more than one step.  The payoff matrix in B is a Boolean reduction that is constructed to have the same best reply structure as the payoff matrix in panel A, but to only have one and zero as its entries. 
}
\label{fig:fig1}
\end{figure}

To see the basic idea consider the game with $N = 4$ shown in Fig. \ref{fig:fig1}A.  Suppose we choose $(1,1)$ as the initial condition.   Assume Column moves first, choosing move $S^C = 2$, which is the best response to Row's move $S^R = 1$.  Then Row's best response is $S^R = 2$, then Column moves $S^C = 1$, etc.  This traps the players in the cycle $(1,1)\rightarrow (1,2) \rightarrow (2,2) \rightarrow (2,1) \rightarrow (1,1)$, corresponding to the red arrows.  We call this a \textit{best reply 2-cycle}, because each player moves twice.   This cycle is an attractor, as can be seen by the fact that starting at $(3,2)$ with a play by Row leads to the cycle.  The first mover can be taken randomly; if the players are on a cycle, this makes no difference, 

\noindent but when off an attractor it can be important.   In fact for this example there are two attractors: If Column had instead gone first, we would have arrived in one step at the \textit{best reply fixed point} at $(3,3)$ (shown in blue).  A  fixed point of the best reply dynamics is a pure strategy Nash equilibrium.   

In Fig. \ref{fig:fig1}B we show a Boolean reduction of the payoff matrix obtained by replacing all best replies by one and all other entries by zero. The Boolean reduction is constructed so that it has the same best reply structure as the matrix it is derived from, but ignores any other aspect of the payoffs.\footnote{The Boolean reduction of the payoff matrix corresponds to a particular class of Boolean networks \cite{kauffman1969metabolic}. We plan to report more details on this correspondence in future work.}

We characterize the set of attractors of best reply dynamics in a given $N \times N$ payoff matrix $\Pi$ by a \textit{best reply vector} $\boldsymbol{v}(\Pi) = (n_N, \ldots, n_2, n_1)$, where $n_1$ is the number of fixed points, $n_2$ the number of 2-cycles, etc.  For instance $\boldsymbol{v} = (0,0,1,1)$ for the example in Fig.~\ref{fig:fig1}.  We define $C=\sum_{k=2}^N{n_k k}$ as the number of moves that are part of cycles. The frequency of non-convergence of best reply dynamics is approximated by the size of the cycles vs. the fixed points, that is $\mathcal{F}(\boldsymbol{v})=C/(C+n_1)$. In Fig.~\ref{fig:fig1}, $\mathcal{F}(0,0,1,1)=2/3$.  This quantity is a rough estimate of the combined size of the basins of attraction of all best reply cycles. It should be regarded as an average rate of non-convergence over multiple realizations of payoff matrices having the same best reply vector but different \textit{best reply configurations}, defined as the unique set of best replies by both players to all possible moves of their opponent. While it is true that best replies that are not on attractors (\textit{free best replies}) may affect the basins of attraction, this tends to average out.\footnote{For example, if all free best replies in Fig. \ref{fig:fig1} were leading to the cycle, the basin of attraction of the cycle would be larger than 2/3. But this is an atypical configuration with $\boldsymbol{v} = (0,0,1,1)$.}

\subsection*{Predictive value}

\begin{figure*}
\centering
\includegraphics[width=.8\textwidth]{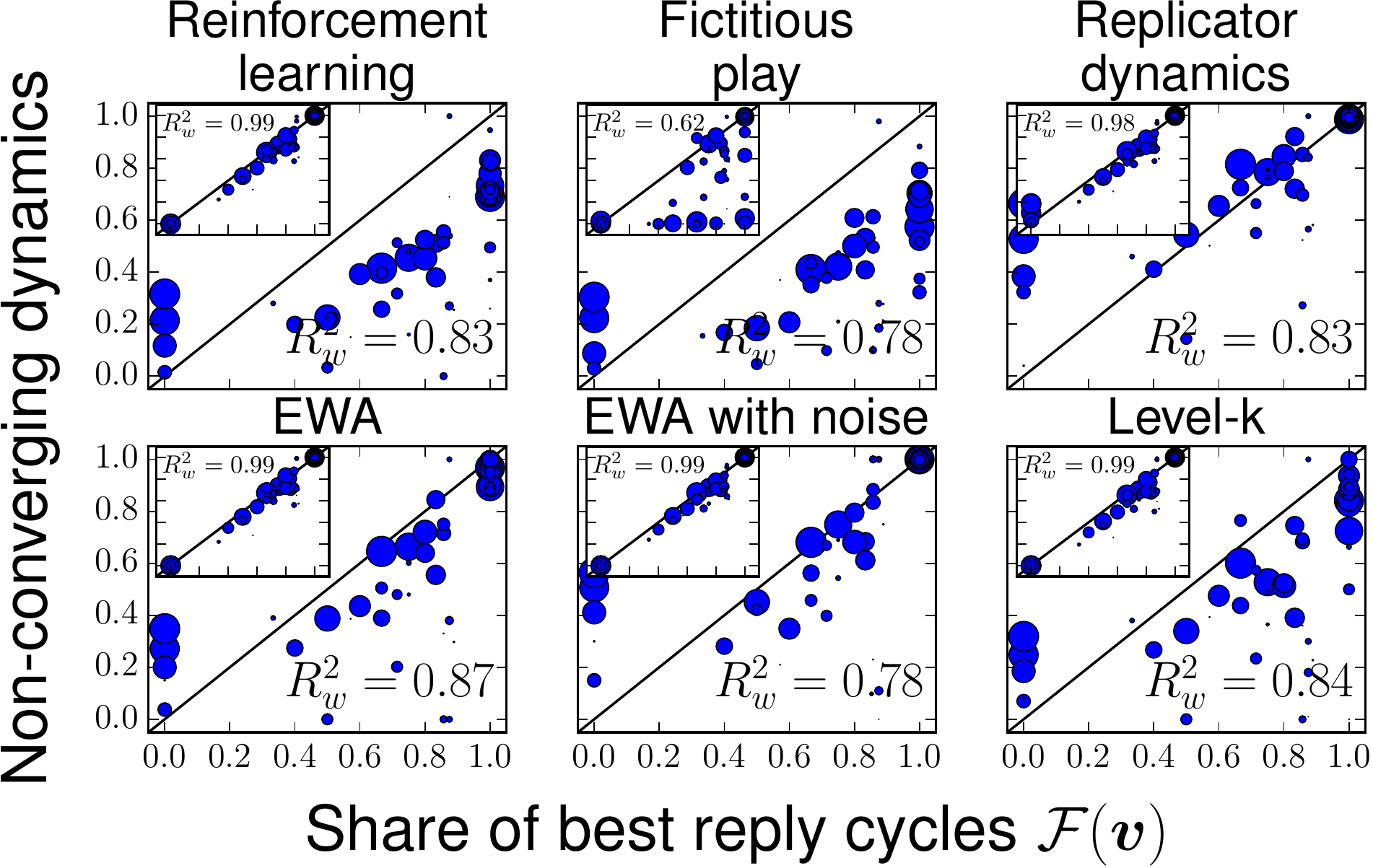}
\caption{Test for how well the best reply structure predicts non-convergence for six learning algorithms. We generate 1000 random payoff matrices describing games with $N=20$ moves and simulate learning from 100 random initial conditions for each one. Each circle corresponds to a specific best reply vector $\boldsymbol{v}$ and its size is the logarithm of the number of times a payoff matrix with $\boldsymbol{v}$ was sampled. The horizontal axis is the frequency of non-convergence under best reply dynamics $\mathcal{F}(\boldsymbol{v})$. For example, the largest circle near $\mathcal{F}(\boldsymbol{v}) = 0.7$ corresponds to $\boldsymbol{v} = (0, \ldots, 0, 1, 1)$, which is frequently sampled. The vertical axis gives the frequency of non-convergence in the simulations, averaged over all payoff matrices and initial conditions having the same $\boldsymbol{v}$.  In the insets simulations are instead based on Boolean reductions of the payoff matrices. The identity line is plotted for reference.}
\label{fig:fig2}
\end{figure*}

We now show that best reply dynamics predicts the convergence frequency of six learning algorithms.  Our goal is to characterize the ensemble of generic games, without constraining their structure.  We do this by using extensive numerical simulations, generating payoff matrices at random, simulating the learning process of the players in a repeated game and then checking convergence to pure and mixed strategy Nash equilibria.  The fact that the behavior of best reply dynamics is strongly correlated to the behavior of other learning algorithms shows that it provides an easy way to study this problem, and that the results about non-convergence for best reply dynamics that we derive in subsequent sections are likely to be indicative of the behavior of a wide variety of different learning algorithms.

We consider six learning algorithms that span different information conditions and levels of rationality. First, reinforcement learning \cite{erev1998predicting} is based on the idea that players are more likely to play moves that yielded a better payoff in the past. It is the standard learning algorithm that is used with limited information and/or without sophisticated reasoning, such as in animal learning. We study the Bush-Mosteller implementation \cite{bush1955stochastic}. 

Fictitious play \cite{robinson1951iterative,brown1951iterative}, our second learning algorithm, requires more sophistication, as it assumes that the players construct a mental model of their opponent. Each player takes the empirical distribution of her opponent's past moves as her mixed strategy, and best responds to this belief. Third, replicator dynamics \cite{hofbauer1998evolutionary} is commonly used in population ecology, but bears a strong connection to learning theory \cite{borgers1997learning}. Fourth, Experience-Weighted Attraction (EWA) has been proposed \cite{camerer1999experience} to generalize several learning algorithms, and has been shown to fit experimental data very well.

 So far we have only considered deterministic approximations of the learning algorithms, resulting from a batch learning assumption: The players observe the moves of their opponent a large number of times before updating their strategies, and so learn based on the actual mixed strategy of their opponent. The deterministic assumption is useful to identify fixed points numerically. As a fifth learning algorithm, we relax this assumption and consider the stochastic version of EWA. In this version, the players update their strategies after observing a single move by their opponent, which is randomly sampled from her mixed strategy. This is also called online learning.
 
 Finally, in level-$k$ learning \cite{nagel1995unraveling}, or anticipatory learning \cite{selten1991anticipatory}, the players try to outsmart their opponent by thinking $k$ steps ahead. For example, here we consider level-2 EWA learning. Both players assume that their opponent is a level-1 learner and update their strategies using EWA. So the players try to preempt their opponent based on her predicted move, as opposed to acting based on the frequency of her historical moves. While the players in the other algorithms are backward-looking, here they are forward-looking. 
 
The details of the learning algorithms and the convergence criteria are listed in the Supplementary Information (SI), Section 1. (We provide a short summary in the Materials and Methods section.) We simulate learning for generic games under each of the six algorithms above.  To define the game we randomly generate payoff matrices for the two players by sampling from a bivariate Gaussian, which is the maximum entropy distribution in this case (see the SI, Section 1.2).  The payoff matrix is held fixed for the duration of each iterated game.  This process is repeated for 1000 randomly generated payoff matrices, testing 100 different initial conditions for each one.  Results with $N=20$ are reported in the main text and results with $N=5$ and $N=50$ are given in the SI, Section 2.
 
In Fig. \ref{fig:fig2} we compare the convergence frequency for best reply dynamics to each of the six learning algorithms.   The circles in each panel correspond to the best reply vectors  $\boldsymbol{v}$, grouping together all payoff matrices with the same $\boldsymbol{v}$. The weight of each best reply vector is the fraction of (1000) times a payoff matrix with $\boldsymbol{v}$ was sampled. This determines the size of the circle and is used for the weighted correlation coefficient $R^2_w$. We place each best reply vector on the horizontal axis according to its frequency of non-convergence under best reply dynamics $\mathcal{F}(\boldsymbol{v})$.  On the vertical axis we plot the frequency of non-convergence for each learning algorithm.  Thus if best reply dynamics perfectly predicts the rate of convergence of the other learning algorithms, all circles should be centered on the identity line. 

There is a strong correlation between the simulations and the predicted values, with weighted correlation coefficient $R^2_w > 0.75$ in every case. In reinforcement learning and fictitious play, $\mathcal{F}(\boldsymbol{v})$ overestimates the frequency of non-convergence.  For fictitious play this is because these algorithms frequently converge to mixed strategy Nash equilibria, whereas best reply dynamics can only converge to pure stategy Nash equilibria.\footnote{For reinforcement learning, the reason is more technical and is discussed in the SI.}   Nevertheless, apart from a constant offset, the rates of non-convergence are proportional.   In contrast, \textit{two-population} replicator dynamics\footnote{Here we consider two-population replicator dynamics and not the more standard one-population version because, focusing on randomly generated games, payoff matrices are asymmetric.} cannot converge to mixed strategy Nash equilibria \cite{gintis2000game}, and so the rate of convergence is lower, and there is no offset from the identity line.  In the SI, Section 2, we show the correlation matrix of the convergence of the six learning algorithms. We find that convergence co-occurs on average 60\% of the times, suggesting a significant degree of heterogeneity among the algorithms.

Although the correlation is good there is not always a detailed correspondence in behavior.  For example, even when best reply cycles are absent convergence is not certain.  The vertical columns of circles above $\mathcal{F}(\boldsymbol{v})=0$ demonstrate this.  This column corresponds to best reply vectors with no cycles, i.e. those of the form $\boldsymbol{v} = (0, \ldots, 0, 0, x)$, where $x = 1, 2, \ldots$ is the number of distinct fixed points and increases from top to bottom.  The circles to the column on the right correspond instead to best reply vectors with cycles and no fixed point ($\mathcal{F}(\boldsymbol{v})=1$), with a higher share of cycles from bottom to top.  The learning algorithms may converge (for example to mixed strategy equilibria) in this situation, but there is a clear trend for less convergence as best reply cycles become more likely. 

The insets show results of simulation runs with Boolean reductions of the payoff matrices.  The correlation is now very strong:  In all cases except fictitious play the weighted correlation is close to unity.  The reason the correlations are so strong for the Boolean reductions is mostly due to the fact that the original payoff matrix has continuous values, so that the learning algorithm may follow what we call {\it quasi-best replies} (see SI, Section 2).  Although the Boolean reduction has precisely the same best reply dynamics as the original matrix, the values of the other payoffs can matter if the learning rule involves history dependence and limited rationality.  For instance, in Fig. \ref{fig:fig1}A, the payoff for Column at (2,3) is 15, while the payoff at (2,1) is 16. The two payoffs are very close and, because of history dependence and limited rationality, player Column might choose move 3 rather than move 1, thereby breaking out of the best reply cycle and reaching the fixed point.
For the case of fictitious play there is also the problem of convergence to mixed strategy Nash equilibria, which is why the correlation for the Boolean reduction is much lower. 

In summary, there exists a robust correlation between the average probability of convergence and the best reply structure.  This is true even though the trajectory of best reply dynamics does not necessarily predict the orbits of the other learning algorithms and the probability of convergence in any specific payoff matrix cannot be exactly calculated from the share of best reply cycles.

\subsection*{Variation of the best reply structure}

We now investigate the prevalence of best reply cycles and fixed points as we vary the properties of the games. Intensively studied classes of games such as coordination, supermodular, dominance solvable and potential games \cite{nachbar1990evolutionary,foster1998nonconvergence,monderer1996fictitious,milgrom1990rationalizability,arieli2016stochastic} are all best reply acyclic.  When is this typical and when is it rare?

In agreement with Galla and Farmer \cite{galla2013complex}, we find that two key parameters of a game are the number of possible moves $N$ and the correlation $\Gamma$ between the payoffs of the two players.  As $N$ increases, it is intuitively clear that the game becomes harder to learn, but it is not obvious how this affects the best reply structure.  To understand how $\Gamma$ affects convergence, we generate payoff matrices so that the expected value of the product of the payoffs to players Row and Column for any given combination of moves is equal to $\Gamma$.  A negative correlation, $\Gamma<0$, implies that the game is competitive, because what is good for one player is likely to be bad for the other. The extreme case is where $\Gamma=-1$, meaning the game is zero-sum.  In contrast $\Gamma>0$ encourages cooperation, in the sense that the payoffs tend to be either good for both players or bad for both players. This intuitively increases the chances for pure strategy Nash equilibria, but it is not clear what it means for best reply cycles.

\begin{figure}[H]
\centering
\includegraphics[width=.45\textwidth]{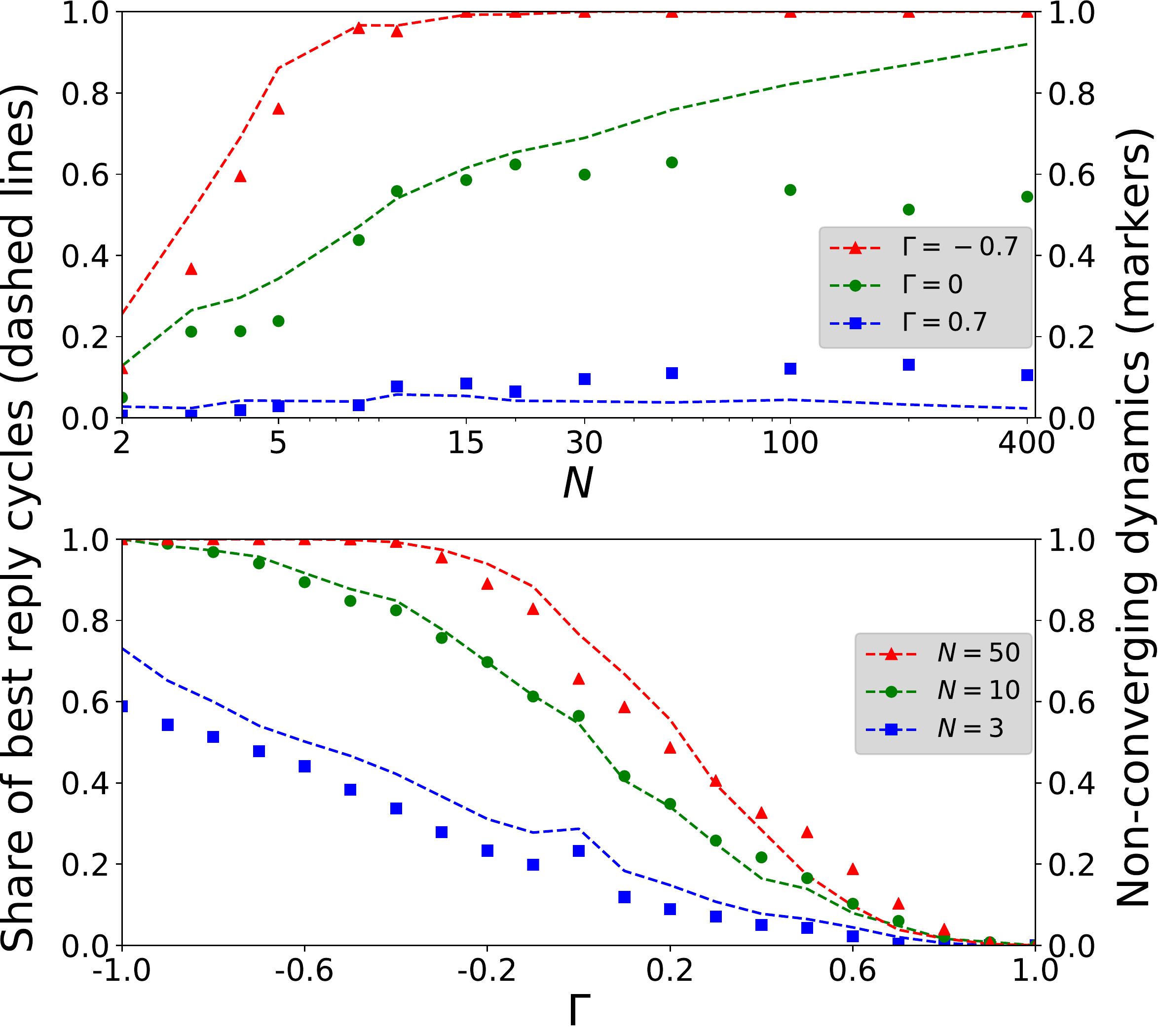}
\caption{How the best reply structure and the rate of convergence of EWA change with the number of moves $N$ and with the average correlation $\Gamma$ between the payoffs of the two players. Dashed lines are the share of best reply cycles $\mathcal{F}(\boldsymbol{v})$ (i.e. the rate of non-convergence of best reply dynamics). Markers are the fraction of simulation runs of EWA that does not converge. A negative correlation $\Gamma$ increases the share of best reply cycles; a positive correlation has the opposite effect. Best reply cycles become predominant as $N$ increases (as long as $\Gamma$ is not large and positive).}
\label{fig:fig3}
\end{figure}

In Fig. \ref{fig:fig3} we show how the share of best reply cycles varies with $N$ and $\Gamma$. For a given value of $N$ and $\Gamma$ we randomly generate payoff matrices and compute the average frequency of non-convergence $\mathcal{F}(\boldsymbol{v})$. We compare this to the average frequency of non-convergence of the EWA learning algorithm.  (We choose EWA because it is the most general learning rule among the six algorithms; its behavior is typical). The good match between the markers and the dashed lines is a confirmation of the

\noindent results in Fig. \ref{fig:fig2} and provides further evidence of the predictive value of the best reply structure. The only exception is for $\Gamma=0$ and $N \geq 30$, where best reply dynamics overestimates the frequency of non-convergence of EWA. 

We find that best reply cycles become prevalent when $\Gamma$ is not positive and $N$ is sufficiently large. In this region of the parameter space acyclic games are extremely rare. Therefore, dominance-solvable, coordination, potential and supermodular games represent a small fraction of all possible payoff matrices that can be created for those $N$ and $\Gamma$. 

\subsection*{Analytical approach}

For $\Gamma=0$ it is possible to derive analytically how the best reply structure varies with $N$.  The total number of possible best reply configurations is $N^{2N}$.  If $\Gamma=0$ all payoff matrices $\Pi$ are equally likely. Therefore we can compute the frequency $\rho (\boldsymbol{v})$ for any set of attractors $\boldsymbol{v}$ by counting the number of best reply configurations leading to $\boldsymbol{v}$. In the jargon of statistical mechanics, we are assuming a micro-canonical ensemble of games. 

Here we just sketch the derivation, referring the reader to the SI (Section 3.1) for a detailed explanation. Because of independence the frequency $\rho (\boldsymbol{v})$ can be written as a product of terms $f$ corresponding to the number of ways to obtain each type of attractor, multiplied by a term $g$ for  free moves (best replies that are not on attractors). We denote by $n$ the number of moves per player which are not already part of cycles or fixed points.

The function $f(n,k)$ counts the ways to have a $k$-cycle (including fixed points, which are cycles of length $k=1$), 
\begin{equation}
f(n,k) = \binom{n}{k}^2 k! (k-1)!,
\end{equation}
where the binomial coefficient means that for each player we can choose any $k$ moves out of $n$ to form cycles or fixed points, and the factorials quantify all combinations of best replies that yield cycles or fixed points with the selected $k$ moves. For instance in Fig.~\ref{fig:fig1}, for each player we can choose any 2 moves out of 4 to form a 2-cycle, and for each of these there are two possible cycles (one clockwise and the other counterclockwise). The number of ways to have a 2-cycle is $f(4,2)=72$. Similarly, for each player we can select any move out of the remaining two to form a fixed point, in $f(2,1)=4$ ways. 

In this example, for both players we can still freely choose one best reply, provided this does not form another fixed point (otherwise the best reply vector would be different). In Fig.~\ref{fig:fig1}, the free best replies are $(3,4)$ for Row and $(4,1)$ for Column. In general, $g_N(n,d)$ counts the number of ways to combine the remaining $n$ free best replies in a $N \times N$ payoff matrix so that they do not form other cycles or fixed points,
\begin{equation}
g_N(n,d) = N^{2n} - \sum_{k=1}^n f(n,k)g_N(n-k,d+1)/(d+1).
\label{eq:g}
\end{equation}
The first term $N^{2n}$ quantifies all possible combinations of the free best replies, and the summation counts the ``forbidden'' combinations, i.e. the ones that form cycles or fixed points.

\begin{figure}[H]
\centering
\includegraphics[width=.45\textwidth]{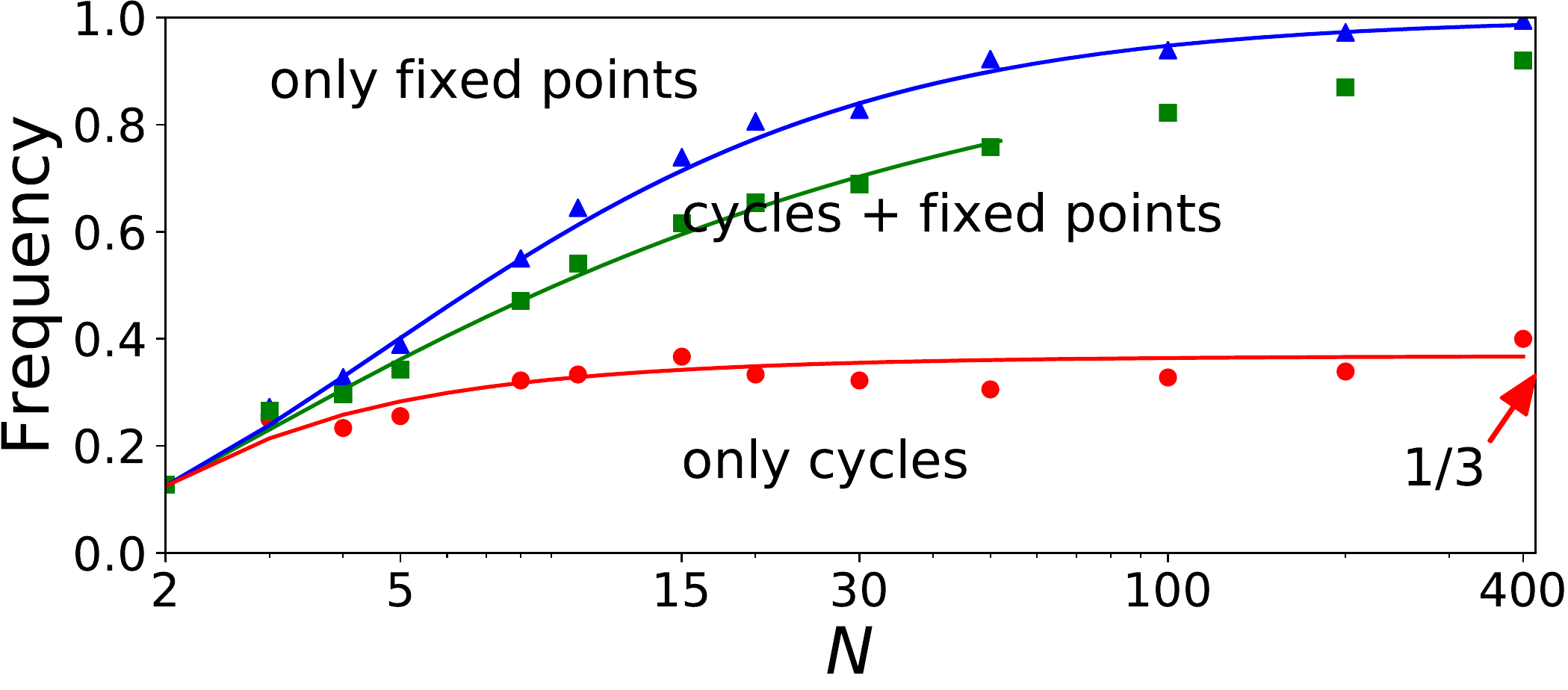}
\caption{Comparison of analytical predictions about best reply cycles to numerical simulations when $\Gamma=0$. Markers are numerical results  and solid lines are analytical results.  Red circles depict the frequency of randomly generated payoff matrices with no fixed points ($\mathcal{F}(\boldsymbol{v})=1$), blue triangles show the frequency with at least one cycle ($\mathcal{F}(\boldsymbol{v})>0$). The text in the figure refers to the area delimited by solid lines, e.g. ``cycles + fixed points'' means that the fraction of payoff matrices with both cycles and fixed points is the distance between the red and blue lines. Finally, green squares represent the average share of best reply cycles $\mathcal{F}_N$; this is discontinued at $N=50$ due to excessive computational cost, see the SI, Section 3.2)}
\label{fig:fig4}
\end{figure}

\noindent This term has a recursive structure. It counts the number of ways to form each type of attractor, and then the number of ways not to have other attractors with the remaining $n-k$ moves. Note that $N$ is a parameter and therefore is indicated as a subscript, while $n$ is a recursion variable. $d$ denotes the recursion depth. Finally, the division by $d+1$ is needed to prevent double, triple, etc. counting of attractors. In the example of Fig.~\ref{fig:fig1}, $g_4(1,0)=15$.

For any given best reply vector $\boldsymbol{v} = (n_N, \ldots, n_2, n_1)$ the general expression for its frequency $\rho$ is
\begin{equation}
\small
\begin{split} 
&\rho(\boldsymbol{v})=\left( \prod_{k=1}^N \prod_{j=1}^{n_k} \frac{f \left(N-\sum_{l=k+1}^N{n_l l} - (j-1)k , k\right)}{j} \right) \\
&\times g_N  \left. \left(N-\sum_{l=1}^N{n_l l},0\right) \middle/ \left( N^{2N} \right)\right..\\
\end{split}
\label{eq:BRvector}
\end{equation}
The product in the first brackets counts all possible ways to have the set of attractors $\boldsymbol{v}$. The first argument of $f$, $N-\sum_{l=k+1}^N{n_l l} - (j-1)k$, iteratively quantifies the number of moves that are not already part of other attractors. The division by $j$, like the division by $d+1$ in Eq. \eqref{eq:g}, is needed to prevent double, triple, etc. counting of attractors. The second term $g_N$ counts all possible ways to position the free best replies so that they do not form other attractors. The first argument of $g_N$ is the count of moves that are not part of attractors, and the initial recursion depth is 0. Finally, we obtain the frequency by dividing by all possible configurations $N^{2N}$. For the payoff matrix in Fig.~\ref{fig:fig1}, $\rho(0,0,1,1)=f(4,2)f(2,1)g_4(1,0)/4^8=0.07$.

Eq. \eqref{eq:BRvector} can then be used to compute the ensemble average of non-convergence of best reply dynamics $\mathcal{F}$ for any given $N$,
\begin{equation}
\label{F}
\mathcal{F}_N = \sum_{\boldsymbol{v}} \rho(\boldsymbol{v}) \mathcal{F}(\boldsymbol{v}),
\end{equation}
summing over all possible $\boldsymbol{v}\text{ s.t.} \sum_{k=1}^N n_k k \leq N$. It is also possible to calculate other quantities, including the fraction of payoff matrices without fixed points ($\mathcal{F}(\boldsymbol{v})=1$) and without cycles ($\mathcal{F}(\boldsymbol{v})=0$).  We provide the expressions and explain their derivation in the SI (Section 3.2).

In Fig. \ref{fig:fig4} we analyze the best reply structure for increasing values of $N$. We report, from bottom to top, the fraction of payoff matrices with no fixed points, the average share of best reply cycles $\mathcal{F}_N$, and the fraction of games with at least one cycle. For instance, for $N=30$, $36\%$ of the payoff matrices have no fixed points, $84\%$ have at least one cycle, (so $16\%$ have no cycles, and $48\%$ have a mixture of cycles and fixed points), with an average $\mathcal{F}_N =0.70$. There is a very good agreement between analytical results (solid lines) and Monte Carlo sampling (markers). The fraction of games with cycles is an increasing function of $N$; it is computationally intractable to compute this for large $N$, but it seems to be tending to one. However, the fraction of games with at least one fixed point seems to reach a fixed value for $N\rightarrow \infty$. In Section 3.3 of the SI we show that this is approximated by $1/3$, in agreement with numerical simulations.

\section*{Discussion}

We have proposed a new formalism that helps understand the conditions under which learning in  repeated games fails to converge to an equilibrium.  For the six learning algorithms we have studied here non-convergence is strongly correlated with the presence of best reply cycles.  When they fail to converge, the trajectories through the strategy space do not closely match best reply cycles.  Instead, as shown by Galla and Farmer for EWA \cite{galla2013complex}, the typical case is chaotic dynamics.\footnote{Bush-Mosteller learning, fictitious play and replicator dynamics all have infinite memory. We observe unstable orbits in which one strategy takes over the others, and this happens periodically with a period increasing exponentially over time. See the SI, Section 1, for examples.}  Why, then, is the presence of best reply cycles closely correlated with non-convergence?  Our hypothesis is that the presence of best reply cycles indicates more complex nonlinear structure in the payoff space that makes convergence to an equilibrium more difficult.  

Mapping out the best reply structure has the advantage of being simple and straightforward---there are no adjustable parameters and no learning takes place. As we have shown here, this makes it possible to use combinatorics to analytically explore the space of all games under the micro-canonical ensemble, using the conceptual framework of statistical mechanics.    

This work can be extended in several directions. It should be possible to account for quasi-best replies, history dependence and limited rationality by studying modified versions of best reply dynamics. For example, we could allow for noisy best replies, in which the players select with a certain probability a move which is not a best reply. We could also allow for level-$k$ reasoning in best reply dynamics, to investigate the role of forward vs. backward looking strategies. 

On a different note, it would also be interesting to characterize the best reply structure in games with more than two players. Our preliminary results indicate that higher-order structures may be relevant. For example, in three-player games two players could be in a best reply cycle, but the remaining player may not. Additionally, we could analyze other ensembles of payoff matrices, for example introducing ordinal constraints.

Finally, the method introduced in this paper can be related to ecology. Generalized Lotka-Volterra equations are equivalent to the replicator dynamics \cite{hofbauer1998evolutionary}, so it may be possible to connect the best reply structure to the network properties of food webs \cite{may1973qualitative}. In Ref. \cite{borrelli2015selection} the authors show that stable subgraphs are statistically overrepresented in empirical food webs, thereby reducing feedback loops. Our preliminary investigations show that loops in food webs are a sufficient but not necessary condition for best reply cycles in the corresponding payoff matrix. 

The main implication of our paper is this:  If real-world situations that can be described as a two-player game are represented to some extent by an ensemble of randomly constructed games and if real players can be approximately described by learning algorithms like the ones we study here, equilibrium is likely to be an unrealistic behavioral assumption when the number of moves is large and the game is competitive. 

\section*{Materials and Methods}
%

We summarize here the protocol that was used to simulate the learning algorithms in Figures \ref{fig:fig2} and \ref{fig:fig3}. We just report the minimal information that would allow replication of the results. A more detailed description, in which we provide behavioral explanations and mention alternative specifications, is given in the Supplementary Information, Section 1. We had to make arbitrary choices about convergence criteria and parameter values, but when testing alternative specifications we found that the correlation coefficients had changed by no more than a few decimal units. This confirms a robust correlation between the rate of convergence of best-reply dynamics and that of the six learning algorithms.

Consider a 2-player, $N$-moves normal form game.  We index the players by $\mu \in \{\text{Row}=R,\quad \text{Column}=C\}$ and their moves by $i,j=1,\ldots N$. Let $x_i^\mu(t)$ be the probability for player $\mu$ to play move $i$ at time $t$, i.e. the $i$-th component of her mixed strategy vector. For notational convenience, we also denote by $x_i(t)$ the probability for player $R$ to play move $i$ at time $t$, and by $y_j(t)$ the probability for player $C$ to play move $j$ at time $t$. We further denote by $s_\mu(t)$ the move which is actually taken by player $\mu$ at time $t$, and by $s_{-\mu}(t)$ the move taken by her opponent.  The payoff matrix for player $\mu$ is $\Pi^\mu$, with $\Pi^\mu(i,j)$ as the payoff $\mu$ receives if she plays move $i$ and the other player chooses move $j$. So if player Row plays move $i$ and player Column plays move $j$, they receive payoffs $\Pi^R(i,j)$ and $\Pi^C(j,i)$ respectively.

\subsection*{Reinforcement learning}

We only describe player Row, because the learning algorithm for Column is equivalent. Player Row at time $t$ has a level of \textit{aspiration} $A^R(t)$ that updates as
\begin{equation}
\begin{split}
A^R(t+1) = (1-\alpha)A^R(t) + \\
\alpha \sum_{i,j} x_i(t) \Pi^R(i,j) y_j(t),
\end{split}
\label{eq:BM0main}
\end{equation}
where $\alpha$ is a parameter. For each move $i$ and at each time $t$ player Row has a level of \textit{satisfaction} $\sigma_i^R(t)$ given by
\begin{equation}
\sigma_i^R(t) = \frac{\sum_{ij} x_i(t) y_j(t) \left( \Pi^R(i,j)-A^R(t) \right)}{\max_{i,j}|\Pi^R(i,j)-A^R(t)|}.
\label{eq:BM1main}
\end{equation}
All components of the mixed strategy vector are updated. The update rule is
\begin{equation}
x_i(t+1)=x_i(t)+ x_i(t) \Delta x_{i}(t) + \sum_{j\neq i} x_j(t) \Delta x_{ij}(t).
\label{eq:BM2main}
\end{equation}
Here, $\Delta x_{i}(t)$ is the contribution due to the choice of move $i$ by player Row (which occurs with probability $x_i(t)$, hence the multiplying term), and $\Delta x_{ij}(t)$ is the contribution on move $i$ due to the choice of another move $j$ (i.e. a normalization update), each occurring with probability $x_j(t)$. We have
\begin{equation}
\Delta x_{i}(t)=\begin{cases}
               \beta \sigma_i^R(t)(1-x_i(t)),& \sigma_i^R(t) > 0,\\
               \beta \sigma_i^R(t)x_i(t),& \sigma_i^R(t) < 0,
            \end{cases}
\label{eq:BM3main}
\end{equation}
and
\begin{equation}
\Delta x_{ij}(t)=\begin{cases}
               - \beta \sigma_j^R(t)x_i(t),& \sigma_j^R(t) > 0,\\
               - \beta \sigma_j^R(t)\frac{x_j(t)x_i(t)}{1-x_j(t)},& \sigma_j^R(t) < 0,
            \end{cases}
\label{eq:BM4main}
\end{equation}
with $\beta$ being a parameter.

Starting from random mixed strategy vectors---the initialization of the mixed strategies will be identical for all learning algorithms that follow---and null levels of aspiration and satisfaction, we iterate the dynamics in Eqs. \eqref{eq:BM0main}-\eqref{eq:BM4main} for 5000 time steps (we set $\alpha=0.2$ and $\beta=0.5$). To identify the simulation run as convergent we only consider the last 20\% of the time steps and the components of the mixed strategy vectors played with average probability greater than $1/N$ in this time interval. If the standard deviation averaged over these components and time steps is larger than 0.01, the simulation run is identified as non-convergent.

\subsection*{Fictitious play}

Player Row calculates the $j$-th component of the expected mixed strategy of Column at time $T$, which we denote by $\tilde{y}_j(T)$, as the fraction of times that $j$ has been played in the past:
\begin{equation}
\tilde{y}_j(T) = \frac{\sum_{t=1}^T I(j,s^C(t))}{T}.
\end{equation}
In the above equation, $I(a,b)$ is the indicator function, $I(a,b)=1$ if $a=b$ and $I(a,b)=0$ if $a\neq b$. Player Row then selects the move that maximizes the expected payoff at time $T$,
\begin{equation}
i(T) = \text{argmax}_k \sum_j \Pi^R(k,j)\tilde{y}_j(T).
\end{equation}
The behavior of Column is equivalent. We use the same convergence criteria and the same length of the simulation runs as in reinforcement learning. There are no parameters in fictitious play.

\subsection*{Replicator dynamics}

We use the discrete-time replicator dynamics
\begin{equation}
\small
\begin{split}
&x_i(t+1) = x_i(t) +  \\
&x_i(t) \delta t \left(\sum_j \Pi^R(i,j) y_j(t) - \sum_{kj} x_k(t) \Pi^R(k,j) y_j(t) \right),\\
&y_j(t+1) = y_j(t) +  \\
&y_j(t) \delta t \left(\sum_i \Pi^C(j,i) x_i(t) - \sum_{ik} y_k(t) \Pi^C(k,i) x_i(t) \right),
\end{split}
\label{eq:RDmain}
\end{equation}
where $\delta t=0.1$ is the integration step. Here the length of the simulation run is endogenously determined by the first component of the mixed strategy vector hitting the machine precision boundary. (Since replicator dynamics is of multiplicative nature, the components drift exponentially towards the faces of the strategy simplex and quickly reach the machine precision boundaries). In order to verify convergence, we check if the largest component of the mixed strategy vector of each player has been monotonously increasing over the last 20\% of the time steps, and if all other components have been monotonously decreasing in the same time interval.

\subsection*{Experience-Weighted Attraction}
Each player $\mu$ at time $t$ has an \textit{attraction} $Q_i^\mu(t)$ towards move $i$. The attractions update as
\begin{equation}
\tiny
	Q_i^\mu(t+1)=\frac{(1-\alpha) \mathcal N (t) Q_i^\mu(t) + (\delta + (1-\delta) x_i^\mu(t)) \sum_j \Pi^\mu(i,j) y_j(t)}{\mathcal N (t+1)},
	\label{eq:EWA1detmain}
\end{equation}
where $\alpha$ and $\delta$ are parameters and $\mathcal N (t)$ is interpreted as \textit{experience}. Experience updates as $\mathcal N (t+1) = (1-\alpha) (1-\kappa) \mathcal N (t) + 1$, where $\kappa$ is a parameter. Attractions map to probabilities through a logit function
\begin{equation}
x_i^\mu(t+1)=\frac{e^{\beta Q_i^\mu(t+1)}}{\sum_j e^{\beta Q_j^\mu(t+1)}},
\label{eq:EWA0main}
\end{equation}
where $\beta$ is a parameter. We simulate Eqs. \eqref{eq:EWA1detmain}-\eqref{eq:EWA0main} for 500 time steps, starting with $\mathcal N(0)=1$. The parameter values are $\alpha=0.18$, $\beta=\sqrt{N}$, $\kappa=1$ and $\delta=1$. If in the last 100 time steps the average log-variation is larger than 0.01, the simulation run is identified as non-convergent. In formula, we check if $1/N \sum_{i=1}^N 5/T \sum_{t=4/5T}^{T} \left( \log x_i(t) \right)^2 > 10^{-2}$, and equivalently for Column.

\subsection*{Experience-Weighted Attraction with noise}
We replace Eq. \eqref{eq:EWA1detmain} by
\begin{equation}
\tiny
	Q_i^\mu(t+1)=\frac{(1-\alpha) \mathcal N (t) Q_i^\mu(t) + (\delta + (1-\delta) I (i, s_\mu(t+1)) \Pi^\mu(i,s_{-\mu}(t+1))}{\mathcal N (t+1)},
	\label{eq:EWA1main}
\end{equation}
i.e. we consider online learning. The parameter values are the same, except $\beta=\sqrt{N}/2$. The convergence criteria are different. Indeed, we run the dynamics for 5000 time steps and---as in reinforcement learning---we consider only the last 20\% of the time steps and only the components of the mixed strategy vectors played with average probability greater than $1/N$ in this time interval. We then we identify the position of the fixed point, and we classify the run as non-convergent if play was farther than 0.02 from the fixed point in more than 10\% of the time steps (i.e. in at least 100 time steps).

\subsection*{Level-k learning}
Let $F^R(\cdot)$ and $F^C(\cdot)$ be the EWA updates for players Row and Column respectively, i.e. if both players use EWA then $x(t+1)=F^R(x(t),y(t))$ and $y(t+1)=F^C(x(t),y(t))$. ($x$ and $y$ without a subscript indicate the full mixed strategy vector.) Then if Column is a level-2 learner, she updates her strategies according to $y^2(t+1)=F^C\left(x(t+1),y(t)\right)=F^C\left(F^R(x(t),y(t)),y(t)\right)$. Row behaves equivalently. In the simulations we assume that both players are level-2 and use the same parameters and convergence criteria as in EWA.

\subsection*{Payoff matrices}
For each payoff matrix, we randomly generate $N^2$ pairs of payoffs---if Row plays $i$ and Column plays $j$, a pair $(a,b)$ implies that Row receives payoff $a$, Column gets payoff $b$. We then keep the payoff matrix fixed for the rest of the simulation. Each pair is randomly sampled from a bivariate Gaussian distribution with mean 0, variance 1 and covariance $\Gamma$.

\end{multicols}
\clearpage

\setcounter{equation}{0}
\setcounter{figure}{0}
\setcounter{table}{0}
\setcounter{section}{0}
\setcounter{page}{1}
\makeatletter
\renewcommand{\theequation}{S\arabic{equation}}
\renewcommand{\thefigure}{S\arabic{figure}}
\renewcommand{\thesection}{S\arabic{section}}
\renewcommand{\thetable}{S\arabic{table}}

\makeatletter
\renewcommand\@bibitem[1]{\item\if@filesw \immediate\write\@auxout
    {\string\bibcite{#1}{S\the\value{\@listctr}}}\fi\ignorespaces}
\def\@biblabel#1{[S#1]}
\makeatother

\begin{center}
\textbf{\huge \textit{Supplemental information for:} \\
 Best reply structure and equilibrium convergence in generic games}
\end{center}

\section{Details of the simulation protocol}
\label{sec:simulation}

We describe here in detail how we produce Figs. 2 and 3 of the main paper.\footnote{The code is available upon request to the corresponding author.} We have to simulate very different learning algorithms over high-dimensional random games and to identify the simulation runs that converge to equilibrium. This leads to unavoidable arbitrary choices in the specification of the learning algorithms, the value of the parameters and the criteria that determine convergence. We have experimented a lot of combinations of design choices, and the overall picture is robust to the specific implementation.\footnote{Unless we choose a parameter setting in which, for example, all learning dynamics converge to fixed points arbitrarily far from Nash equilibria irrespective of the payoff matrix. See below.} Only the weighted correlation coefficients change by a few decimal units.

We describe all these issues in detail in Section \ref{sec:learningalg}. In Section \ref{sec:payoffmatrix} we explain how we generate the payoff matrices. 

\subsection{Learning algorithms}
\label{sec:learningalg}

We analyze six learning algorithms: reinforcement learning, fictitious play, replicator dynamics, Experience-Weighted Attraction (EWA), EWA with noise and level-k learning. For each of these, we provide a high-level qualitative description, we define them formally, and we specify the convergence criteria and the value of the parameters. We also explain the numerical issues we need to address.

For example in the case of reinforcement learning, fictitious play and replicator dynamics the algorithms have infinite memory and so cannot reach a fixed point in finite simulation time. In order to cope with this we need to introduce approximations that we detail here. Another example of a challenging problem is the loss of normalization due to numeric approximations and rounding errors. In the case of EWA, EWA with noise and level-k learning the memory is finite, so it is easier to identify fixed points. However, if memory is too short, some algorithms converge to fixed points in the center of the simplex in which the players randomize among their moves, independently of the payoff matrix. These fixed points can be arbitrarily far from Nash equilibria. Therefore, we need to choose parameter values that make the structure of the game potentially determine convergence.

One important general point is that in real experiments learning algorithms are stochastic, in the sense that at each round of the game the players sample one move with probability determined by their mixed strategy vector. However, we wish to take a deterministic approximation, as it is much easier to identify whether the learning dynamics converge to a fixed strategy. This approximation is usually achieved by assuming that the players observe a large sample of moves by their opponent before updating their mixed strategies \cite{crawford1974learning}.\footnote{This assumption was justified by Conlisk \cite{conlisk1993adaptation} in terms of \textit{two-rooms experiments}: the players are in two separate rooms and need to play against each other many times before they know the outcome of the stage game. Bloomfield \cite{bloomfield1994learning} implemented this idea in an experimental setup.} In the jargon of machine learning, the deterministic approximation corresponds to \textit{batch learning}, while the stochastic version is \textit{online learning}. We consider batch learning in five cases, but we also study one instance of online learning (EWA with noise) and show that the results are robust to stochasticity. 

Another important general point is that we check convergence to fixed points, but these may or may not correspond to Nash equilibria. For example, if fictitious play converges to a fixed point then this is a Nash equilibrium \cite{robinson1951iterativesupp},\footnote{Furthermore, the only stable fixed points of two-population replicator dynamics are pure strategy Nash equilibria \cite{gintis2000gamesupp}.} but as mentioned above EWA with very short memory might converge to fixed points which are arbitrarily far from Nash equilibria. Unfortunately, calculating the full set of Nash equilibria and then checking the distance from the simulated fixed points is computationally unfeasible in games with a large number of moves. In the specific case of $2\times 2$ games and EWA, with sufficiently long memory the fixed points are very close to Nash equilibria (e.g. at a distance of $10^{-6}$ or less) \cite{pangallo2017taxonomy}. As the frequency of convergence of EWA, EWA with noise, level-k learning and reinforcement learning shows very similar properties (cf. Fig. 2 in the main paper) to fictitious play and replicator dynamics -- which reach Nash equilibria exactly -- we believe that the lack of perfect correspondence between fixed points and Nash equilibria is not a major issue. If anything, convergence to Nash equilibrium would be even more unlikely, strengthening the main message of our paper.

\subsubsection{Notation}

Consider a 2-player, $N$-moves normal form game.  We index the players by $\mu \in \{\text{Row}=R,\text{Column}=C\}$ and their moves by $i,j=1,\ldots N$. Let $x_i^\mu(t)$ be the probability for player $\mu$ to play move $i$ at time $t$, i.e. the $i$-th component of his mixed strategy vector. For notational simplicity, we also denote by $x_i(t)$ the probability for player $R$ to play move $i$ at time $t$, and by $y_j(t)$ the probability for player $C$ to play move $j$ at time $t$. We further denote by $s_\mu(t)$ the move which is actually taken by player $\mu$ at time $t$, and by $s_{-\mu}(t)$ the move taken by his opponent.  
The payoff matrix for player $\mu$ is $\Pi^\mu$, with $\Pi^\mu(i,j)$ as the payoff $\mu$ receives if he plays move $i$ and the other player chooses move $j$. So if player Row plays strategy $i$ and player Column plays strategy $j$, they receive payoffs $\Pi^R(i,j)$ and $\Pi^C(j,i)$ respectively.

\subsubsection{Reinforcement Learning}
\label{sec:simRL}

As an example of reinforcement learning, we study the Bush-Mosteller learning algorithm \cite{bush1955stochasticsupp}, using the specification in Refs. \cite{macy2002learning} and \cite{galla2013complexsupp}. This is not the only possible choice for reinforcement learning. For example, other algorithms have been proposed by Erev and Roth \cite{erev1998predictingsupp}. We focus on the Bush-Mosteller algorithm because it is the most different learning rule from the other algorithms we consider.\footnote{On the contrary, the Erev-Roth algorithm can be viewed as a special case of EWA, see Section \ref{sec:simEWA}.}

In the Bush-Mosteller version of reinforcement learning, each player has a certain level of \textit{aspiration}, i.e. his discounted average payoff. This leads to a \textit{satisfaction} for each move -- positive if the payoff the player gets as a consequence of choosing that move is larger than the aspiration level, negative otherwise. The probability to repeat a certain move is increased if the satisfaction was positive, decreased if it was negative.

\paragraph{Formal definition}

More formally, let $A^\mu(t)$ be the aspiration level for player $\mu$ at time $t$. It evolves according to
\begin{equation}
A^\mu(t+1) = (1-\alpha)A^\mu(t) + \alpha \Pi^\mu(s_\mu(t),s_{-\mu}(t)).
\end{equation}
Aspiration is a weighted average of the payoff received at time $t$, $\Pi^\mu(s_\mu(t),s_{-\mu}(t))$, and past aspiration levels. Therefore, payoffs received in the past are discounted by a factor $1-\alpha$. Here $\alpha$ represents the rate of \textit{memory loss}. Satisfaction is defined by
\begin{equation}
\sigma_i^\mu(t) = \frac{\Pi^\mu\left(i, s_{-\mu}(t)\right)-A^\mu(t)}{\max_{i,j}|\Pi^\mu(i,j)-A^\mu(t)|}.
\label{eq:BMsatisfaction}
\end{equation}
After taking move $i$ at time $t$, player $\mu$ has a positive satisfaction if the payoff he received is higher than his aspiration. Note that $\alpha$ is also called \textit{habituation}, as a repeated choice of move $i$ by player $\mu$ leads the aspiration level to correspond to the payoff for move $i$. Satisfaction would then approach zero, as the player gets habituated. In Eq. \eqref{eq:BMsatisfaction}, the denominator is a normalization factor that keeps $\sigma$ within -1 and +1.\cite{macy2002learning} The probability to play move $i$ again is updated as 
\begin{equation}
x_i^\mu(t+1)=\begin{cases}
               x_i^\mu(t) + \beta \sigma_i^\mu(t)(1-x_i^\mu(t)),& \sigma_i^\mu(t) > 0,\\
               x_i^\mu(t) + \beta \sigma_i^\mu(t)x_i^\mu(t),& \sigma_i^\mu(t) < 0.
            \end{cases}
\label{eq:BMupdate1}
\end{equation}
In the above equation $\beta$ is the learning rate. Positive satisfaction leads to an increase of the probability (but habituation slows and eventually stops the rise, because habituation decreases the satisfaction), negative satisfaction has the opposite effect. The probabilities for the moves that were not taken are updated from the normalization condition. Denoting them by $j \neq i$, we have 
\begin{equation}
x_j^\mu(t+1)=\begin{cases}
               x_j^\mu(t) - \beta \sigma_i^\mu(t)x_j^\mu(t),& \sigma_i^\mu(t) > 0,\\
               x_j^\mu(t) - \beta \sigma_i^\mu(t)\frac{x_i^\mu(t)x_j^\mu(t)}{1-x_i^\mu(t)},& \sigma_i^\mu(t) < 0.
            \end{cases}
\label{eq:BMupdate2}
\end{equation}

The learning algorithm described so far is stochastic. As mentioned before, we wish to take a deterministic limit in which the players observe a large sample of moves by their opponent before updating their mixed strategies. We assume that the sample is large enough so that it can be identified with the mixed strategy vector. For simplicity, we switch to the notation in which $x_i^R(t)\equiv x_i(t)$ and $x_j^C(t)\equiv y_j(t)$. We also only consider player Row, because the learning algorithm for Column is equivalent. Aspiration updates as
\begin{equation}
A^R(t+1) = (1-\alpha)A^R(t) + \alpha \sum_{i,j} x_i(t) \Pi^R(i,j) y_j(t).
\end{equation}
Satisfaction is calculated for all moves $i$ which are played with positive probability:
\begin{equation}
\sigma_i^R(t) = \frac{\sum_{ij} x_i(t) y_j(t) \left( \Pi^R(i,j)-A^R(t) \right)}{\max_{i,j}|\Pi^R(i,j)-A^R(t)|}.
\end{equation}
Finally, all components of the mixed strategy vector are updated both as if they were played, or as if they were not played, depending on the probabilities $x_i(t)$. The update rule is
\begin{equation}
x_i(t+1)=x_i(t)+ x_i(t) \Delta x_{i}(t) + \sum_{j\neq i} x_j(t) \Delta x_{ij}(t).
\label{eq:BMupdate0bis}
\end{equation}
Here, $\Delta x_{i}(t)$ is the contribution due to the choice of move $i$ by player Row (which occurs with probability $x_i(t)$, hence the multiplying term), and $\Delta x_{ij}(t)$ is the contribution on move $i$ due to the choice of another move $j$ (i.e. the normalization update), each occurring with probability $x_j(t)$. Following Eqs. \eqref{eq:BMupdate1} and \eqref{eq:BMupdate2}, we have
\begin{equation}
\Delta x_{i}(t)=\begin{cases}
               \beta \sigma_i^R(t)(1-x_i(t)),& \sigma_i^R(t) > 0,\\
               \beta \sigma_i^R(t)x_i(t),& \sigma_i^R(t) < 0,
            \end{cases}
\end{equation}
and\footnote{Note the small notational clutter between Eq. \eqref{eq:BMupdate1} and Eq. \eqref{eq:BMupdate1bis}. In Eq. \eqref{eq:BMupdate1} move $j$ was updated as a consequence of playing move $i$. In Eq. \eqref{eq:BMupdate1bis} move $i$ is updated as a consequence of playing move $j$, with probability $x_j(t)$. }
\begin{equation}
\Delta x_{ij}(t)=\begin{cases}
               - \beta \sigma_j^R(t)x_i(t),& \sigma_j^R(t) > 0,\\
               - \beta \sigma_j^R(t)\frac{x_j(t)x_i(t)}{1-x_j(t)},& \sigma_j^R(t) < 0.
            \end{cases}
\label{eq:BMupdate1bis}
\end{equation}

\paragraph{Convergence criteria}

\begin{figure}
\centering
\includegraphics[width=.70\textwidth]{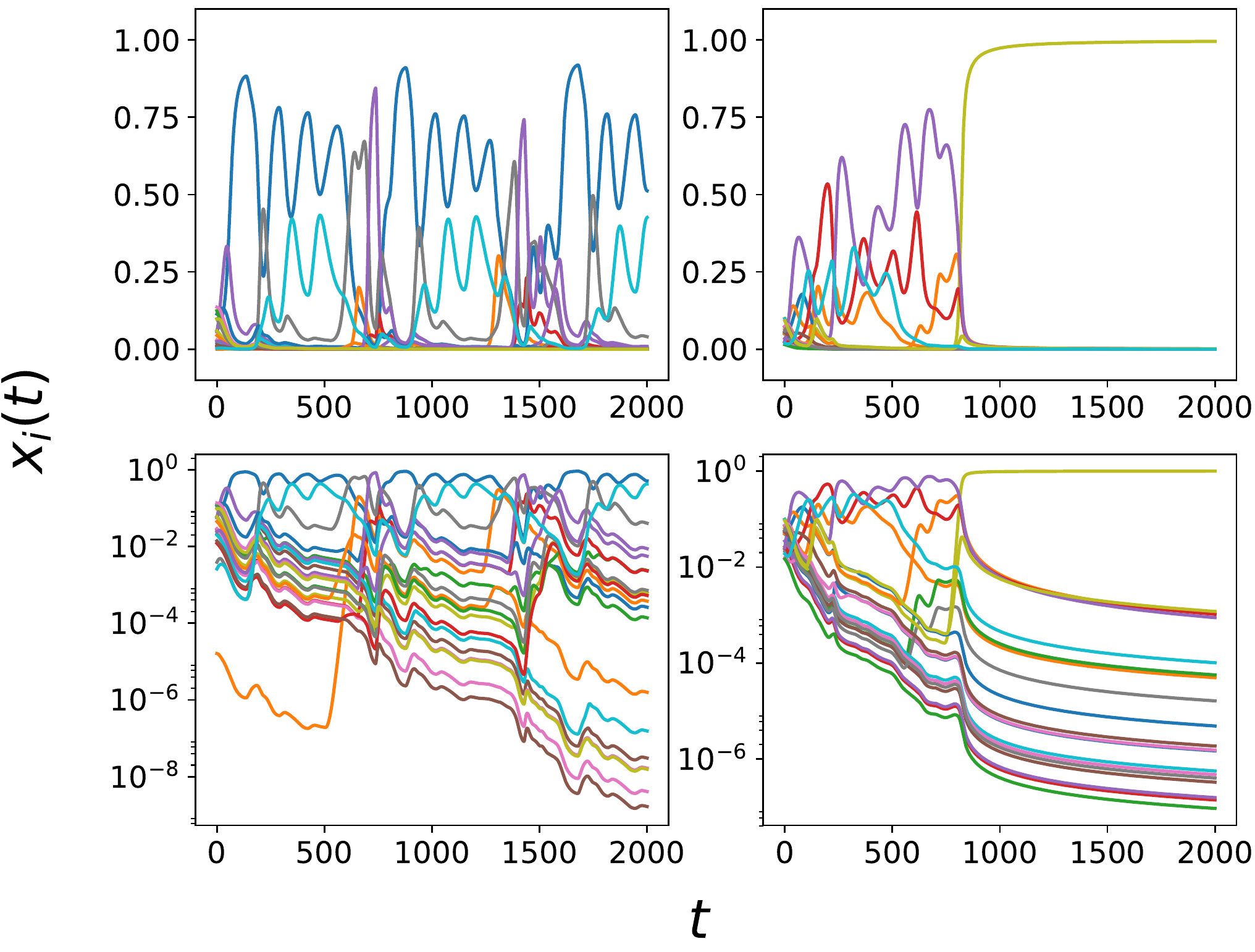}
\caption{Instances of simulation runs of the Bush-Mosteller reinforcement learning algorithm with $N=20$. Each line is a component of the mixed strategy vector of Row (not all components are visible, as they overlap). Left panels: non-converging simulation run. Right panels: converging simulation run. Top panels: linear scale. Bottom panels: logarithmic scale on the vertical axis. Only the first 2000 out of 5000 time steps are shown.}
\label{fig:exampleBM}
\end{figure}

In Figure \ref{fig:exampleBM} we show instances of both converging and non-converging simulation runs. As is clear from the logarithmic plots in the bottom panels, no components of the mixed strategy vector ever reach a fixed point within simulation time. The reason is simple: Eqs. \eqref{eq:BMupdate0bis} do not have a memory loss term, so the probability for unsuccessful strategies keeps decreasing over time. Only numeric approximations would yield a fixed point, but under most parameter settings the Bush-Mosteller dynamics takes very long to reach the machine precision boundary. 

Therefore, we choose a simple heuristic to determine if the learning dynamics has reached a fixed point:
\begin{enumerate}
\item Only consider the last 20\% time steps.
\item Only keep the moves that have been played with a frequency larger than $1/N$.
\item If the average standard deviation (i.e. averaged over the most frequent moves) is larger than 0.01, identify the simulation run as non-convergent. Otherwise, identify it as convergent.
\end{enumerate}
We experimented with slightly different specifications, with no significant effects on the results.

\paragraph{Parameter values}

If the aspiration memory loss $\alpha$ and/or the learning rate $\beta$ are very small, the learning dynamics always reaches fixed points at the center of the strategy simplex, irrespective of the payoff matrix. In this fixed point the players simply randomize between all moves. In a certain sense, they are not learning from playing the game. Except from this unrealistic situation, we do not observe much sensitivity to the parameter values.\footnote{If $\beta$ is too large, we get numerical problems, as the learning dynamics overshoots the strategy simplex boundaries.} We perform the simulations with $\alpha=0.2$ and $\beta = 0.5$.

We simulate the learning dynamics by iterating Eqs. \eqref{eq:BMupdate0bis} for 5000 time steps.\footnote{For $N=5$, numeric approximations make the dynamics lose normalization after 2000 time steps. In this case we simulate for 2000 time steps only.}

\subsubsection{Fictitious Play}
\label{sec:simFP}

Fictitious play was first proposed as an algorithm to calculate the Nash equilibria of a game, and later interpreted as a learning algorithm \cite{brown1951iterativesupp,robinson1951iterativesupp}. It is an example of \textit{belief learning}. Instead of learning based on the experienced payoffs, as in reinforcement learning, the players update their beliefs on what move could be taken by their opponent, and react to their beliefs.

In fictitious play, each player takes the empirical distribution of moves by her opponent as an estimate of his mixed strategy, calculates the expected payoff of her moves given this belief, and chooses the move that maximizes her expected payoff. Here we study the standard fictitious play algorithm, in which the players weigh all past moves equally, and choose the best performing move with certainty. Variants include \cite{fudenberg1998theorysupp} \textit{weighted fictitious play}, in which the players discount the past moves of their opponent and give higher weight to the more recent moves, and \textit{stochastic fictitious play}, in which the players select the best performing move with a certain probability, and potentially all other moves with a smaller probability.

We focus on the standard fictitious play algorithm because the other versions are simply a special case of EWA (see Section \ref{sec:simEWA}).

\paragraph{Formal definition}

Player Row calculates the $j$-th component of the expected mixed strategy of Column at time $T$, which we denote by $\tilde{y}_j(T)$, simply as the fraction of times that $j$ has been played in the past:
\begin{equation}
\tilde{y}_j(T) = \frac{\sum_{t=1}^T I(j,s^C(t))}{T}.
\end{equation}
In the above equation, $I(a,b)$ is the indicator function, $I(a,b)=1$ if $a=b$ and $I(a,b)=0$ if $a\neq b$. Player Row then selects the move that maximizes the expected payoff at time $T$,\footnote{Because we study payoff matrices with random coefficients, it is almost impossible that two moves yield the same payoff. If that was the case, usually the player selects among such moves with equal probability.}
\begin{equation}
i(T) = \argmax_k \sum_j \Pi^R(k,j)\tilde{y}_j(T).
\end{equation}
The behavior of Column is equivalent.

\paragraph{Convergence criteria}

\begin{figure}
\centering
\includegraphics[width=.7\textwidth]{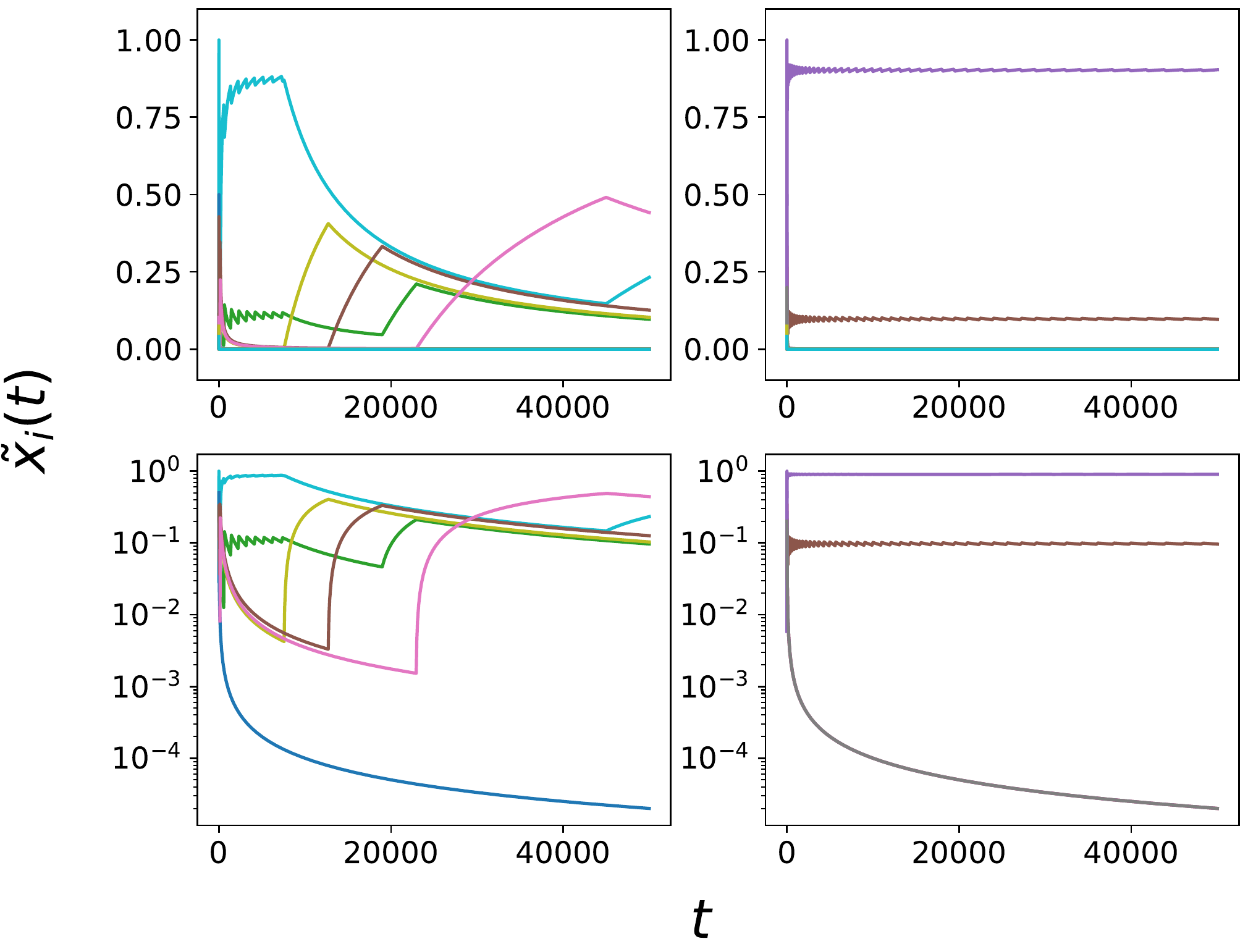}
\caption{Instances of simulation runs of fictitious play with $N=20$. Each line is a component of the mixed strategy vector of Row (not all components are visible, as they overlap). Left panels: non-converging simulation run. Right panels: converging simulation run. Top panels: linear scale. Bottom panels: logarithmic scale on the vertical axis. We show here 50000 time steps, although 5000 iterations turned out to be enough to measure the convergence rate almost as accurately.}
\label{fig:exampleFP}
\end{figure}

We look at the convergence of the estimated mixed strategy vectors at time $t$, $\tilde{x}_i(t)$ and $\tilde{y}_j(t)$. As it is clear from Figure \ref{fig:exampleFP}, the behavior of fictitious play is very similar to that of Bush-Mosteller dynamics. Therefore, we use the same convergence criteria. Note that changing the expected strategies takes more and more time as $t$ increases. In a certain sense, the behavior of the players becomes more set, as they need more sampling evidence to change their expectations.

\paragraph{Parameter values}

Fictitious play has no parameters. We only need to choose the maximum number of iterations, which we take as 5000. We experimented with longer time series (50000 time steps), but the tradeoff between accuracy and speed was unfavorable.

\subsubsection{Replicator Dynamics}
\label{sec:simRD}

Replicator dynamics \cite{smith1982evolutionsupp} is the standard tool used in evolutionary game theory \cite{hofbauer1998evolutionarysupp}. It is a stylized model representing the evolution of individuals with certain traits in a population. The fitness of each trait depends on the population shares of the other traits, and on the average fitness. Although it is mostly used in population biology, the replicator dynamics has also been studied as a learning algorithm in game theory. The key connection is through the \textit{population of ideas} \cite{borgers1997learningsupp}. Each move can be viewed as a trait, and the evolution of the population shares of each trait corresponds to the dynamics of the components of the mixed strategy vector. 

The most typical form of replicator dynamics only concerns one population. If the payoff matrix is symmetric, the game can be seen as between a focal player and the rest of the population. However, being concerned with generic and randomly determined two-player games, the payoff matrix is typically asymmetric. This naturally leads to \textit{two-population} replicator dynamics. The dynamical properties of the two-population version are different from those of the one-population algorithm. For our purposes, the most important difference is that one-population replicator dynamics typically converges to mixed strategy Nash equilibria, whereas two-population replicator dynamics only converges to \textit{strict} Nash equilibria (i.e. pure strategy Nash equilibria in which the payoff at equilibrium is strictly larger than any other payoff that can be obtained if the opponent does not change his move) \cite{gintis2000game}.

\paragraph{Formal definition}

Letting $x_i$ and $y_j$ denote the population shares of individuals with traits $i$ and $j$ respectively, two-population replicator dynamics reads
\begin{equation}
\begin{split} 
&\dot{x_i}(t) = x_i(t) \left(\sum_j \Pi^R(i,j) y_j(t) - \sum_{kj} x_k(t) \Pi^R(k,j) y_j(t) \right),\\
&\dot{y_j}(t) = y_j(t) \left(\sum_i \Pi^C(j,i) x_i(t) - \sum_{ik} y_k(t) \Pi^C(k,i) x_i(t) \right).
\end{split}\label{eq:RDcontinuous}
\end{equation}
The shares of trait $i$ in population Row and trait $j$ in population Column evolve according to the fitness of that trait (as given by the expected payoff) compared to the average fitness in the respective population \cite{hofbauer1998evolutionary}.   

Replicator dynamics needs to be discretized for simulation. We use the Euler discretization
\begin{equation}
\begin{split}
&x_i(t+1) = x_i(t) +  x_i(t) \delta t \left(\sum_j \Pi^R(i,j) y_j(t) - \sum_{kj} x_k(t) \Pi^R(k,j) y_j(t) \right),\\
&y_j(t+1) = y_j(t) +  y_j(t) \delta t \left(\sum_i \Pi^C(j,i) x_i(t) - \sum_{ik} y_k(t) \Pi^C(k,i) x_i(t) \right),
\end{split}
\label{eq:RD}
\end{equation}
where $\delta t$ is the integration step.

\paragraph{Convergence criteria}

\begin{figure}
\centering
\includegraphics[width=.66\textwidth]{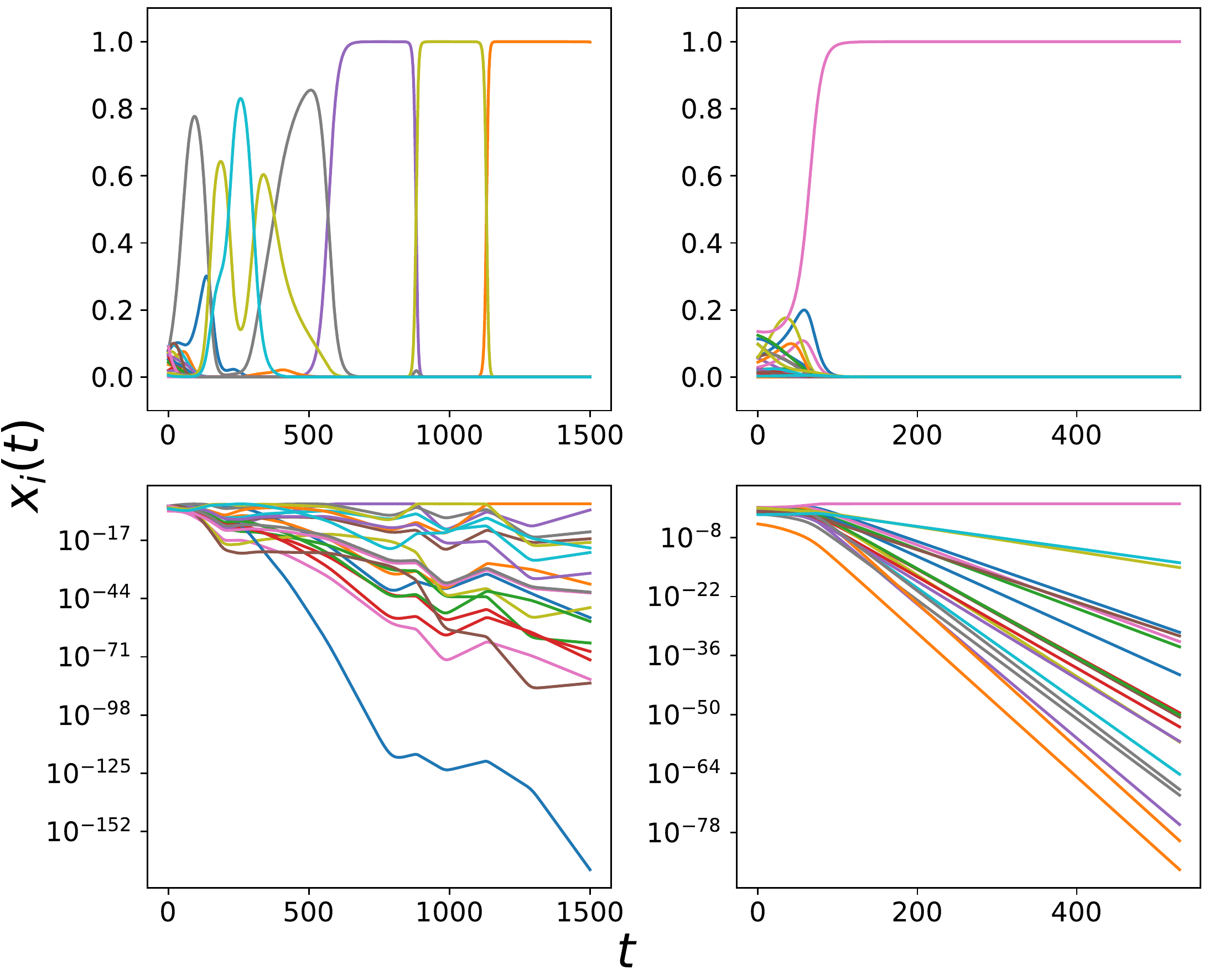}
\caption{Instances of simulation runs of replicator dynamics with $N=20$. Each line is a component of the mixed strategy vector of Row (not all components are visible, as they overlap). Left panels: non-converging simulation run. Right panels: converging simulation run. Top panels: linear scale. Bottom panels: logarithmic scale on the vertical axis. The maximum simulation time (1500 or 500 time steps) is endogenously determined by the first component of the mixed strategy vector hitting the machine precision boundary.}
\label{fig:exampleRD}
\end{figure}

In Figure \ref{fig:exampleRD} we can see the technical problems associated with simulating the replicator dynamics. First, because only strict Nash equilibria are stable, all stable fixed points sit at the boundaries of the probability simplex and cannot be reached in finite simulation time. Second, the period of cycles increases over time (due to the infinite memory of the replicator equations), and even unstable dynamics drifts towards the edges of the probability simplex. 

Third, while in the cases of Bush-Mosteller reinforcement learning and fictitious play the components of the mixed strategy vector were changing by relatively few orders of magnitude, the functional form of the replicator dynamics \eqref{eq:RDcontinuous} implies an exponential change.\footnote{As can be seen from the straight lines in the bottom panels of Figure \ref{fig:exampleRD}.} Therefore, the map \eqref{eq:RD} can be reliably simulated only for a limited \textit{confidence time interval}: we stop the simulation run as soon as one component $x_i$ or $y_j$ reaches the machine precision limits.\footnote{We experimented with arbitrary precision numbers, using the Python package \texttt{decimal}. This is not very helpful, as it takes exponentially more time for the players to switch to other moves as the simulation goes on. Moreover, it is extremely computationally expensive, so that one simulation run with arbitrary precision numbers can last more than 100 times than the equivalent with floating point numbers.}

This precaution is necessary because, if the dynamics is following a cycle, a certain move may not be played for a long time interval, with its probability decreasing over time. At some point, it may become convenient for the player to choose that move again, so the probability would start increasing again. But if the probability had hit the precision limits of the computer beforehand, it would be stuck at zero, falsely identifying the simulation run as having reached a fixed point.

Another problem concerns rounding approximations, which imply that normalization may be lost. If that happens, we stop the simulation run and discard the results.

With the integration step we choose, the confidence time interval is on average of the order of 1000 time steps (but can vary considerably, as can be seen in Figure \ref{fig:exampleRD}). We could use the same convergence criteria as for Bush-Mosteller dynamics and fictitious play, but the short simulation time and the shape of the cycles -- in linear scale, the dynamics is constant for a long time, and then suddenly changes -- suggest to use a different heuristic. We check whether in the last 20\% of the time steps the probabilities of the most used move for both players are monotonically increasing, while all other probabilities are monotonically decreasing. In other words
\begin{enumerate}
\item Only consider the last 20\% time steps.
\item For each player, find the move with the highest probability, and verify whether this probability has been increasing for the full time interval. 
\item Check that the probabilities for all other moves have been decreasing. 
\item If conditions 2-3 are satisfied for both players, identify the simulation run as convergent.
\end{enumerate}
These criteria simply reflect what we observe in Figure \ref{fig:exampleRD}. While we cannot conclude that this heuristic works in general, a direct inspection of over 100 simulation runs for several values of $N$ confirms that convergence to pure strategy Nash equilibria or failure to converge has been correctly identified in the vast majority of cases.

Finally, we would like to add a word of caution on the seemingly stronger instability of replicator dynamics as compared to other learning algorithms. Because of infinite memory and depending on the initial condition, it might take long to ``find'' a pure strategy Nash equilibrium, meaning that the replicator dynamics can hit the machine precision limits first, when it is still in a ``transient''. In other words, it may not be in the basin of attraction determined by a cycle, but it may also have not reached a pure strategy Nash equilibrium within the confidence time interval. This is especially the case for large payoff matrices, $N \geq 50$. 

\paragraph{Parameter values}

We simulate the replicator dynamics by choosing an integration step of $\delta t = 0.1$ (small enough so to prevent overshooting of the boundaries of the probability simplex), and a simulation time of 3000 time steps maximum. However, as discussed before, the simulation time is typically shorter and determined by the first strategy hitting the machine precision boundary.

\subsubsection{Experience-Weighted Attraction}
\label{sec:simEWA}

Experience-Weighted Attraction (EWA) has been proposed by Camerer and Ho \cite{camerer1999experiencesupp} to generalize reinforcement and belief learning algorithms (such as fictitious play, or best reply dynamics). The key insight is that real players use information about experienced payoffs, as in reinforcement learning. But they also try and predict the next moves of their opponent, as in belief learning. The authors report a better experimental out-of-sample goodness-of-fit than with simple reinforcement learning or fictitious play, showing evidence in favor of their theory.

The connection between reinforcement and belief learning lies in the update of the moves that were not played, i.e. in considering the \textit{foregone payoffs}. If only the probabilities of the moves that are played are updated, EWA reduces to a simple version of reinforcement learning (not to the Bush-Mosteller implementation described in Section \ref{sec:simRL}). If all probabilities are updated with the same weight, EWA reduces to fictitious play or best reply dynamics, depending on the parameters.

Finally, note that EWA also reduces to replicator dynamics by taking the limits of some parameters (e.g., by taking the limit of infinite memory).\cite{sato2005stability}

\paragraph{Formal definition}

In EWA, the mixed strategies are determined from the so-called \textit{attractions} or \textit{propensities} $Q_i^\mu(t)$. These are real numbers that quantify the level of appreciation of player $\mu$ for move $i$ at time $t$. Attractions are not normalized, so the probability for player Row to play move $i$ is given by a logit,
\begin{equation}
x_i(t+1)=\frac{e^{\beta Q_i^R(t+1)}}{\sum_j e^{\beta Q_j^R(t+1)}},
\label{eq:EWA0}
\end{equation}
 where $\beta$ is the payoff sensitivity or \textit{intensity of choice}\footnote{The larger $\beta$, the more the players consider the attractions in determining their strategy. In the limit $\beta \rightarrow \infty $ the players choose with certainty the move with the largest attraction. In the limit $\beta \rightarrow 0$ they choose randomly, disregarding the attractions.} and a similar expression holds for $y_j(t+1)$. The propensities update as follows:

\begin{equation}
	Q_i^\mu(t+1)=\frac{(1-\alpha) \mathcal N (t) Q_i^\mu(t) + (\delta + (1-\delta) I (i, s_\mu(t+1)) \Pi^\mu(i,s_{-\mu}(t+1))}{\mathcal N (t+1)},
	\label{eq:EWA1}
\end{equation}

 where 

\begin{equation}
	\mathcal N (t+1) = (1-\alpha) (1-\kappa) \mathcal N (t) + 1.
	\label{eq:Nt}
\end{equation}
 
Here $\mathcal N (t)$ represents \textit{experience} because it increases monotonically with the number of rounds played; the more it grows, the smaller becomes the influence of the received payoffs on the attractions (as the denominator increases). The propensities change according to the received payoff when playing move $i$ against move $s_{-\mu}$ by the other players, i.e. $\Pi^\mu(i,s_{-\mu}(t+1))$. The indicator function $I (i, s_\mu(t+1))$ is equal to 1 if $i$ is the actual move that was played by $\mu$ at time $t+1$, that is $i=s_\mu(t+1)$, and equal to 0 otherwise. All attractions (those corresponding to strategies that were and were not played) are updated with weight $\delta$, while an additional weight $1-\delta$ is given to the specific attraction corresponding to the move that was actually played. Finally, the memory loss parameter $\alpha$ determines how quickly previous attraction and experience are discounted and the parameter $\kappa$ interpolates between cumulative and average reinforcement learning \cite{camerer1999experience}.

As with the other learning algorithms, we take a deterministic limit. Under the assumption of batch learning, Eq. \eqref{eq:EWA1} reads
\begin{equation}
	Q_i^R(t+1)=\frac{(1-\alpha) \mathcal N (t) Q_i^R(t) + (\delta + (1-\delta) x_i(t)) \sum_j \Pi^R(i,j) y_j(t)}{\mathcal N (t+1)},
	\label{eq:EWA1det}
\end{equation}
and a similar expression holds for Column.

\paragraph{Convergence criteria}

\begin{figure}[ht]
\centering
\includegraphics[width=.70\textwidth]{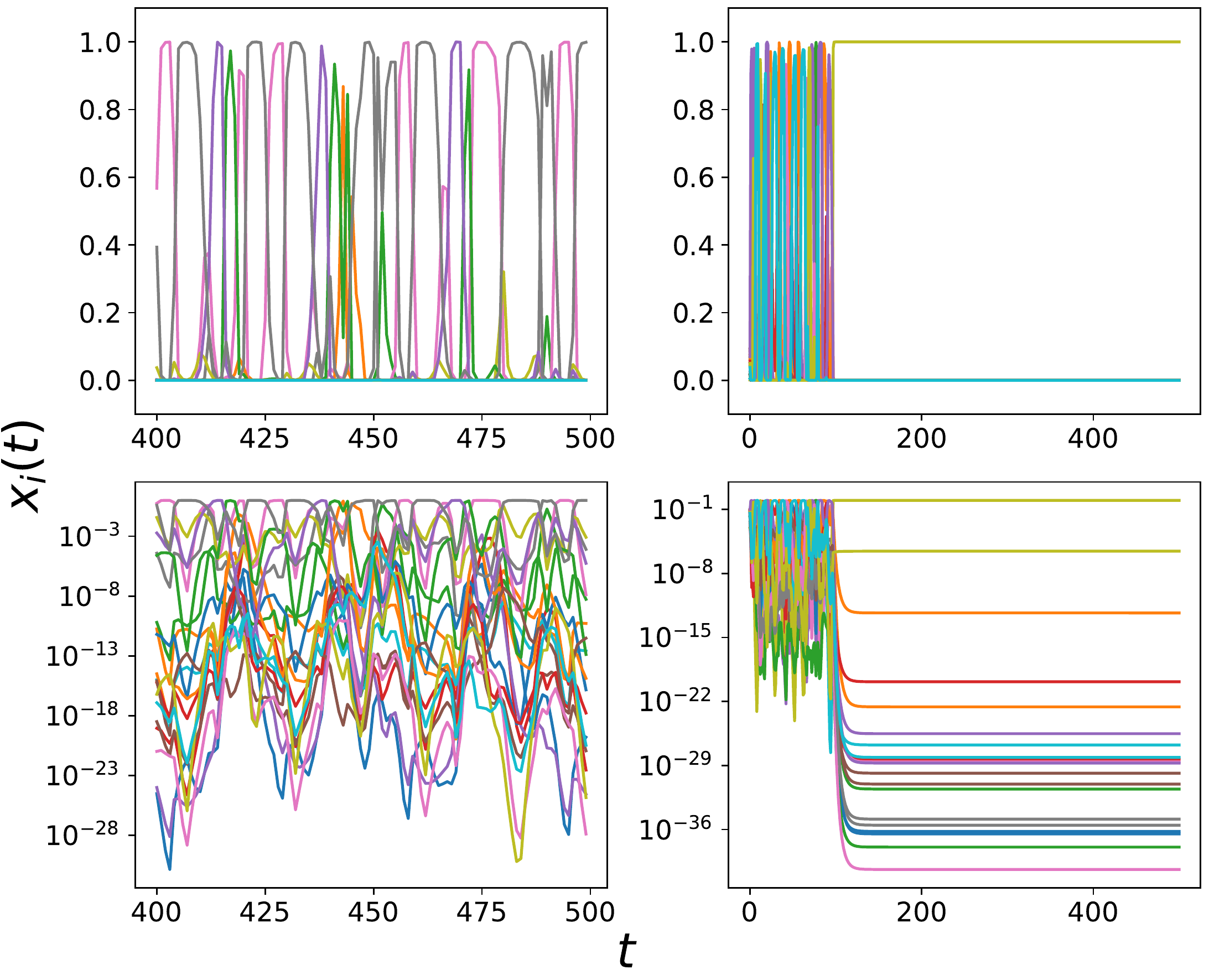}
\caption{Instances of simulation runs of Experience-Weighted Attraction with $N=20$. Each line is a component of the mixed strategy vector of Row (not all components are visible, as they overlap). Left panels: non-converging simulation run. The range of the horizontal axes has been chosen to ease visualization of the dynamics. Right panels: converging simulation run. Top panels: linear scale. Bottom panels: logarithmic scale on the vertical axis.}
\label{fig:exampleEWA}
\end{figure}

Consider Figure \ref{fig:exampleEWA}, right panels. All components of the EWA dynamical system reach a fixed point, differently from the other learning algorithms, so it is easier to identify convergence. We run the EWA dynamics for 500 time steps and we consider the last 20\% time steps to determine convergence. With the parameter values we choose for $\alpha$, $\beta$, $\kappa$ and $\delta$, the transient is usually of the order of 100 time steps, so 500 steps is enough to identify convergence. We then check that the average variance of the logarithms of the components of the mixed strategy vectors does not exceed a certain (very small) threshold. We look at the logarithms because the probabilities following the EWA dynamics vary on an exponential scale and can be of the order of, e.g., $10^{-100}$. In formula, if $1/N \sum_{i=1}^N 5/T \sum_{t=4/5T}^{T} \left( \log x_i(t) \right)^2 > 10^{-2}$ or $1/N \sum_{j=1}^N 5/T \sum_{t=4/5T}^{T} \left( \log y_j(t) \right)^2 > 10^{-2}$, with $T=500$, we identify the simulation run as non-convergent.

\paragraph{Parameter values}

EWA has two main advantages from a computational point of view. First, if the memory loss parameter is positive ($\alpha>0$), all stable attractors of the EWA system lie \textit{within} the probability simplex. This means that no moves are ever given null or unit probability and makes it possible to reliably simulate the EWA map for arbitrarily long time, since for a sufficiently large value of $\alpha$ the machine precision limits are never reached. The intuition for this property is simple: the performance of very successful or very unsuccessful moves is forgotten exponentially over time, so even a very small value of $\alpha$ prompts the players to choose unsuccessful moves with positive probability.  The second advantage is that the EWA system is explicitly normalized every time step, making numerical errors unlikely.

EWA also has a computational disadvantage: because it uses exponential functions to map attractions into probabilities, if the value of the payoff sensitivity $\beta$ is too large, the components of the mixed strategy vector may vary by too many orders of magnitude, and therefore overshoot the boundaries of the mixed strategy simplex. 

So care should be taken in choosing the values of $\alpha$ and $\beta$. This is the case also because of an additional feature of the EWA system: with large memory loss or small payoff sensitivity, the learning dynamics converges to the center of the strategy simplex. In the limit where $\beta=0$ the players just choose uniformly at random between their possible moves, irrespective of the payoff matrix. In Ref. \cite{galla2013complex} it was observed that for sufficiently large values of $\alpha/\beta$ a unique fixed point was always stable. Such a fixed point can be arbitrarily far from mixed strategy Nash equilibria, and so by changing their strategy the players can improve their payoff. We are not interested in this ``trivial'' attractor as we want to focus on the effect of the best reply structure of the payoff matrix on the learning dynamics. Therefore, we choose parameter values for $\alpha$ and $\beta$ that prevent convergence to this fixed point.

A final important technical remark is that we rescale the payoff sensitivity $\beta$ by $\sqrt{N}$ as the payoff matrix gets larger.   
The reason is that the expected payoffs $\sum_j \Pi^R(i,j) y_j$ and $\sum_i \Pi^C(j,i) x_i$ scale as $1/\sqrt{N}$. Indeed, focusing on the expected payoff of player Row, $\sum_j{\Pi^R(i,j)}$ scales as $\sqrt{N}$ due to the Central Limit Theorem (recall that the payoffs are generated randomly, see below for the precise rule), while the components $y_j$ scale as $1/N$ due to the normalization constraint. So  $\sum_j \Pi^R(i,j) y_j$ scales as $1/\sqrt{N}$. The same argument applies to the expected payoff of player Column. Now, note that $\beta$ multiplies the expected payoff from Eqs. \eqref{eq:EWA0} and \eqref{eq:EWA1det}. Therefore, increasing the size of the payoff matrix has the same effect as decreasing $\beta$, until the attractor at the center of the strategy simplex becomes stable again. To prevent this from happening, we rescale $\beta$ by $\sqrt{N}$, so that $\beta \sum_j \Pi^R(i,j) y_j$ and $\beta \sum_i \Pi^C(j,i) x_i$ do not scale with $N$.

For all simulations we choose $\alpha=0.18$, $\beta=\sqrt{N}$, $\kappa=1$ and $\delta=1$, which ensure that the EWA dynamics stays within the probability simplex, that it does not overshoot the simplex boundaries and that it does not reach the trivial attractor in the center of the simplex.

\subsubsection{Experience-Weighted Attraction with noise}
\label{sec:simEWAnoise}

So far we have assumed batch learning. Here we consider online learning, i.e. the players update their mixed strategies after observing a single move by their opponent. The players choose a move with probability given by their mixed strategy vector. We focus on EWA because of its superior numerical properties (as compared to the other algorithms). Given that introducing noise makes identifying convergence more challenging, we choose the algorithm for which identifying convergence has been simplest.

\paragraph{Formal definition}

EWA with noise is simply given by Eqs. \eqref{eq:EWA0}, \eqref{eq:EWA1} and \eqref{eq:Nt}. At time $t$, player Row selects move $i$ with probability $x_i(t)$ and player Column selects move $j$ with probability $y_j(t)$.

\paragraph{Convergence criteria}

\begin{figure}[ht]
\centering
\includegraphics[width=.75\textwidth]{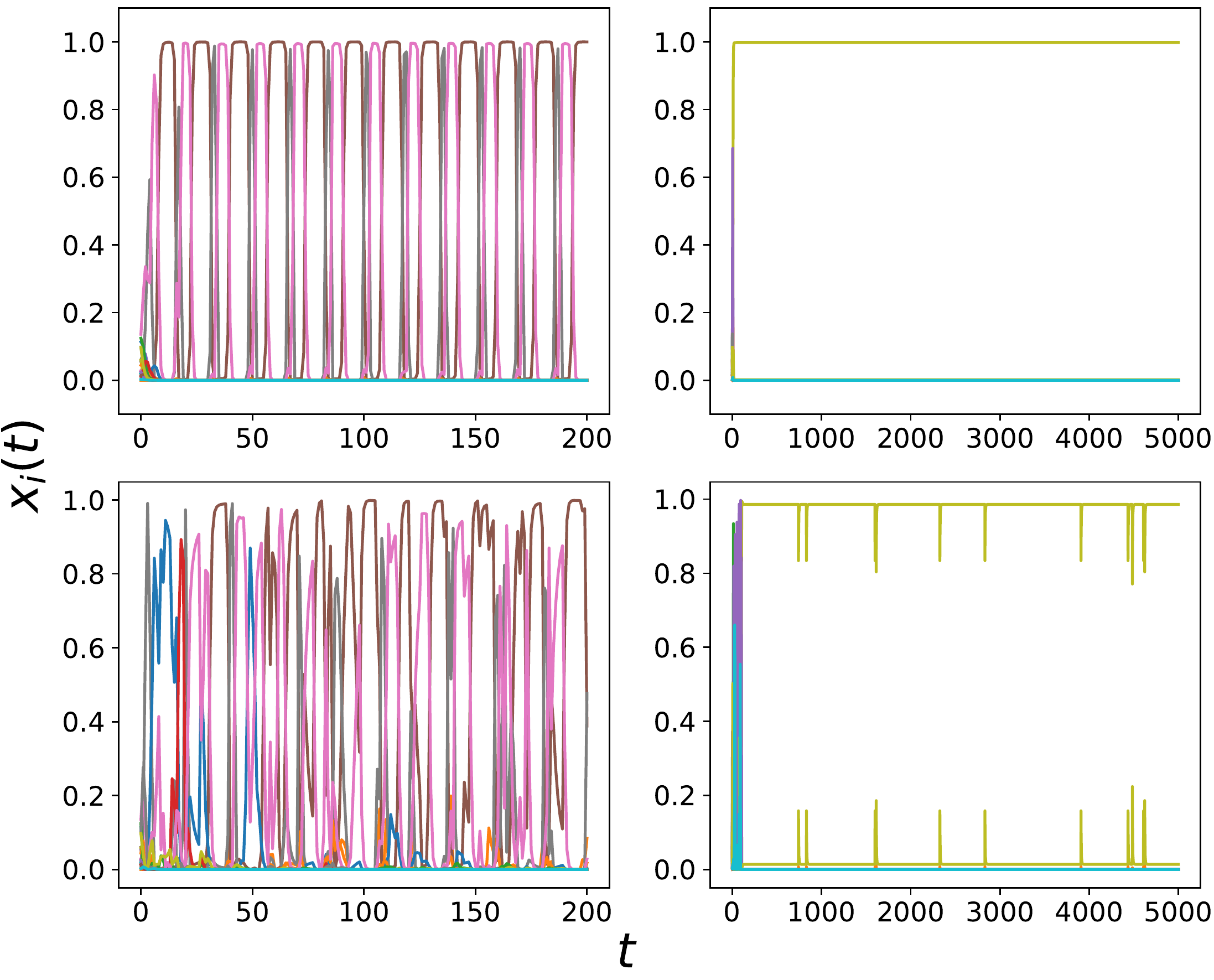}
\caption{Instances of simulation runs of EWA and EWA with noise with $N=20$. Each line is a component of the mixed strategy vector of Row (not all components are visible, as they overlap). Left panels: non-converging simulation run. Right panels: converging simulation run. Top panels: deterministic approximation of EWA. Bottom panels: EWA with noise.}
\label{fig:exampleEWAwithnoise}
\end{figure}

As can be seen in Figure \ref{fig:exampleEWAwithnoise}, the deterministic approximation of EWA and the noisy version are generally very similar. In the convergent example a move which is not the most commonly played one (i.e. the light green line) is selected from time to time, and this potentially pulls player Row away from equilibrium. What usually occurs instead is that the player returns to equilibrium after a short time.

We use the following convergence heuristic:
\begin{enumerate}
\item Only consider the last 20\% time steps.
\item Only keep the moves that have been played with a frequency larger than $1/N$.
\item Find the most common value of the probabilities, i.e. the fixed point.
\item Count the occurrences in which the probabilities are farther than 0.02 from the most common value.
\item If the occurrences are more than 10\% of the considered time interval, identify the simulation run as non-convergent. Otherwise, identify it as convergent.
\end{enumerate}

\paragraph{Parameter values}

Differently from the case of deterministic EWA, we need to consider a longer time interval for the dynamics to settle down to an attractor. We take 5000 iterations as a maximum, as for Bush-Mosteller dynamics and fictitious play. The values of the parameters are the same, except for the intensity of choice: we take $\beta=\sqrt{N}/2$. The reason why we reduce the intensity of choice is that $\beta=\sqrt{N}$ leads the dynamics too close to the boundaries of the strategy simplex, and noise almost disappears. Indeed, if the dominant strategy is played with probability e.g. $x_i(t) = 0.99995$, deviations from equilibrium are extremely unlikely, and we recover the deterministic case. 

\subsubsection{Level-k learning}
\label{sec:simLk}

We refer to level-$k$ learning as a generalization of anticipatory learning (proposed by Selten \cite{selten1991anticipatorysupp}). Selten assumed that player Row does not believe that Column would behave as she did in the past. He rather tries to outsmart her by best replying to the strategy that he thinks she will play on the following time step. Row needs a forecast of her strategy, and obtains it by assuming that Column is an EWA learner.

This idea can be generalized by assuming that the players can think $k$ steps ahead \cite{lecutier2013stochastic,evans2013klevel}. In level-$k$ thinking \cite{nagel1995unravelingsupp,crawford2013structural} $k$-players assume that the other players are level $k-1$, and the process is iterated down to level 1. Level-1 players choose randomly. Level-2 players know that level-1 players choose randomly, and select the strategy that yields the highest payoff given this piece of information. Level-3 players know how level-2 players behave, and react accordingly, and so on. 

In our case, level-1 players are EWA learners. Level-2 players know that level-1 players update their strategies using EWA, and try to get a better payoff by pre-empting their opponent's move. Level-3 players would know how level-2 players choose their strategy, and select the best possible strategy in response. Here we will assume that both players are level-2, as we did not find a substantial difference with larger values of $k$ (which quickly become behaviorally implausible).

\paragraph{Formal definition}

For convenience, we combine Eqs. \eqref{eq:EWA0} and \eqref{eq:EWA1det}:
\begin{equation}
x_i^1(t+1)=\frac{x_i(t)^{(1-\alpha)\mathcal N (t)/\mathcal N (t+1)}\exp{\left(\beta(\delta + (1-\delta) x_i(t)) \sum_j \Pi^R(i,j) y_j(t)/\mathcal N (t+1)\right)}}{Z_x(t+1)},
\label{eq:levelk}
\end{equation}
with $Z_x(t+1)=\sum_l x_l(t)^{(1-\alpha)\mathcal N (t)/\mathcal N (t+1)}\exp{\left(\beta(\delta + (1-\delta) x_l(t)) \sum_j \Pi^R(l,j) y_j(t)/\mathcal N (t+1)\right)}$. We are using superscript 1 to indicate that player Row is a level-1 (i.e. an EWA) learner. A similar expression holds for Column.

We denote the right-hand side in Eq. \eqref{eq:levelk} by $F(y(t))$, with $y(t)=(y_1(t),\ldots y_N(t))$. So, $x_i^1(t+1)=F(y(t))$. Player Row learns based on the past mixed strategy vector of Column. We define
\begin{equation}
y_j^2(t+1)=F(x^1(t+1)).
\label{eq:level2}
\end{equation}
Here Column is a level-2 player as she believes that Row is a level-1 player, and therefore updates his strategies using Eq. \eqref{eq:levelk}. In general,
\begin{equation}
y_j^k(t+1)=F(x^{k-1}(t+1)).
\end{equation}

\paragraph{Convergence criteria}
The dynamics is qualitatively very similar to EWA, so we use the same convergence criteria.

\paragraph{Parameter values}
We also use the same parameter values. Both Row and Column are level-2 players.

\subsection{Initialization of the payoff matrices}
\label{sec:payoffmatrix}

In order to study generic payoff matrices, we sample the space of all possible payoff matrices by generating the payoff elements at random. 
Following Ref. \cite{galla2013complex}, at initialization we randomly generate $N^2$ pairs of payoffs (i.e., if Row plays $i$ and Column plays $j$, a pair $a,b$ implies that Row gets $a$, Column gets $b$), and we keep the payoff matrix fixed for the rest of the simulation (so the system described by the payoff matrix can be thought of as \textit{quenched}). We consider an \textit{ensemble} of payoff matrices constrained by the mean, variance and correlation of the pairs. The Maximum Entropy distribution that obeys these constraints is a bivariate Gaussian \cite{galla2013complex}, which we parametrize with zero mean, unit variance and correlation $\Gamma$. Therefore, $\Gamma<0$ implies that the game is competitive (zero-sum in the extreme case where $\Gamma=-1$), while $\Gamma>0$ encourages cooperation (see the main text).  If $\Gamma=0$ all best reply configurations are equiprobable because the payoffs are chosen \textit{independently} at random, so we shall consider this as a benchmark case where we sample the space of all possible games with equal probability. 

Fig. 2 of the main paper: We generate 1000 payoff matrices at random with $\Gamma=0$ and $N=20$, starting from 100 random initial conditions for each payoff matrix. 
 
Fig. 3 of the main paper, top panel: We generate 180 payoff matrices at random with $\Gamma=0$, starting from 10 random initial conditions for each payoff matrix, for the following numbers of moves: $N=\{2,3,4,5,8,10,15,20,30,$ $50,100,200,400\}$. We sensibly reduce the number of simulation runs per value of $N$ because the random generation of the payoff matrix, the identification of the best reply structure and the simulations of the dynamics are time consuming for $N \geq 50$.

Fig. 3 of the main paper, bottom panel: Same as top panel, but we consider correlations $\Gamma=\{-1.0,-0.9,-0.8,\ldots,$ $0.0,0.1,\ldots, 0.9, 1.0\}$ and only 50 payoff matrices for each value of $\Gamma$. 

Fig. 4 of the main paper: same as Fig. 3, top panel.

\section{Supplementary numerical results}
\label{sec:suppnum}

\begin{figure}[ht]
\centering
\includegraphics[width=.9\textwidth]{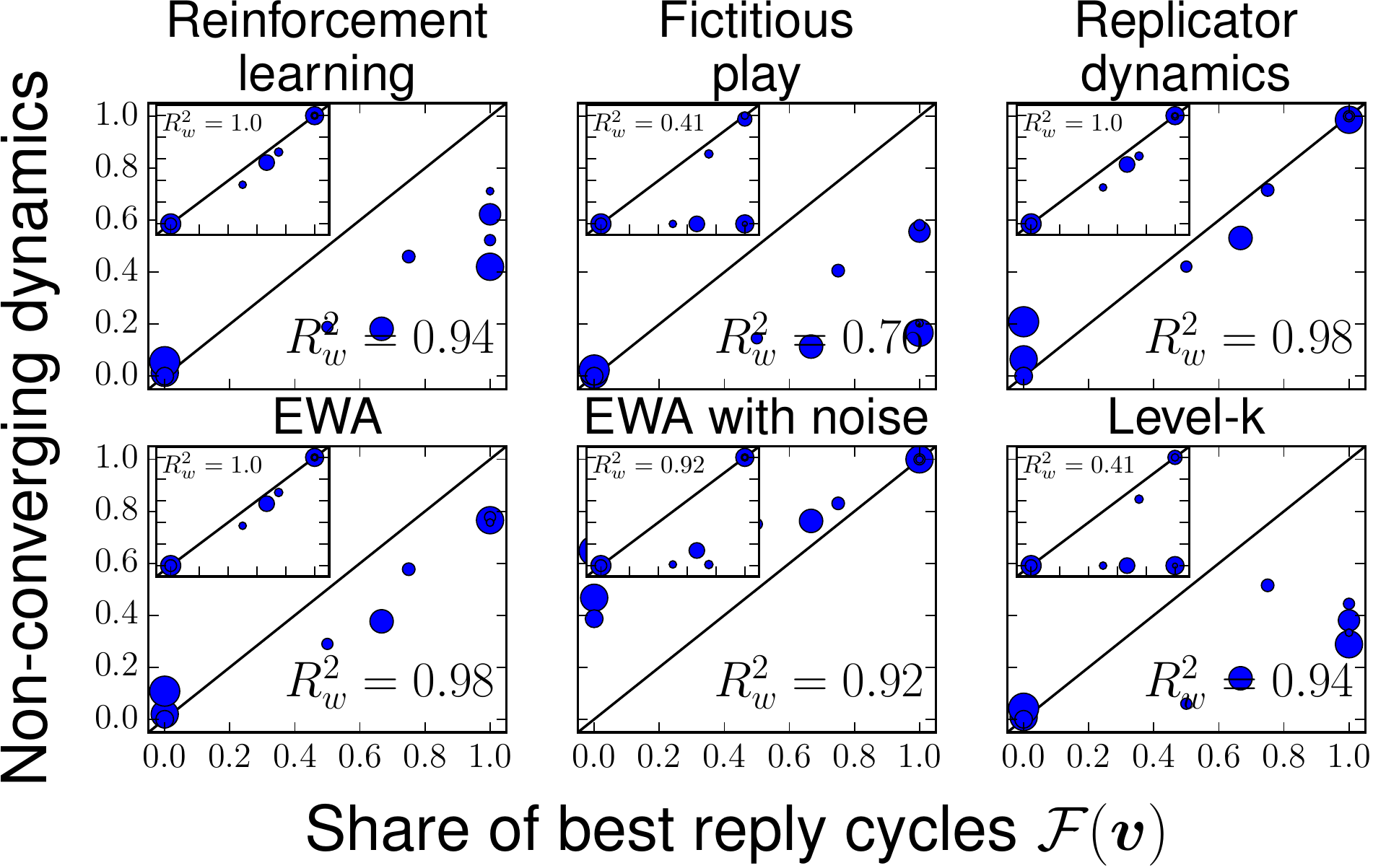}
\caption{Same as Fig. 2 of the main paper, but with $N=5$ instead of $N=20$.}
\label{fig:fig2suppN5}
\end{figure}

\begin{figure}[ht]
\centering
\includegraphics[width=.9\textwidth]{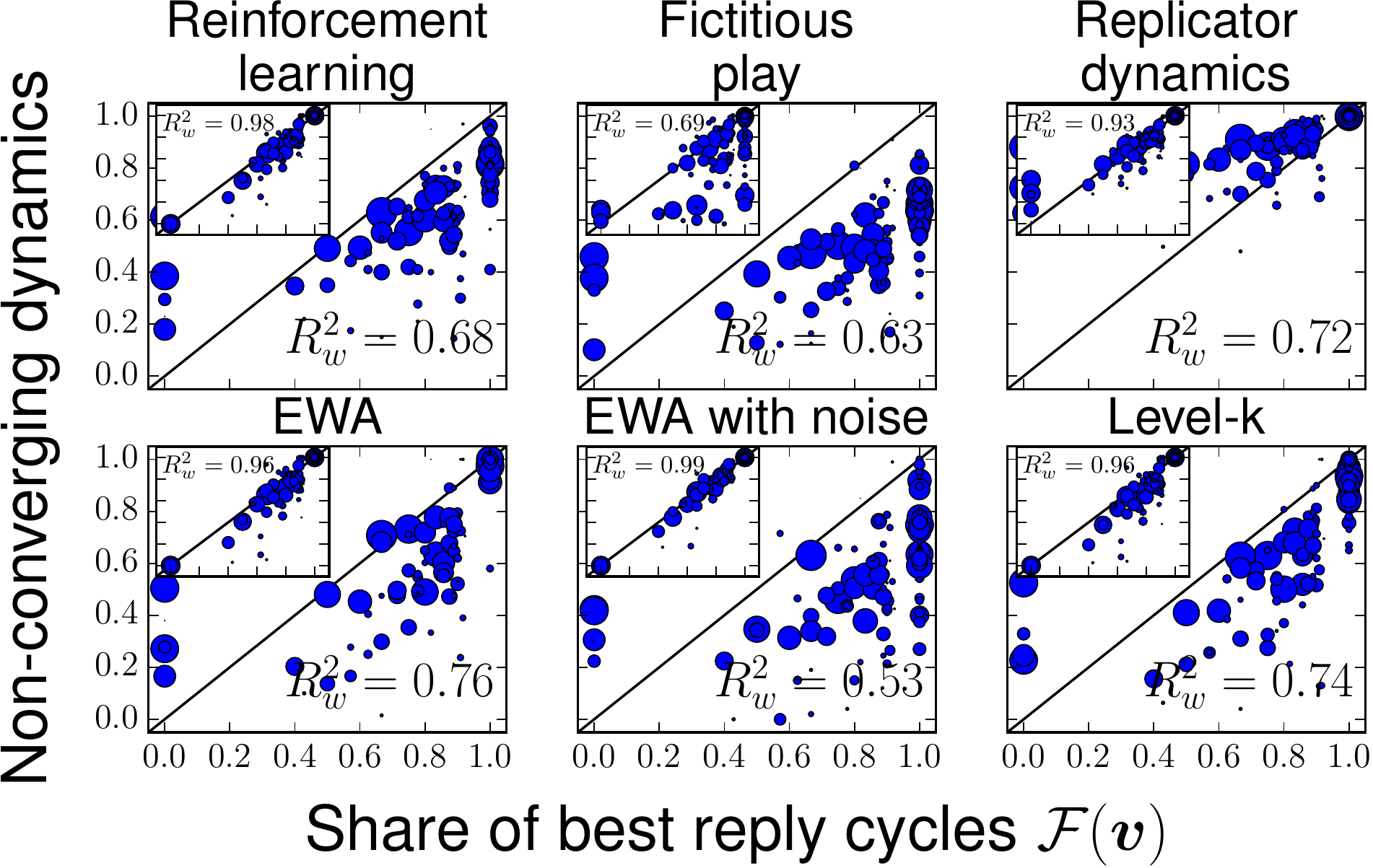}
\caption{Same as Fig. 2 of the main paper, but with $N=50$ instead of $N=20$.}
\label{fig:fig2suppN50}
\end{figure}

In this section we first perform a few robustness tests with respect to the numerical findings in the main paper. We then present a few additional results regarding the heterogeneity of the learning algorithms and the correlation between Boolean and non-Boolean payoff matrices.

For what concerns the robustness tests, we check whether we get the same results as in Fig. 2 of the main paper, once we consider a different number of moves $N$. As can be seen in Figures \ref{fig:fig2suppN5} and \ref{fig:fig2suppN50}, the overall pattern is similar, but there are some differences. We are plotting the fraction of non-convergence of best reply dynamics, as given by the relative share of best reply cycles $\mathcal{F}(\boldsymbol{v})$, on the horizontal axis. The fraction of non-converging simulation runs for the six learning algorithms we have been considering is on the vertical axis. 

For $N=5$ the correlation is stronger than with $N=20$, and the values of the weighted correlation coefficient are even larger than 0.9 in non-Boolean payoff matrices. We conjecture that this is due to a higher share of the moves that are part of cycles and fixed points. Indeed, for $N=5$ the most common best reply vector with cycles is  $\boldsymbol{v}=(0,0,0,1,0)$, so the moves that are part of the cycle are 2/5. On the other hand, in a best reply vector with a 2-cycle and $N=20$ the moves that are part of the cycle are 2/20, so the payoffs that are not best replies have more importance and the issue of \textit{quasi-best replies} is more severe.

An interesting detail is that level-$k$ learning converges in most cases. Inspection of individual simulation runs suggests that by anticipating the moves of their opponent, the players are less likely to get stuck in periodic cycles and converge instead to mixed strategy equilibria.

For $N=50$ we observe the opposite pattern than with $N=5$: the correlation becomes weaker (but still larger than 0.6 in most cases). This effect is most likely caused by a smaller share of moves that are part of cycles or fixed points (the most common best reply vector is $\boldsymbol{v}=(0,\ldots 0, 1,1)$, involving only 3/50 of the moves). Quasi-best replies probably play a more important role. However, we cannot exclude measurement error.

\begin{figure}[ht]
\centering
\includegraphics[width=.6\textwidth]{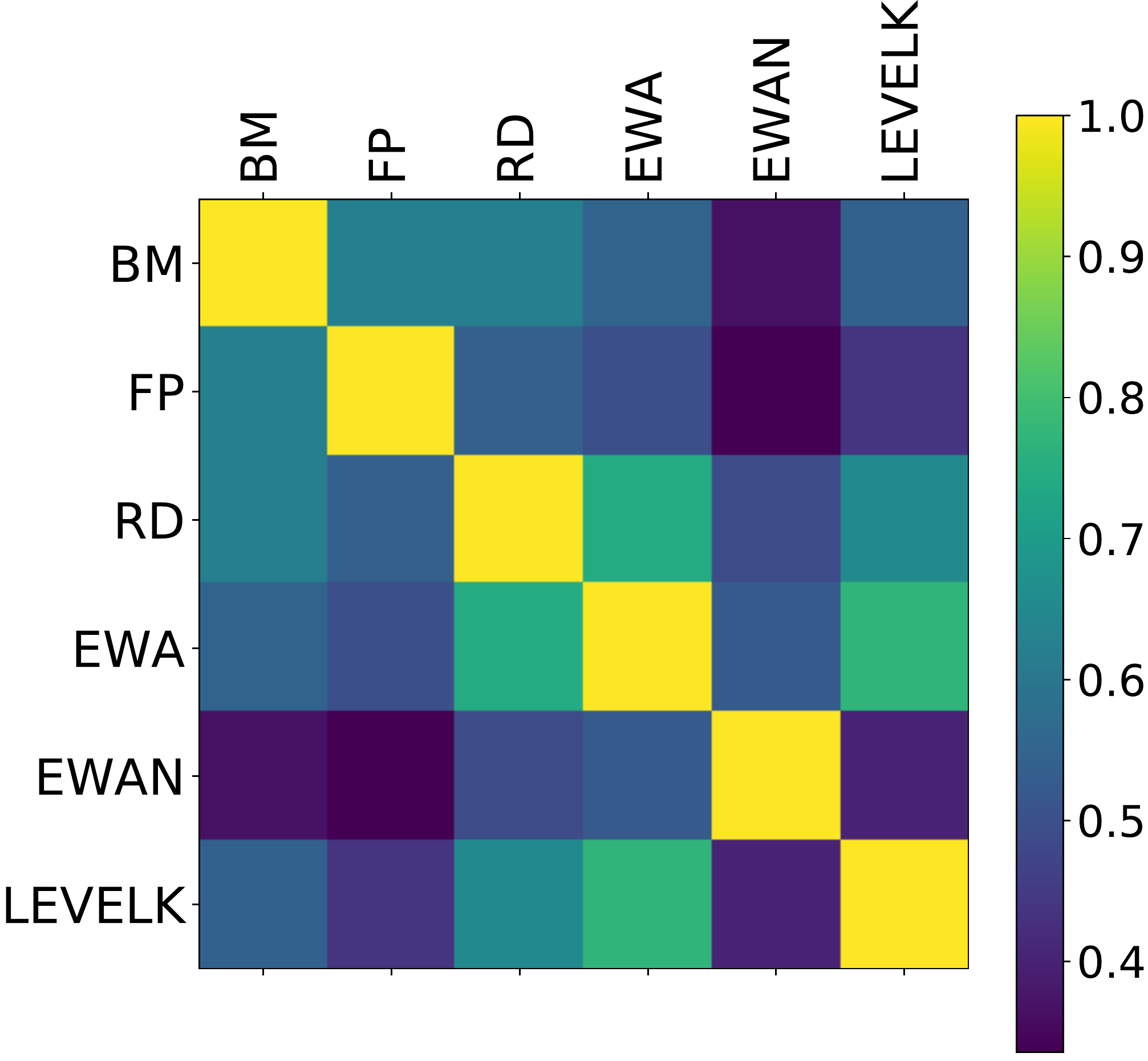}
\caption{Correlation matrix of the co-occurrence of non-convergence in any of the 1000 payoff matrices among Bush-Mosteller (BM) learning, Fictitious Play (FP), Replicator Dynamics (RD), Experience-Weighted Attraction (EWA), EWA with Noise (EWAN) and level-k learning (LEVELK).}
\label{fig:corr_learning_algorithms}
\end{figure}

In Figure \ref{fig:corr_learning_algorithms} we show the correlation matrix of the co-occurrence of convergence of the six learning algorithms we have considered. For each of the 1000 payoff matrices that were sampled for $N=20$, and for each learning algorithm, we calculate the frequency of non-convergence. Therefore, we have six vectors of 1000 components, and we consider the correlation among them. Perfect correlation would mean that for each payoff matrix the non-convergence rate is identical. 

We find that the three most correlated algorithms are replicator dynamics, Experience-Weighted Attraction (EWA) and level-k learning. The two least correlated algorithms are fictitious play and EWA with noise. The correlation ranges between 0.35 and 0.85, suggesting a relatively strong heterogeneity between the six algorithms.

Finally, in Table \ref{tab:corr_boolean_nonboolean} we show the correlation between the co-occurrence of convergence in Boolean and non-Boolean payoff matrices. As before, we consider vectors of 1000 components, in which each component is the frequency of non-convergence in a specific payoff matrix. The correlations are obtained from the pairwise comparison between the vectors referring to Boolean and non-Boolean payoff matrices.

As Boolean payoff matrices are constructed to have the same best reply structure as their non-Boolean counterpart, lack of perfect correlation is due to the details of the payoffs. Interestingly, correlation is very low in the case of fictitious play, whereas it is relatively high with replicator dynamics and EWA.

\begin{table}[htbp]
	\centering
		\begin{tabular}{|c|c|c|c|c|c|c|}
			\hline
			 BM & FP & RD & EWA & EWAN & LEVELK & mean \\
			 0.49 & 0.35 & 0.65 & 0.61 & 0.46 & 0.52 & 0.51  \\
			\hline						
		\end{tabular}
	\caption{Correlation between the co-occurrence of non-convergence in Boolean and non-Boolean payoff matrices, for the six learning algorithms we have considered.}
		\label{tab:corr_boolean_nonboolean}
\end{table}

\FloatBarrier

\section{Details of the analytical calculations}
\label{sec:analytical}

First, we provide a thorough derivation of the expression for the frequency of best reply vectors, and use it on some examples. Second, we obtain additional expressions that quantify the fraction of payoff matrices with at least one cycle of any given length (including fixed points, which are cycles of length one), and use these equations to find the share of payoff matrices with no fixed points or at least one cycle. Third, we derive asymptotic estimates for the frequency of cycles and fixed points in infinite dimensional payoff matrices.

\subsection{Frequency of best reply vectors}
\label{sec:freqbrv}

 We first discuss the count of the ways to form $k$-cycles and fixed points of best reply dynamics, and then we count the ways to place the free best replies (i.e. those that are not part of either cycles or fixed points). Finally we show how we combine these numbers together to obtain the count of best reply configurations that correspond to a specific set of attractors. 

\begin{figure}[htbp]
        \centering
        \begin{subfigure}[b]{0.3\textwidth}
                \includegraphics[width=\textwidth]{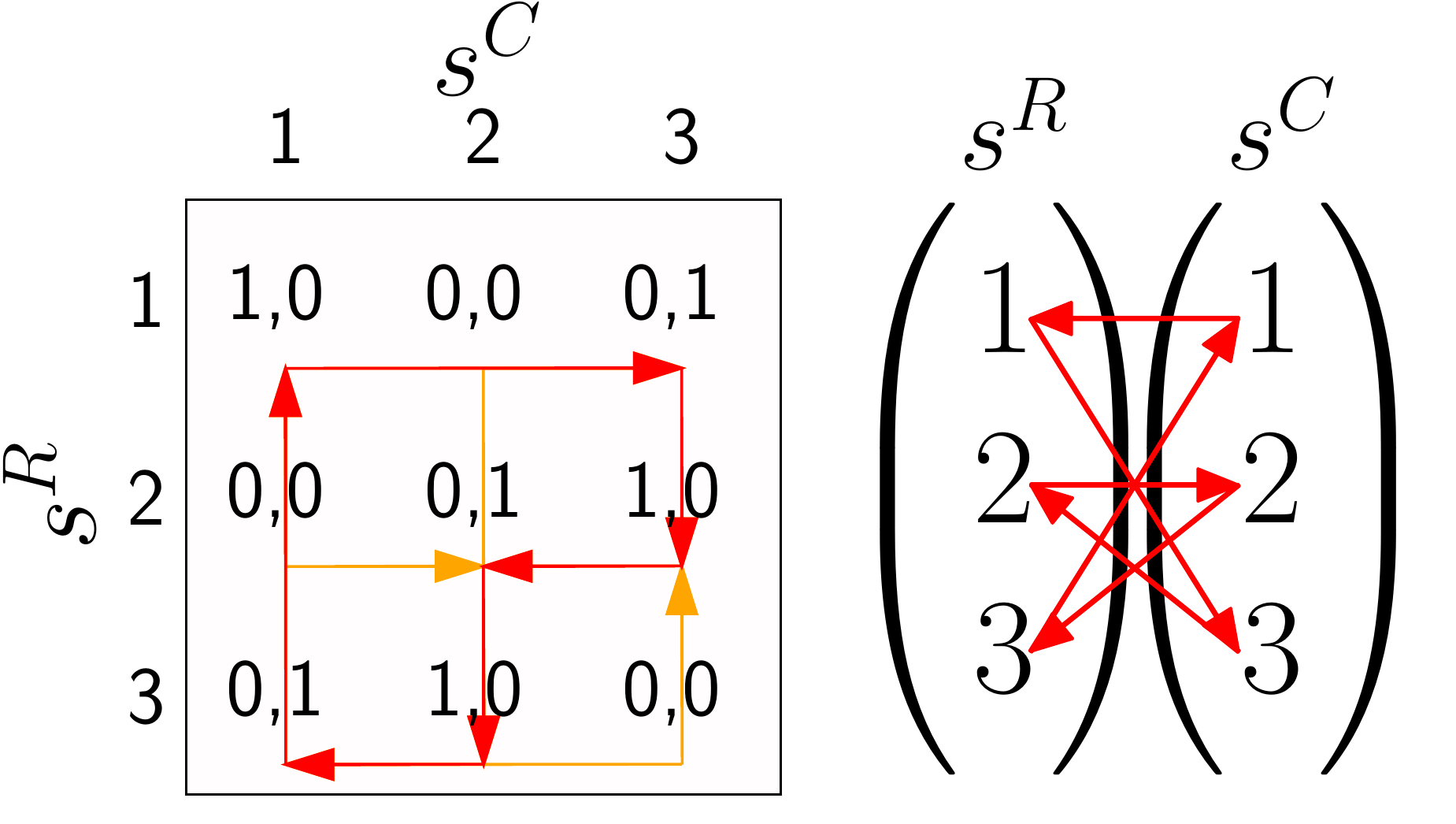}
                \label{fig:supp1a}
        \end{subfigure}
        ~ 
        \begin{subfigure}[b]{0.3\textwidth}
                \includegraphics[width=\textwidth]{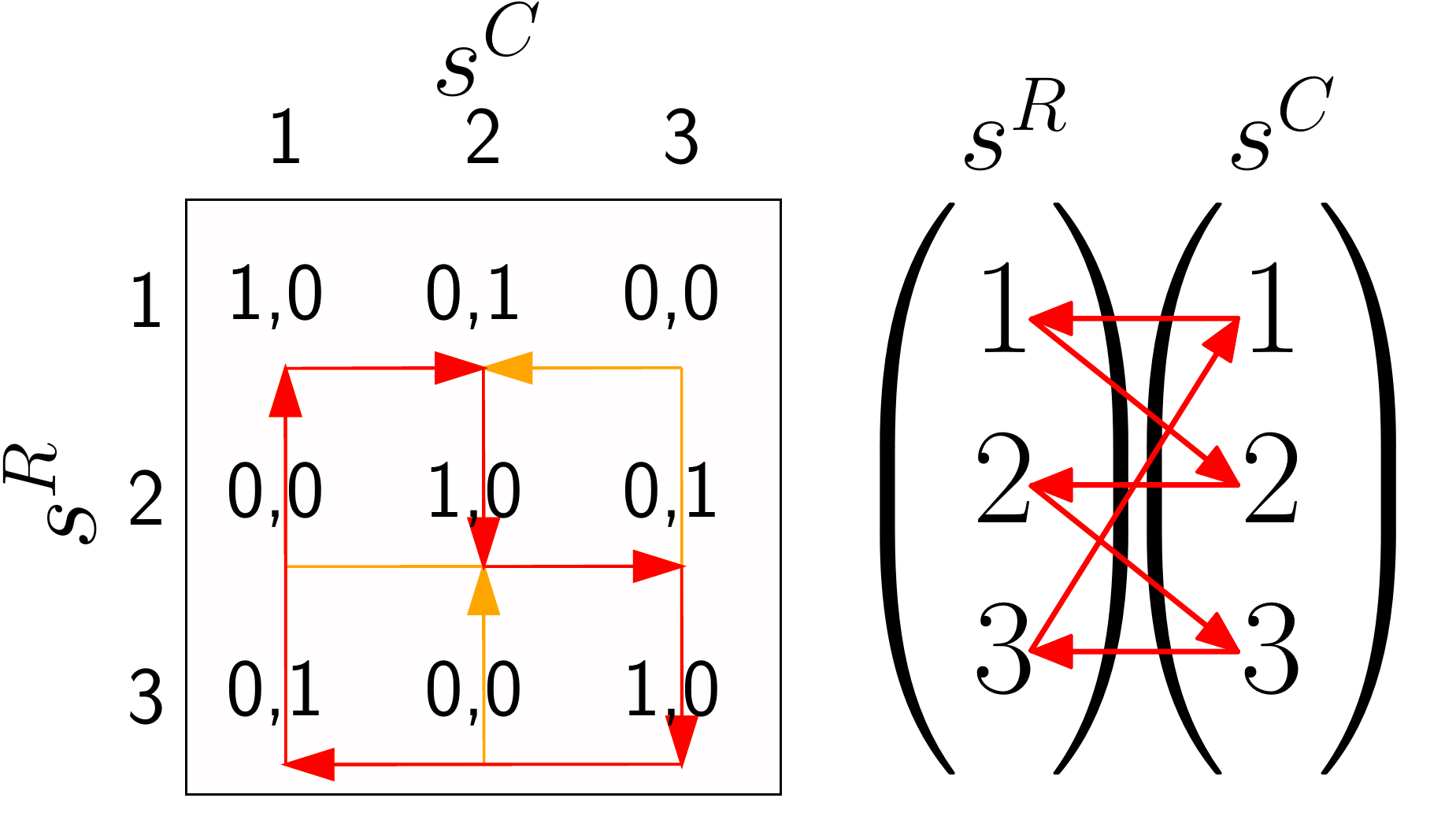}
                \label{fig:supp1b}
        \end{subfigure}  
        ~ 
        \begin{subfigure}[b]{0.3\textwidth}
                \includegraphics[width=\textwidth]{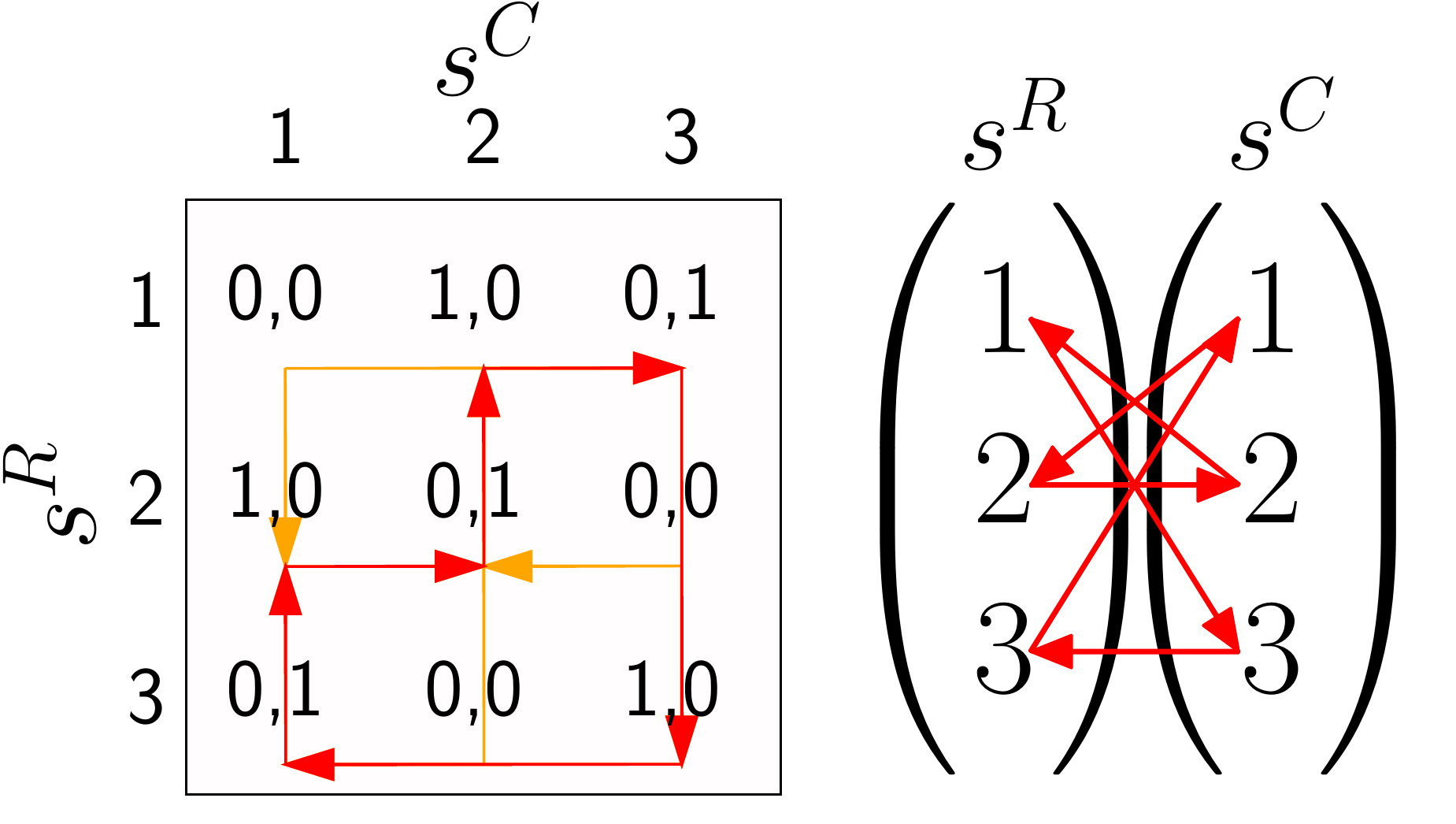}
                \label{fig:supp1c}
        \end{subfigure} 
        ~
        \begin{subfigure}[b]{0.3\textwidth}
                \includegraphics[width=\textwidth]{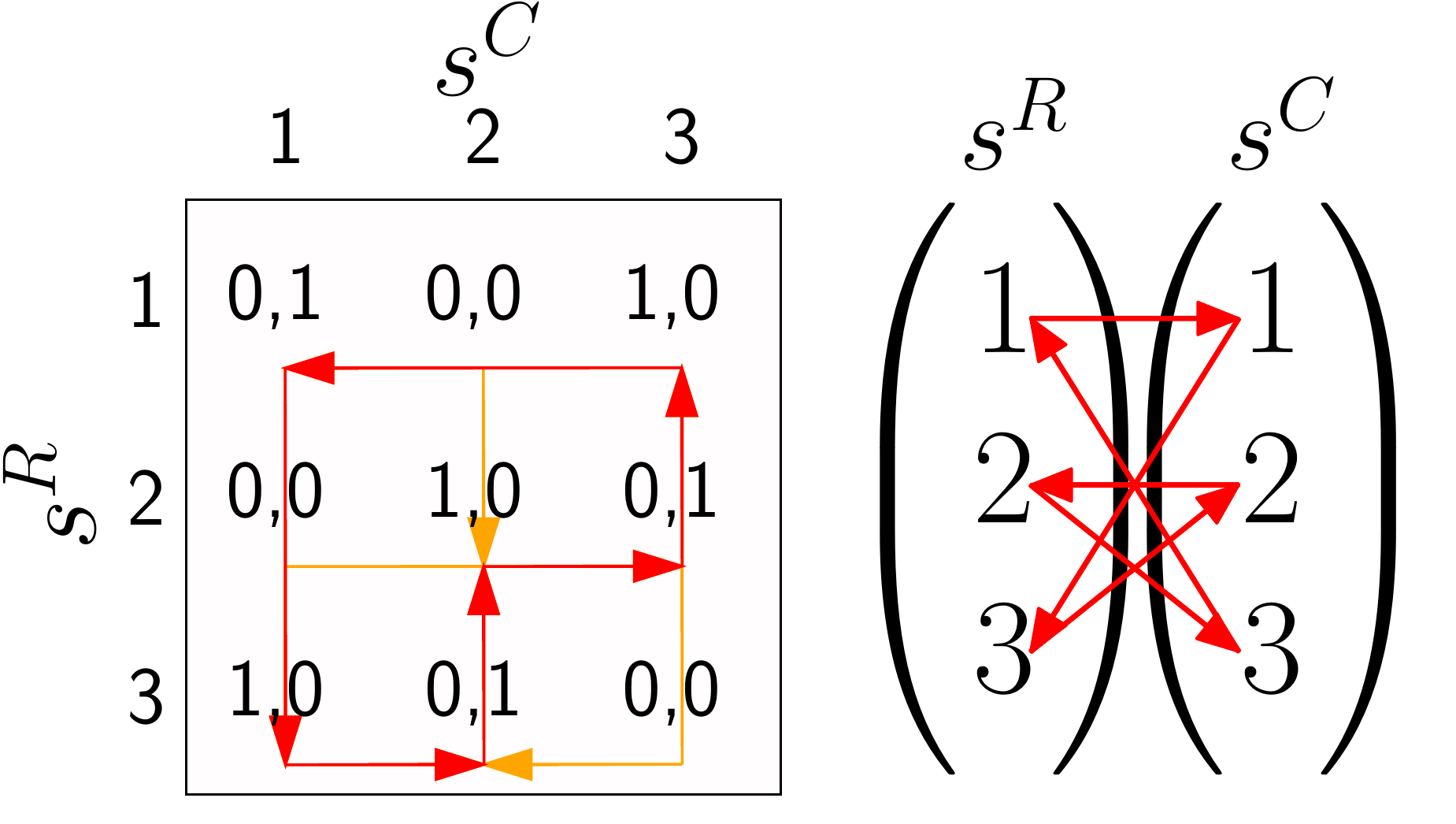}
                \label{fig:supp1d}
        \end{subfigure}
        ~ 
        \begin{subfigure}[b]{0.3\textwidth}
                \includegraphics[width=\textwidth]{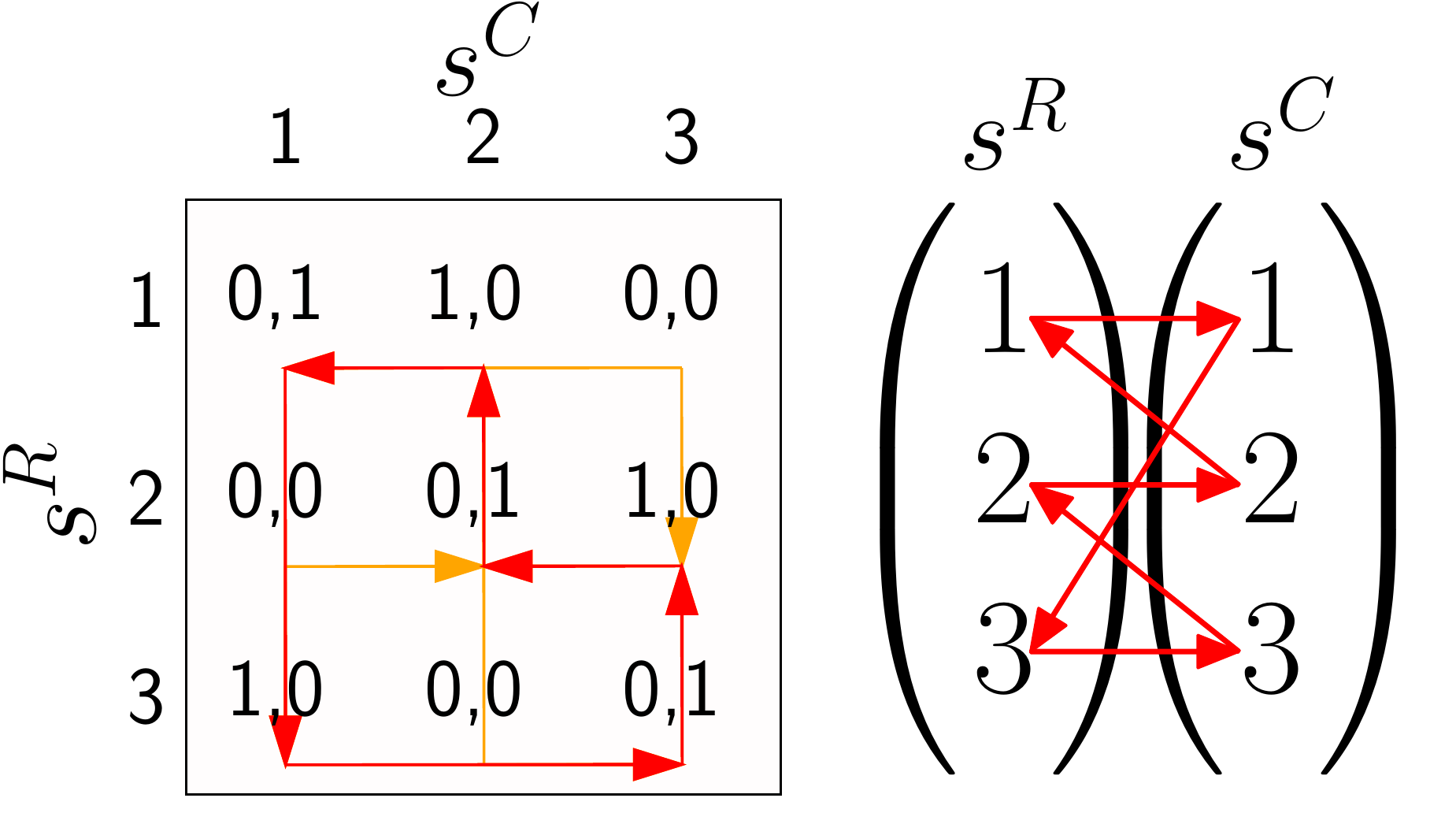}
                \label{fig:supp1e}
        \end{subfigure}  
        ~ 
        \begin{subfigure}[b]{0.3\textwidth}
                \includegraphics[width=\textwidth]{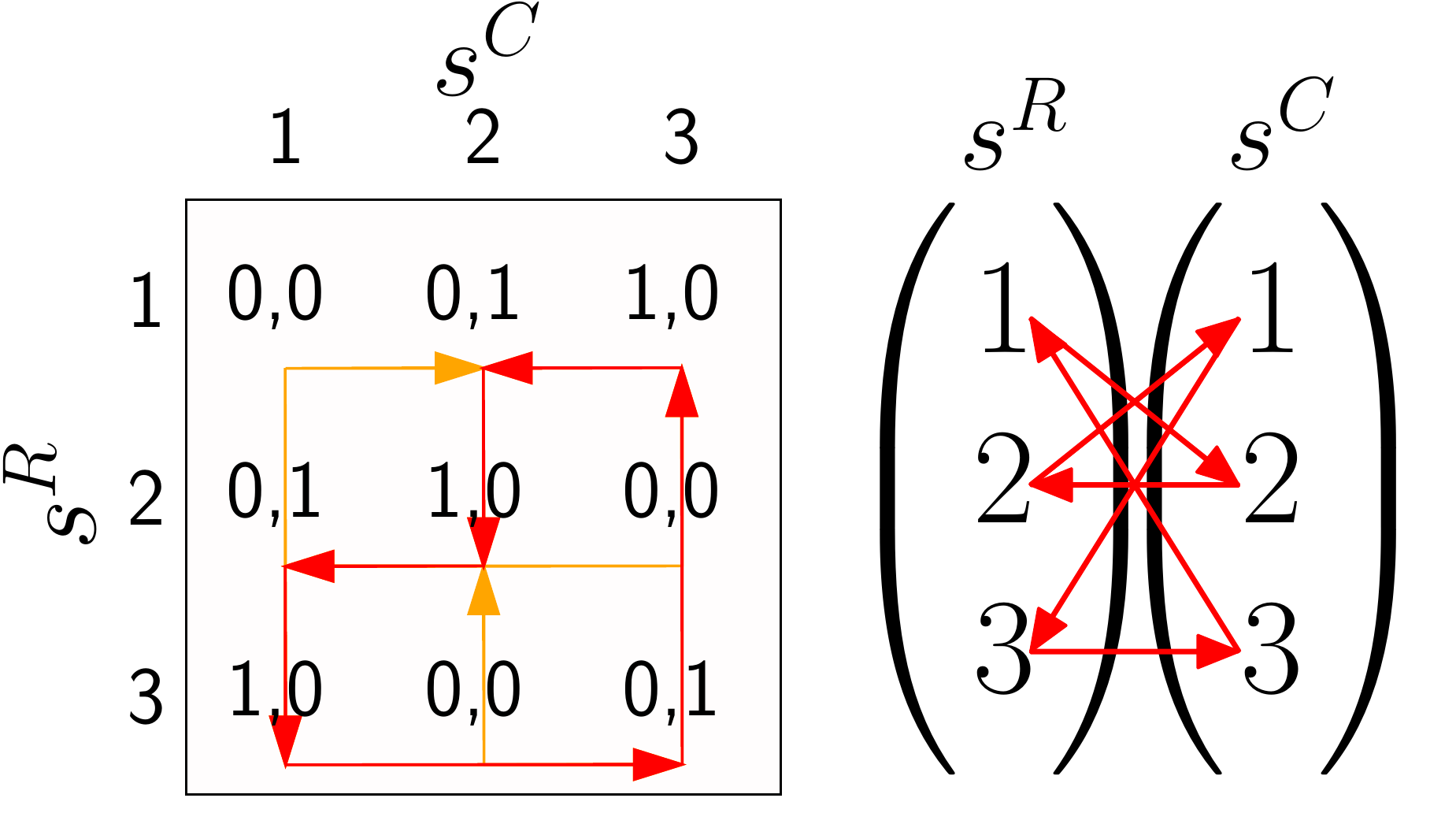}
                \label{fig:supp1f}
        \end{subfigure} 
        ~
        \begin{subfigure}[b]{0.3\textwidth}
                \includegraphics[width=\textwidth]{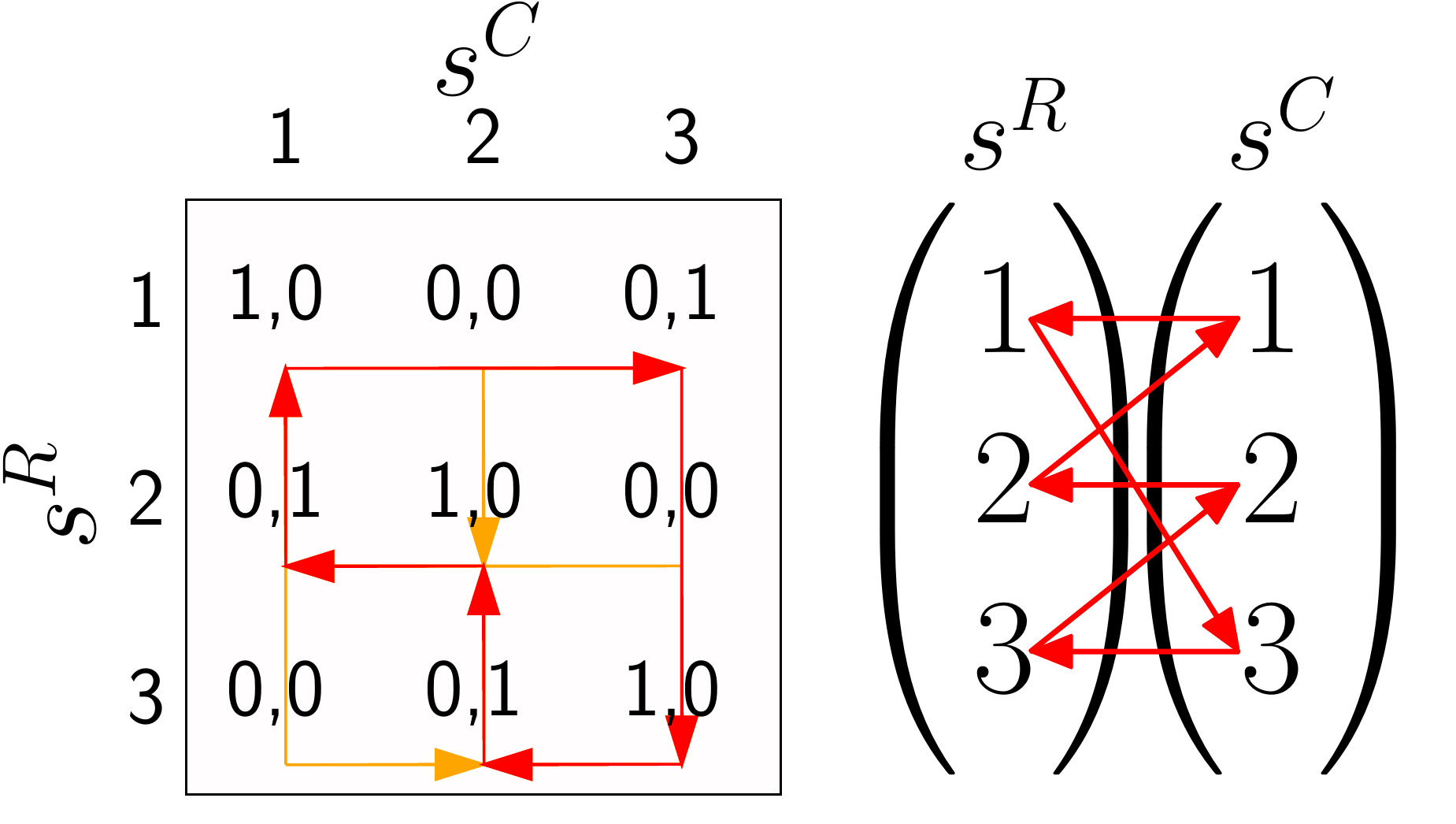}
                \label{fig:supp1g}
        \end{subfigure}
        ~ 
        \begin{subfigure}[b]{0.3\textwidth}
                \includegraphics[width=\textwidth]{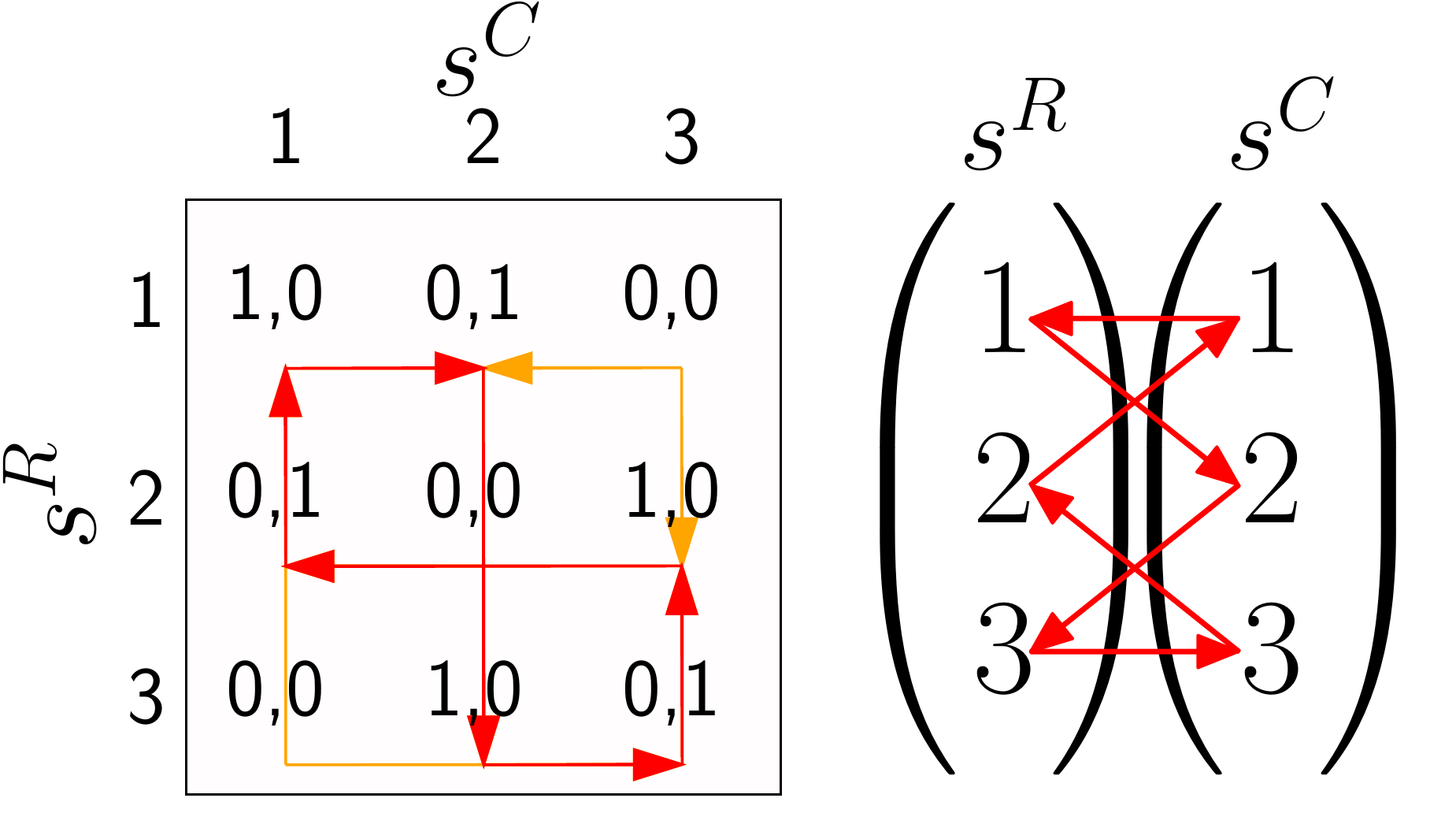}
                \label{fig:supp1h}
        \end{subfigure}  
        ~ 
        \begin{subfigure}[b]{0.3\textwidth}
                \includegraphics[width=\textwidth]{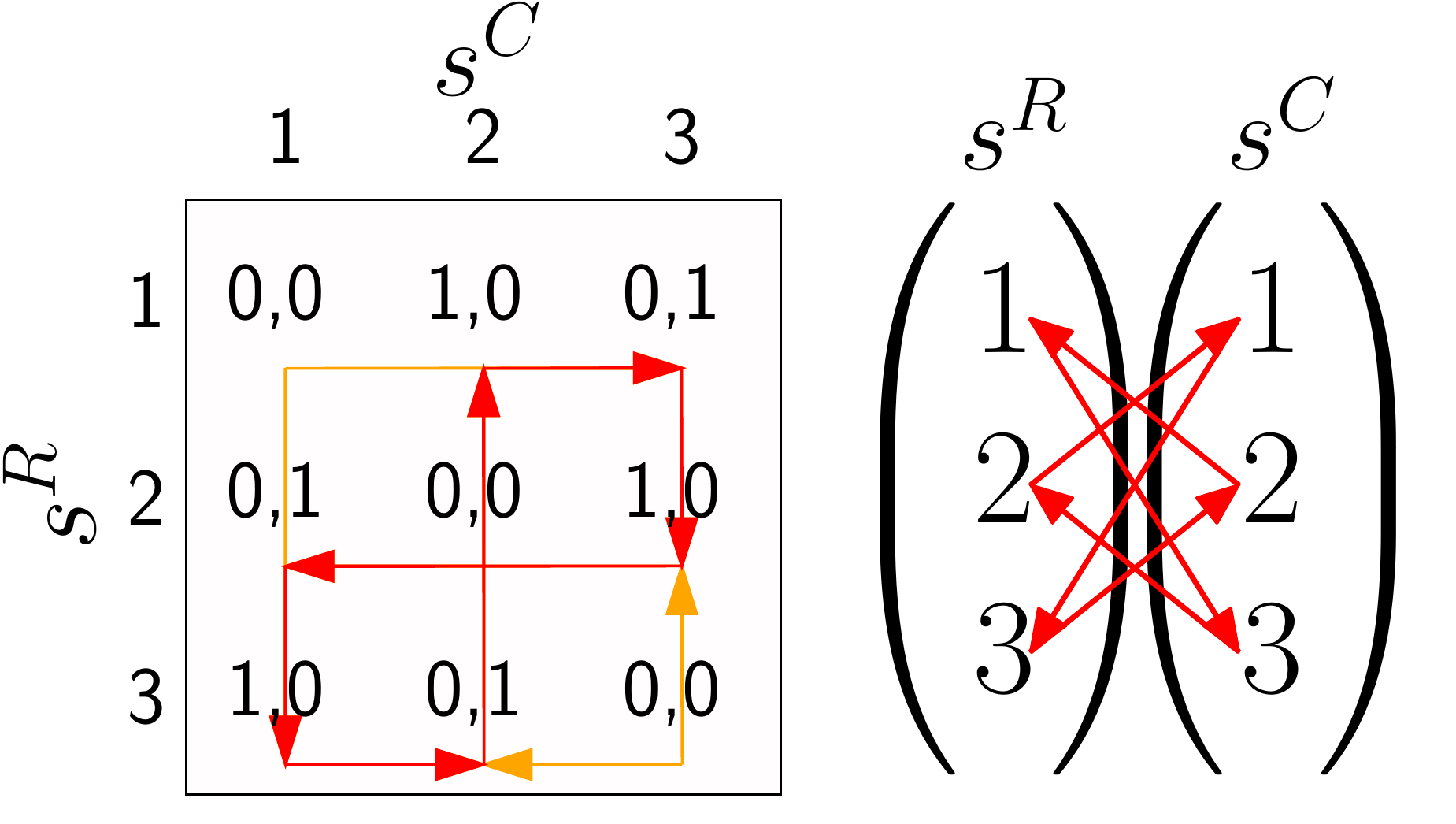}
                \label{fig:supp1i}
        \end{subfigure} 
        ~
        \begin{subfigure}[b]{0.3\textwidth}
                \includegraphics[width=\textwidth]{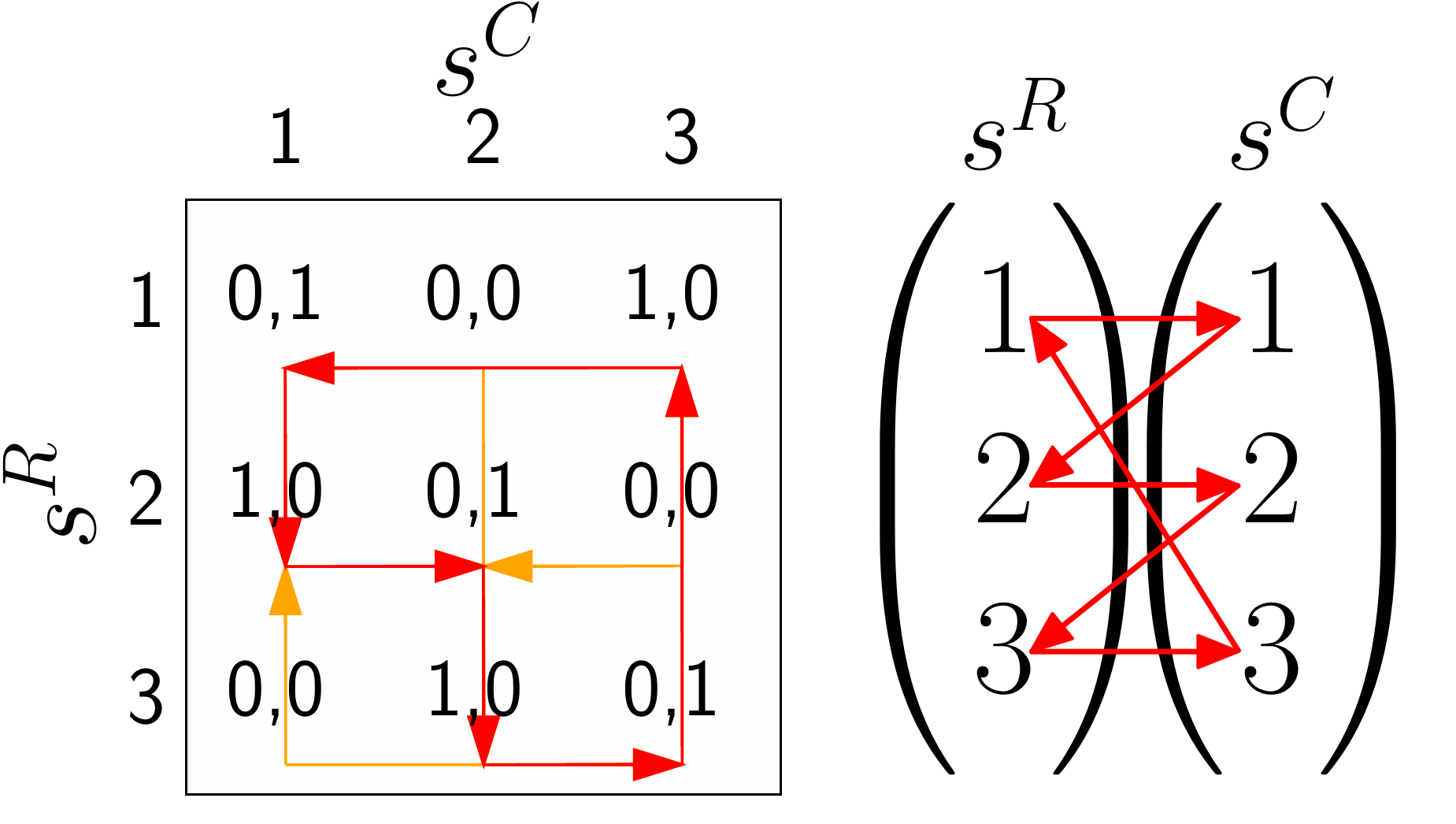}
                \label{fig:supp1l}
        \end{subfigure}
        ~ 
        \begin{subfigure}[b]{0.3\textwidth}
                \includegraphics[width=\textwidth]{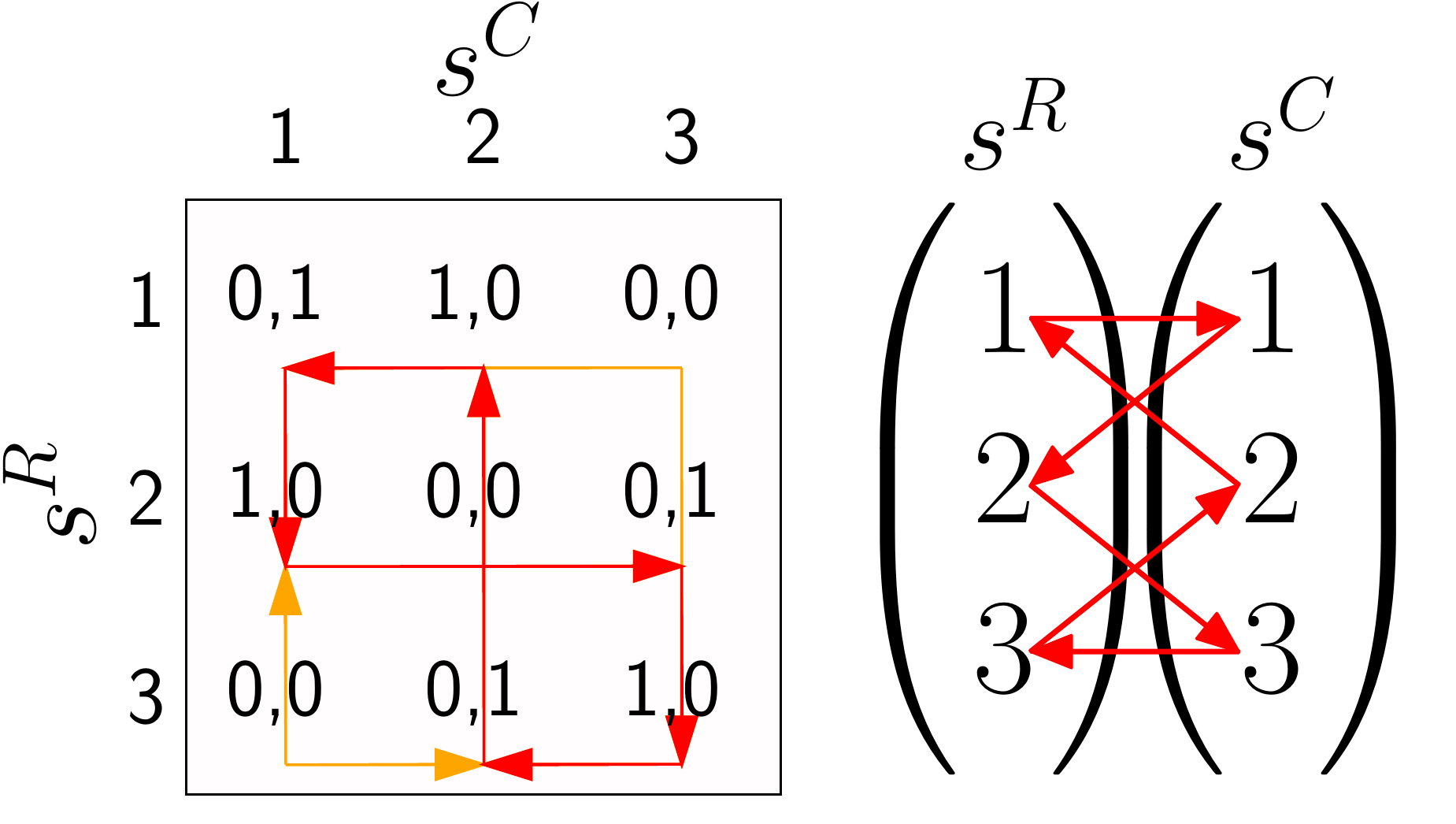}
                \label{fig:supp1m}
        \end{subfigure}  
        ~ 
        \begin{subfigure}[b]{0.3\textwidth}
                \includegraphics[width=\textwidth]{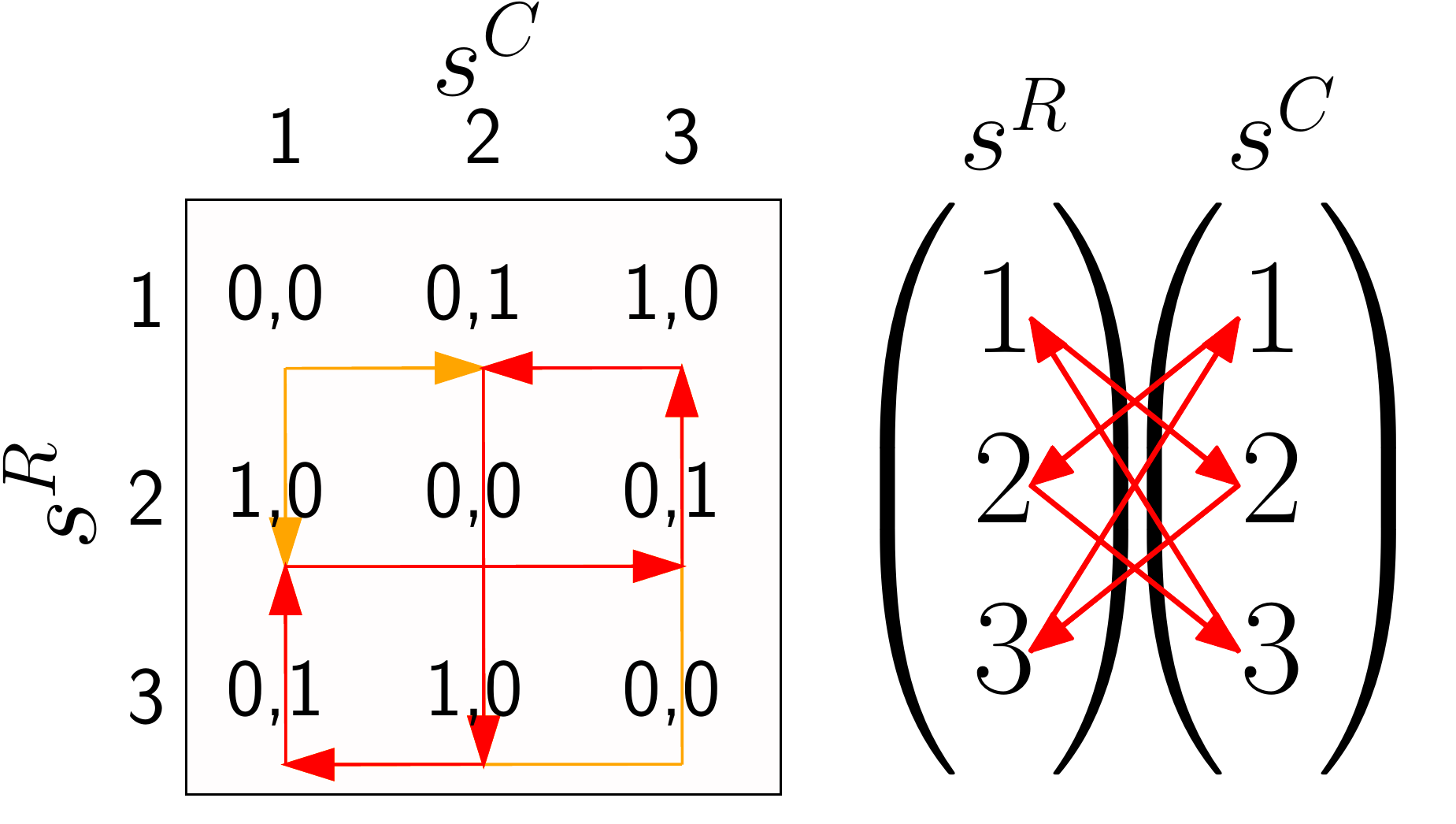}
                \label{fig:supp1n}
        \end{subfigure} 
				\caption{All possible 3! 2! = 12 ways to combine 3 moves per player to form 3-cycles. The color code has been kept consistent with the main text. The (1,2,3) vertical arrays contain the labels of the moves and the arrows represent the best replies. A cycle is a closed loop of best replies. These 12 combinations are also all best reply configurations featuring a 3-cycle in payoff matrices with $N=3$. Using Eq. \eqref{eq:f}, $f(3,3)=12$.}\label{fig:supp1}
\end{figure}

We start the count of $k$-cycles by example. In Fig. \ref{fig:supp1} we exhaustively report all possible ways to form 3-cycles in a payoff matrix with $N=3$. The vertical $(1,2,3)$ arrays and the arrows that connect the labels of the moves illustrate the main intuition: we find all possible best reply sequences that form a closed loop. We arbitrarily start at $s^R=1$ (because this is a cycle, the starting point does not matter), we look at the best reply by player Column, $s^C \in \{1,2,3\}$, and we connect $s^R=1$ with $s^C$. In the top left panel, we connect $s^R=1$ to $s^C=3$. The first choice can be done in $k=3$ ways. Once we have determined the first best reply by Column, we continue constructing the cycle by choosing a second best reply by Row. The second choice can only be done in $k-1=2$ ways. In the top left panel, we connect $s^C=3$ to $s^R=2$. We then select a second best reply by Column.  Again, we have $k-1=2$ possibilities. In the top left panel, we connect $s^R=2$ to $s^C=2$. The third and last best replies for Row and Column are constrained, there is only one ($k-2=1$) way to choose the remaining BR. In the top left panel, we connect $s^C=2$ to $s^R=3$ and $s^R=3$ to $s^C=1$. We have $3\cdot 2\cdot 2\cdot 1\cdot 1 = 12$ ways to form 3-cycles with $n=3$ available moves. Recall that $n$ denotes the number of moves per player which are not already part of cycles or fixed points.  In general $n$ might be smaller than $N$, but in Fig. \ref{fig:supp1} all moves are part of the cycle, so $N=n=k=3$.

It is possible to generalize this argument and to conclude that there are $k! (k-1)! $ ways to form $k$-cycles, once we determine which moves of players Row and Column are involved. Any $k$ moves out of $n$ can be chosen (by both players), so there are $\binom{n}{k}^2$ possibilities.  We define
\begin{equation}
f(n,k)=\binom{n}{k}^2 k! (k-1)! , 
\label{eq:f}
\end{equation}
with $2 \leq k \leq n$, as the count of the ways to have a $k$-cycle with $n$ available moves per player. In the above example, $f(3,3)=12$.

\begin{figure}[htbp]
        \centering
        \begin{subfigure}[b]{0.145\textwidth}
                \includegraphics[width=\textwidth]{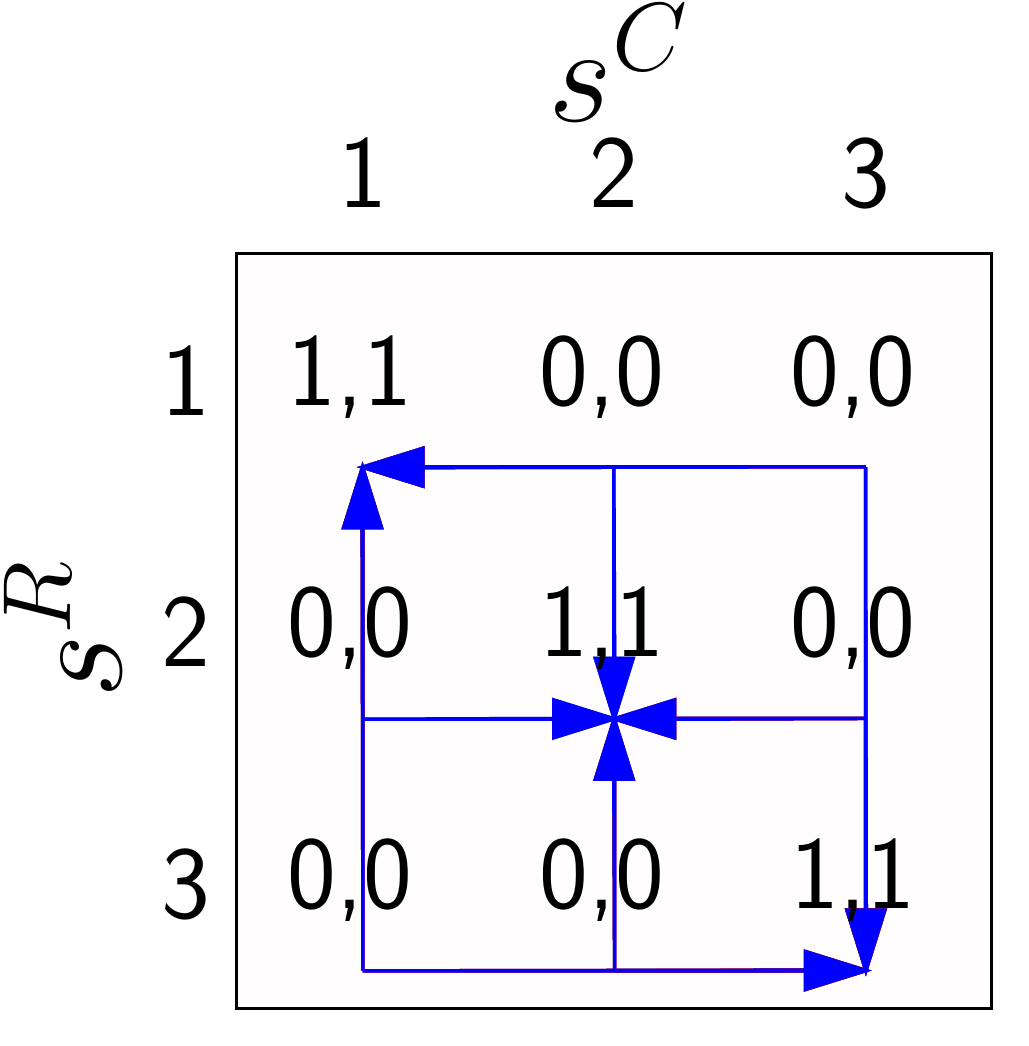}
                \label{fig:supp2a}
        \end{subfigure}
        ~ 
        \begin{subfigure}[b]{0.145\textwidth}
                \includegraphics[width=\textwidth]{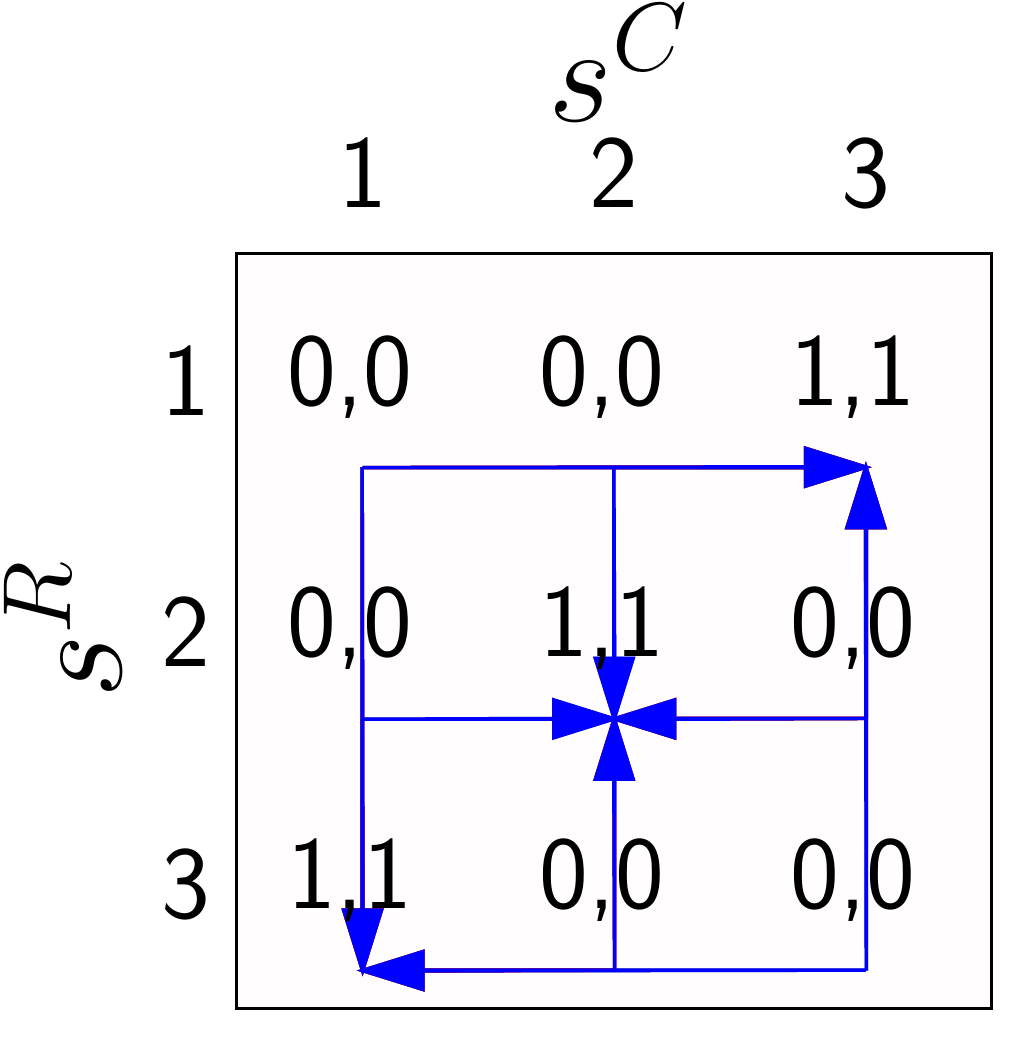}
                \label{fig:supp2b}
        \end{subfigure}  
        ~ 
        \begin{subfigure}[b]{0.145\textwidth}
                \includegraphics[width=\textwidth]{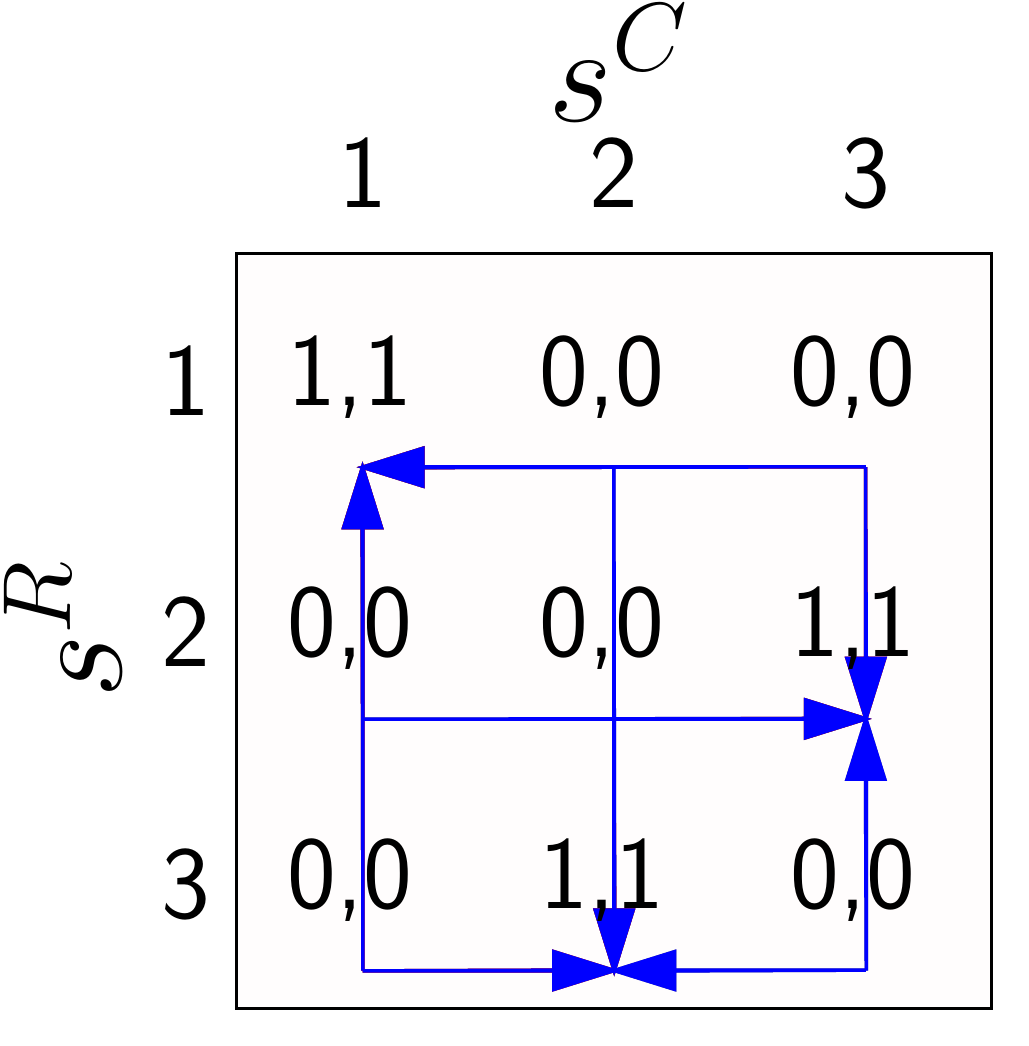}
                \label{fig:supp2c}
        \end{subfigure} 
        ~
        \begin{subfigure}[b]{0.145\textwidth}
                \includegraphics[width=\textwidth]{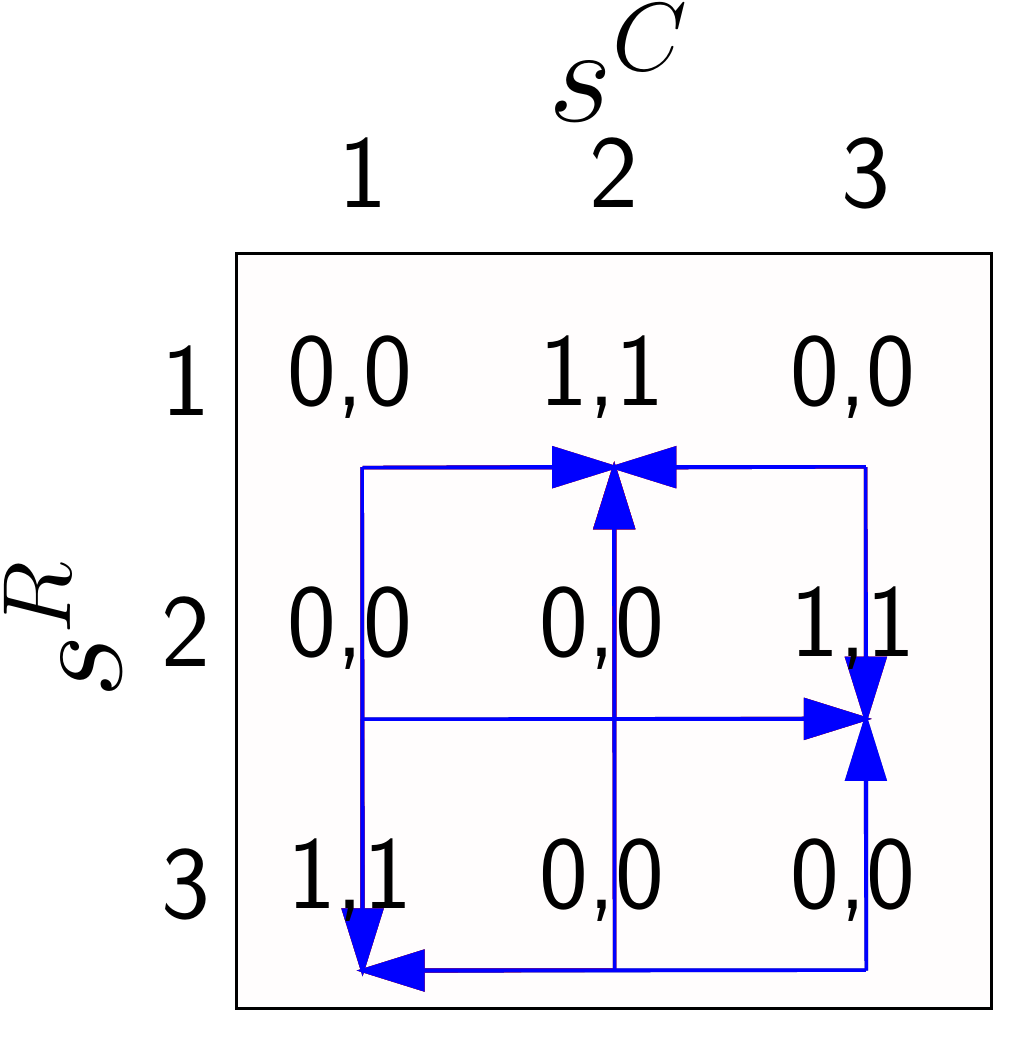}
                \label{fig:supp2d}
        \end{subfigure}
        ~ 
        \begin{subfigure}[b]{0.15\textwidth}
                \includegraphics[width=\textwidth]{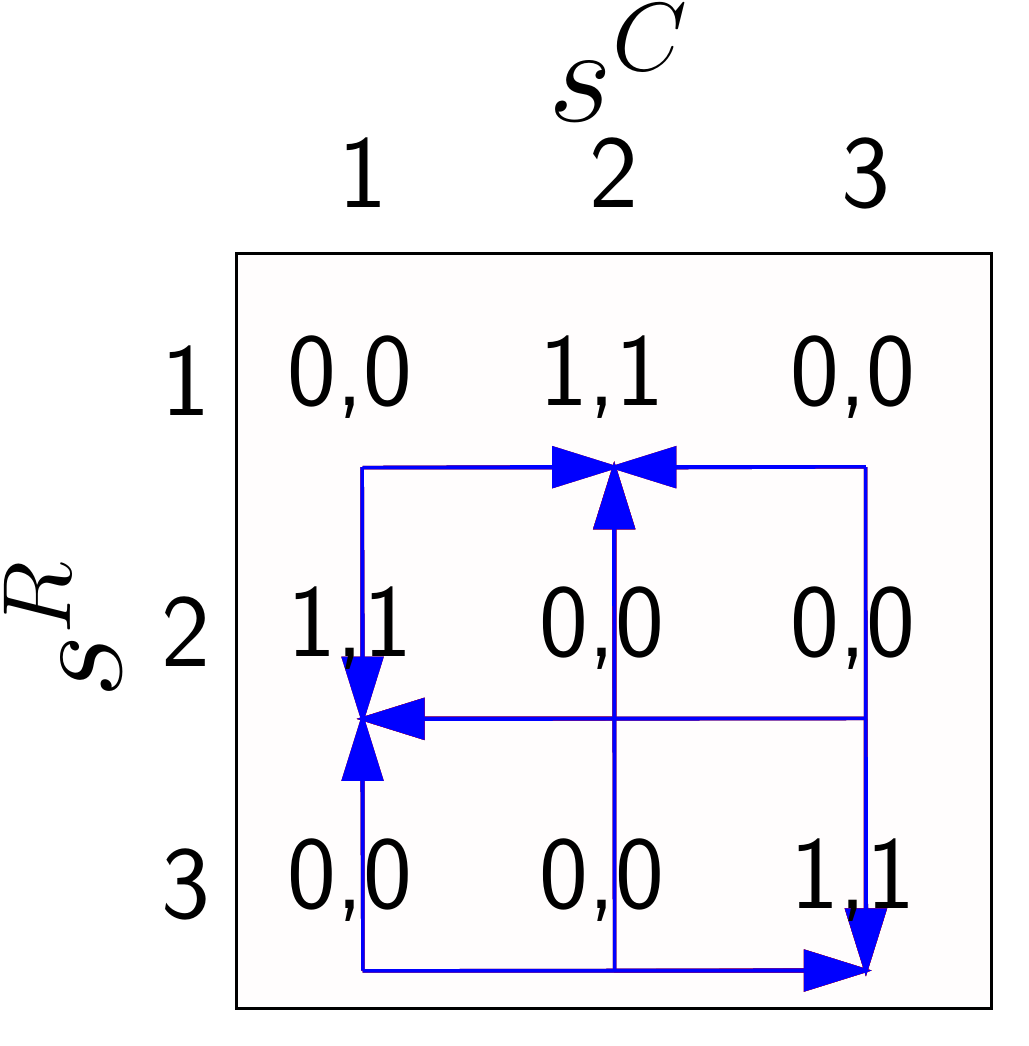}
                \label{fig:supp2e}
        \end{subfigure}  
        ~ 
        \begin{subfigure}[b]{0.145\textwidth}
                \includegraphics[width=\textwidth]{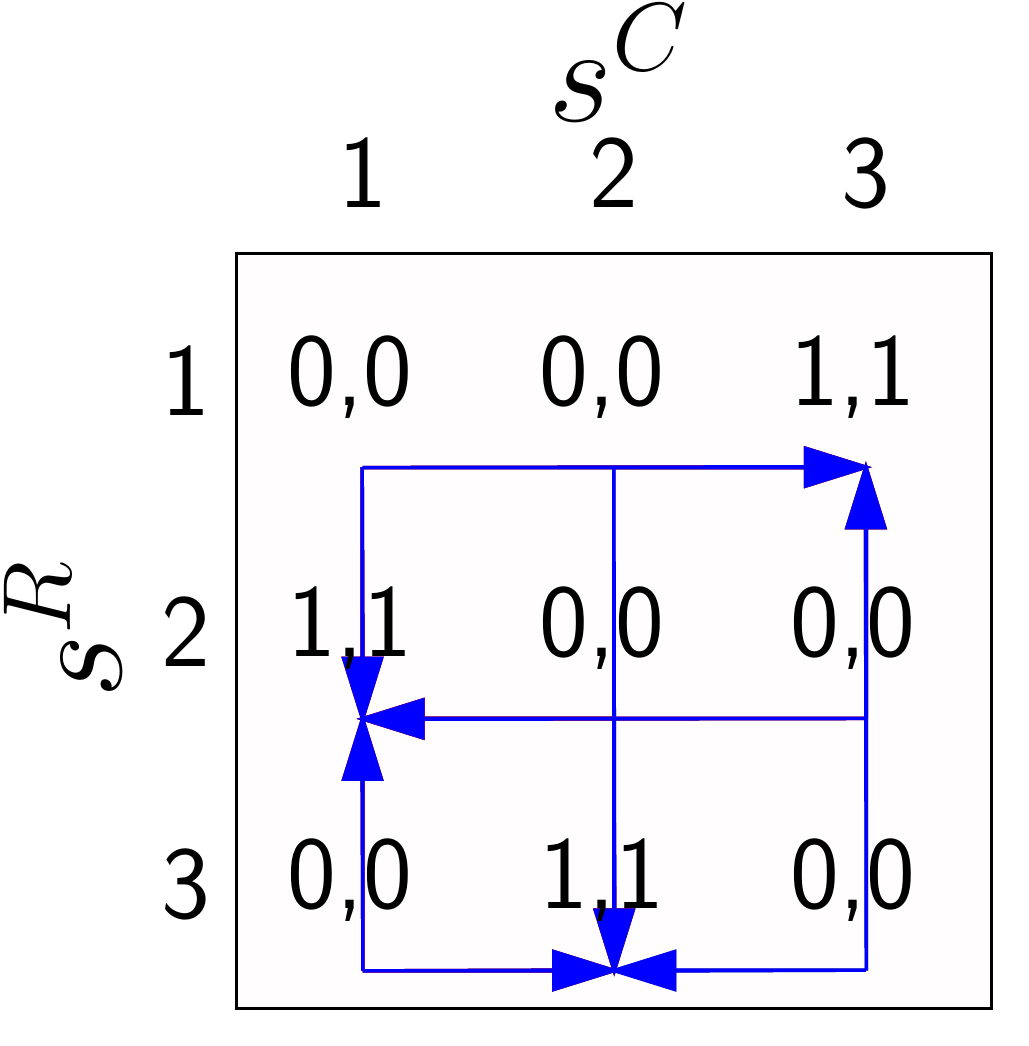}
                \label{fig:supp2f}
        \end{subfigure} 
				\caption{All possible 3! = 6 ways to combine 3 moves per player to form 3 fixed points. The color code has been kept consistent with the main text. Note that these are also all best reply configurations featuring 3 fixed points in payoff matrices with $N=3$. Using Eq. \eqref{eq:f}, $f(3,1)\frac{f(2,1)}{2}\frac{f(1,1)}{3}=9\frac{4}{2}\frac{1}{3}=6$.}\label{fig:supp2}
\end{figure}

We now look at the ways to form fixed points, and we begin again by example. In Fig. \ref{fig:supp2} we report all possible ways to form 3 fixed points in a payoff matrix with $N=3$. Once we determine which moves are part of the fixed points (all, in this case), we form all possible combinations of fixed points by picking pairs of moves from the lists of available moves by both players. For convenience, we start again from $s^R=1$. We form a fixed point by choosing any move $s^C \in \{1,2,3\}$, so that $\left( s^R, s^C \right)$ is a fixed point. In the left panel, we choose (1,1) as the first fixed point. We then consider $s^R=2$. There are only two moves available from player Column to form a second fixed point. In the left panel, (2,2) is the second fixed point. Finally, for $s^R=3$ only one move by Column is available. By process of elimination, in the left panel (3,3) is the third and last fixed point.

This example illustrates that the computation of the number of fixed points is very similar to the case of cycles, and indeed fixed points are just cycles of length one. In order to get the number of ways to form fixed points, we can apply Eq. \eqref{eq:f} iteratively and consider the double, triple etc. counting of fixed points. We get
\begin{equation}
\prod_{j=1}^{n_1} \frac{f (n+1-j,1)}{j}
\label{eq:fg}
\end{equation}
as the count of the ways to have $n_1$ fixed points with $n$ available moves per player. In the above example, $f(3,1)\frac{f(3,2)}{2}\frac{f(3,3)}{3}=9\frac{4}{2}\frac{1}{3}=6$.

\begin{figure}[htbp]
        \centering
        \begin{subfigure}[b]{0.22\textwidth}
                \includegraphics[width=\textwidth]{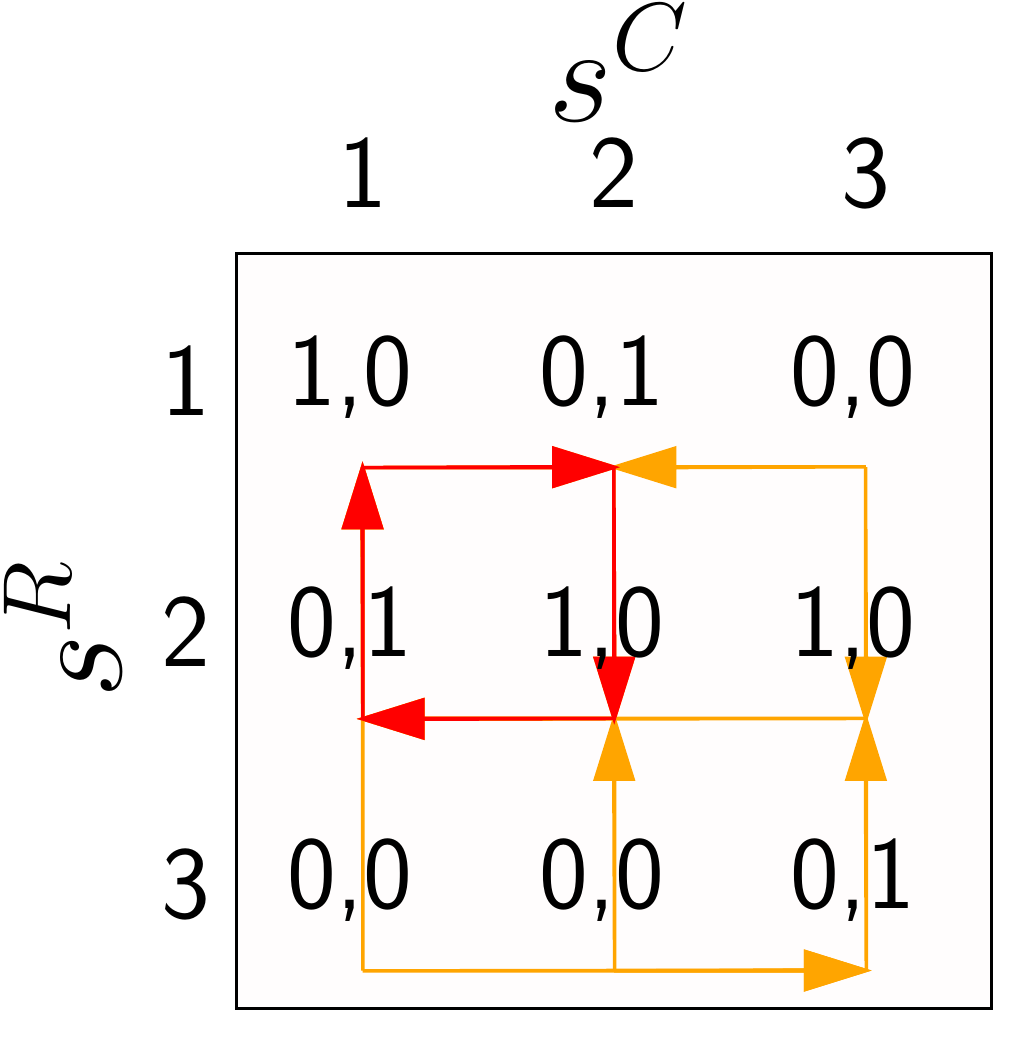}
                \label{fig:supp3a}
        \end{subfigure}
        ~ 
        \begin{subfigure}[b]{0.22\textwidth}
                \includegraphics[width=\textwidth]{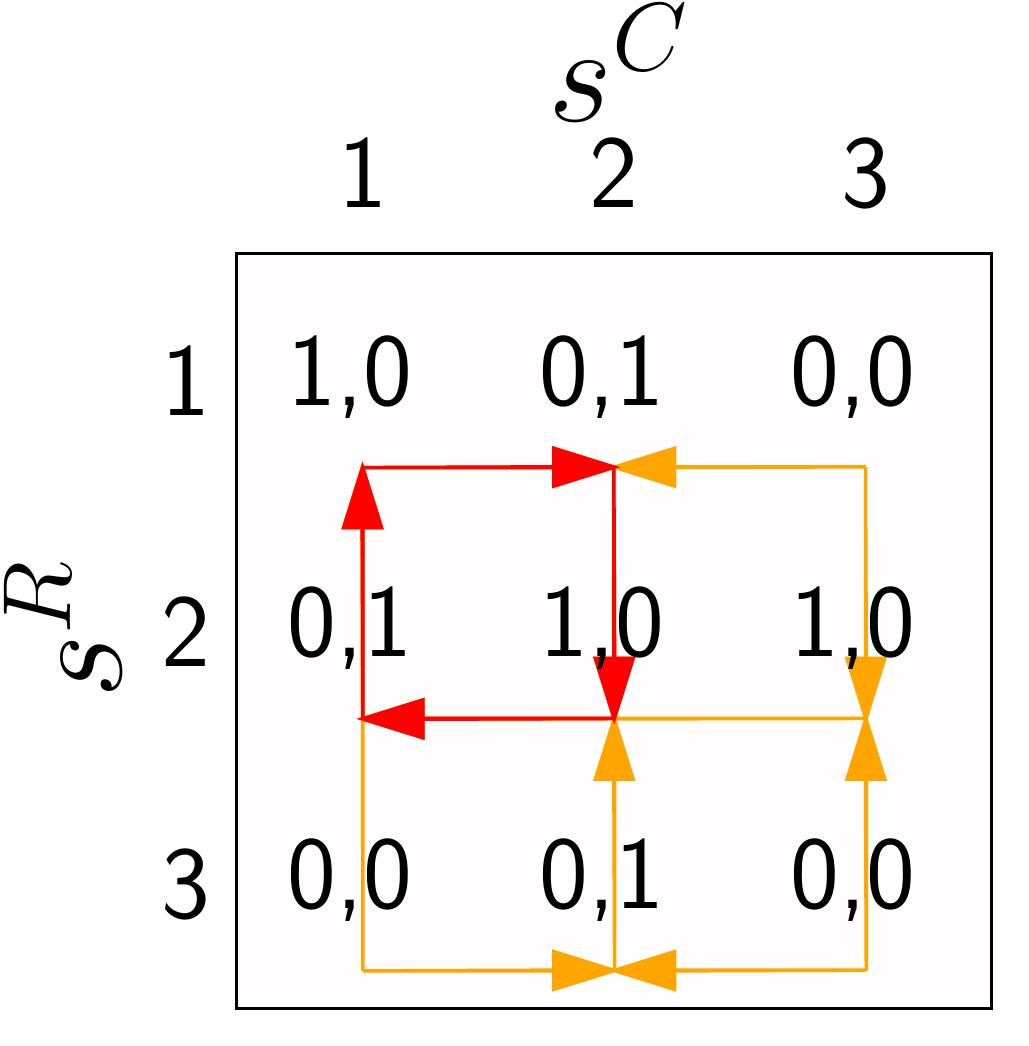}
                \label{fig:supp3b}
        \end{subfigure}  
        ~ 
        \begin{subfigure}[b]{0.22\textwidth}
                \includegraphics[width=\textwidth]{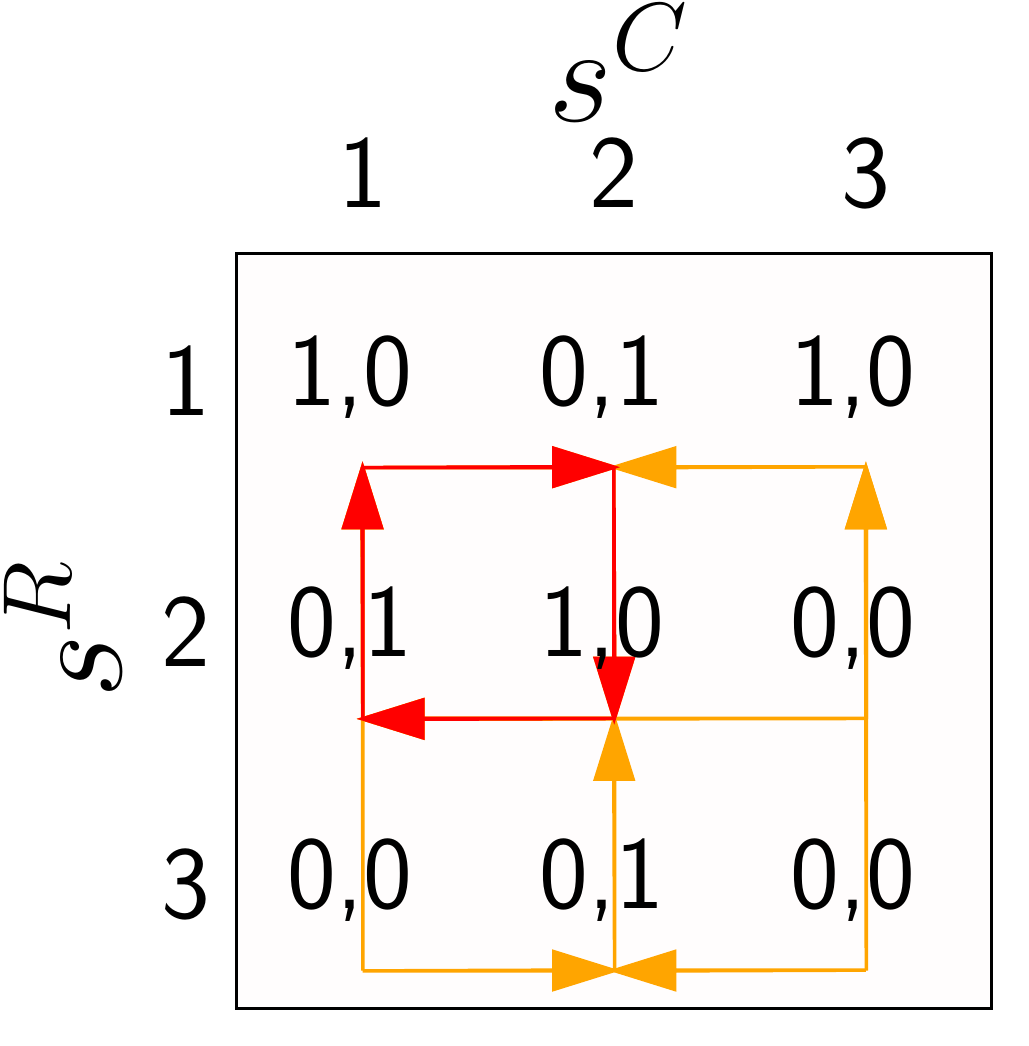}
                \label{fig:supp3c}
        \end{subfigure} 
        ~
        \begin{subfigure}[b]{0.22\textwidth}
                \includegraphics[width=\textwidth]{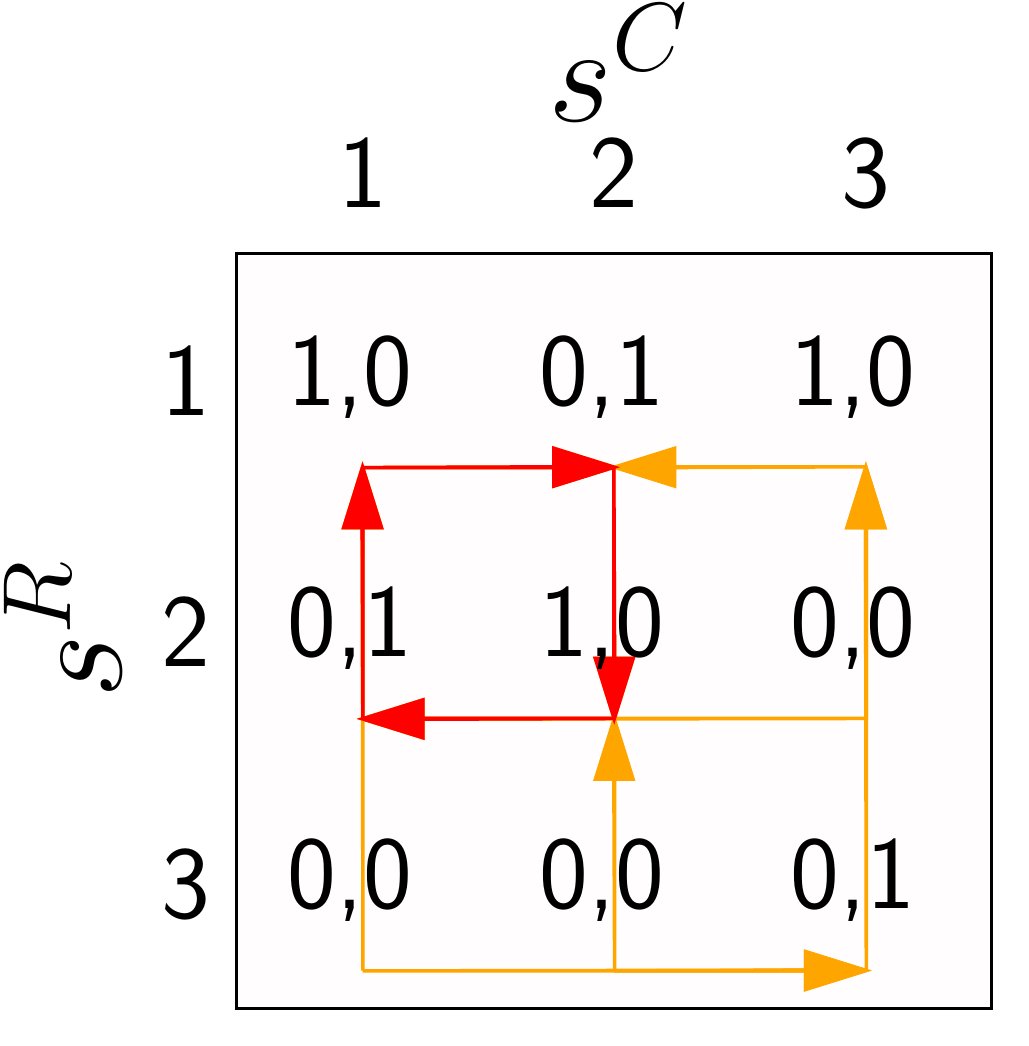}
                \label{fig:supp3d}
        \end{subfigure}
        ~ 
        \begin{subfigure}[b]{0.22\textwidth}
                \includegraphics[width=\textwidth]{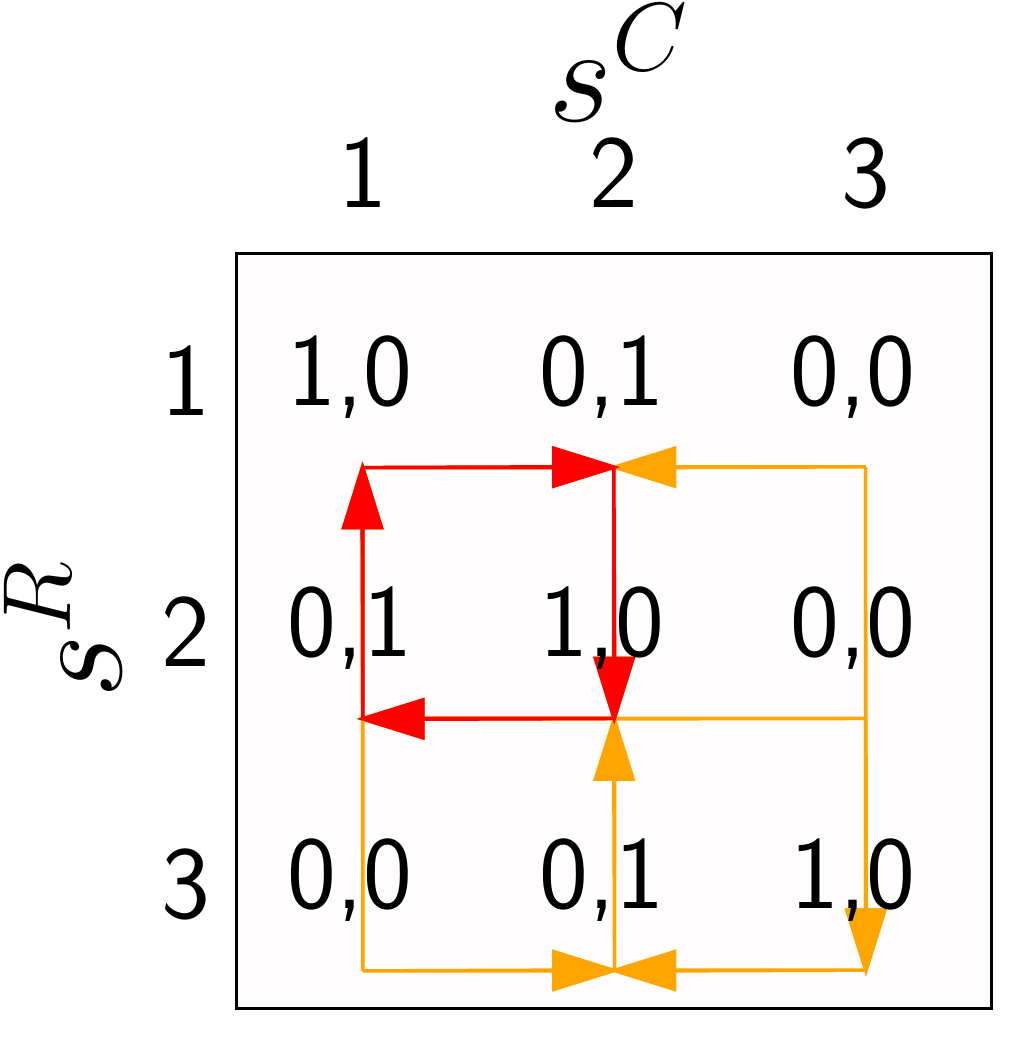}
                \label{fig:supp3e}
        \end{subfigure}  
        ~ 
        \begin{subfigure}[b]{0.22\textwidth}
                \includegraphics[width=\textwidth]{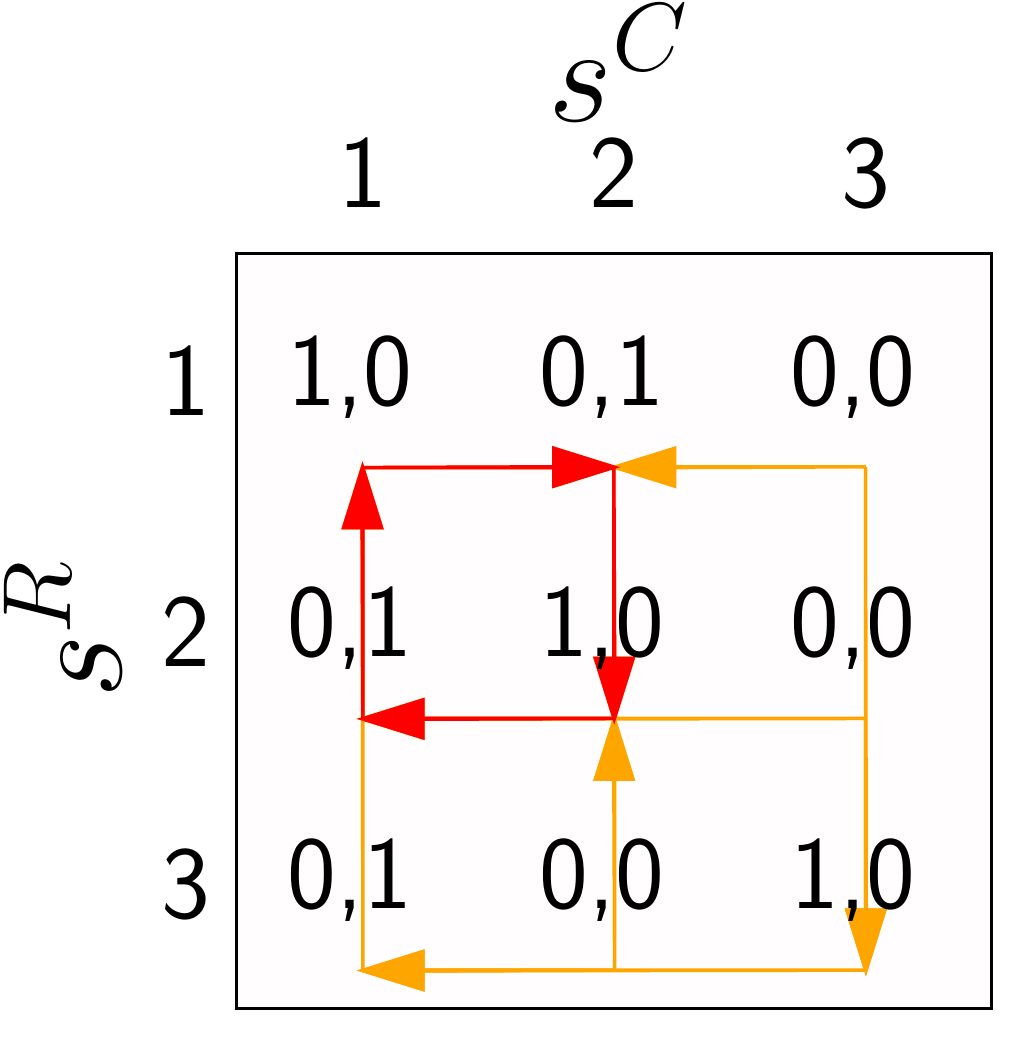}
                \label{fig:supp3f}
        \end{subfigure} 
        ~ 
        \begin{subfigure}[b]{0.22\textwidth}
                \includegraphics[width=\textwidth]{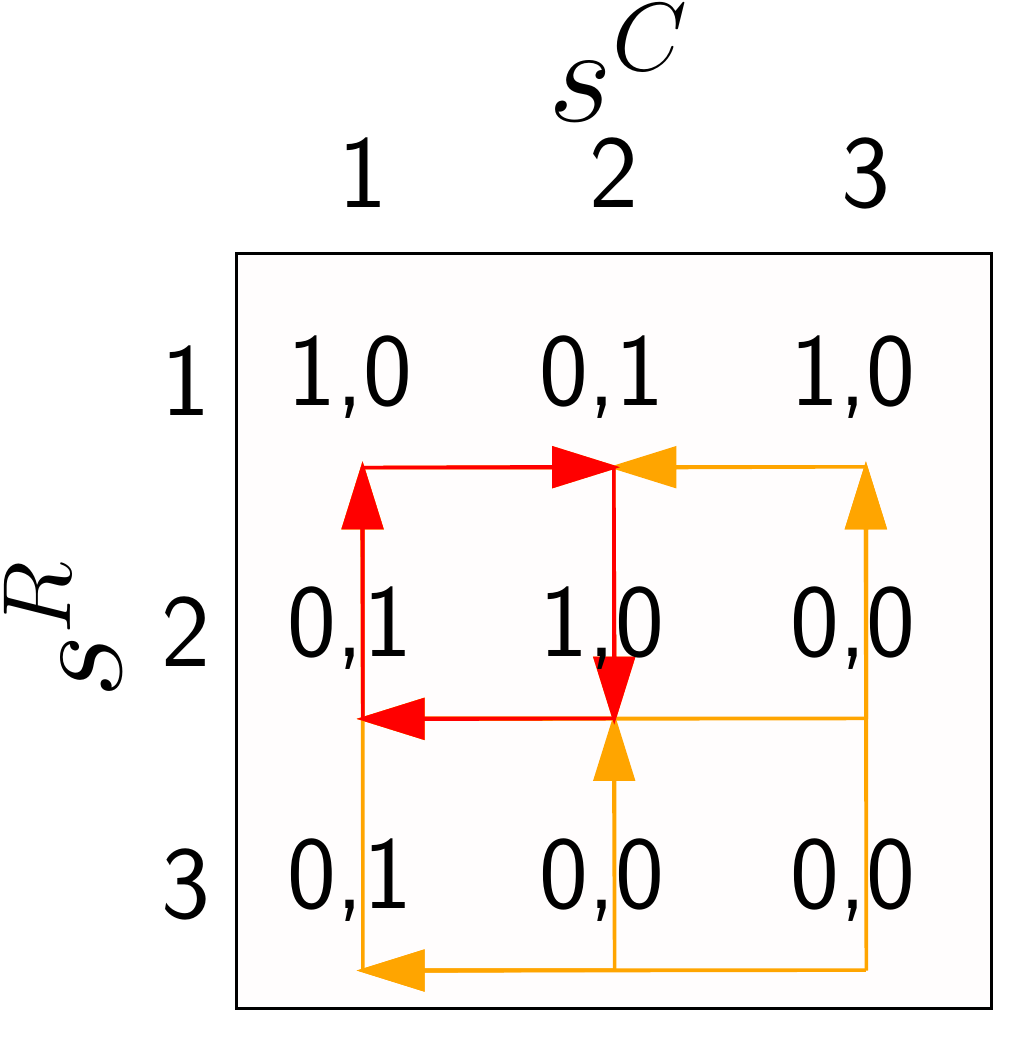}
                \label{fig:supp3g}
        \end{subfigure}  
        ~ 
        \begin{subfigure}[b]{0.22\textwidth}
                \includegraphics[width=\textwidth]{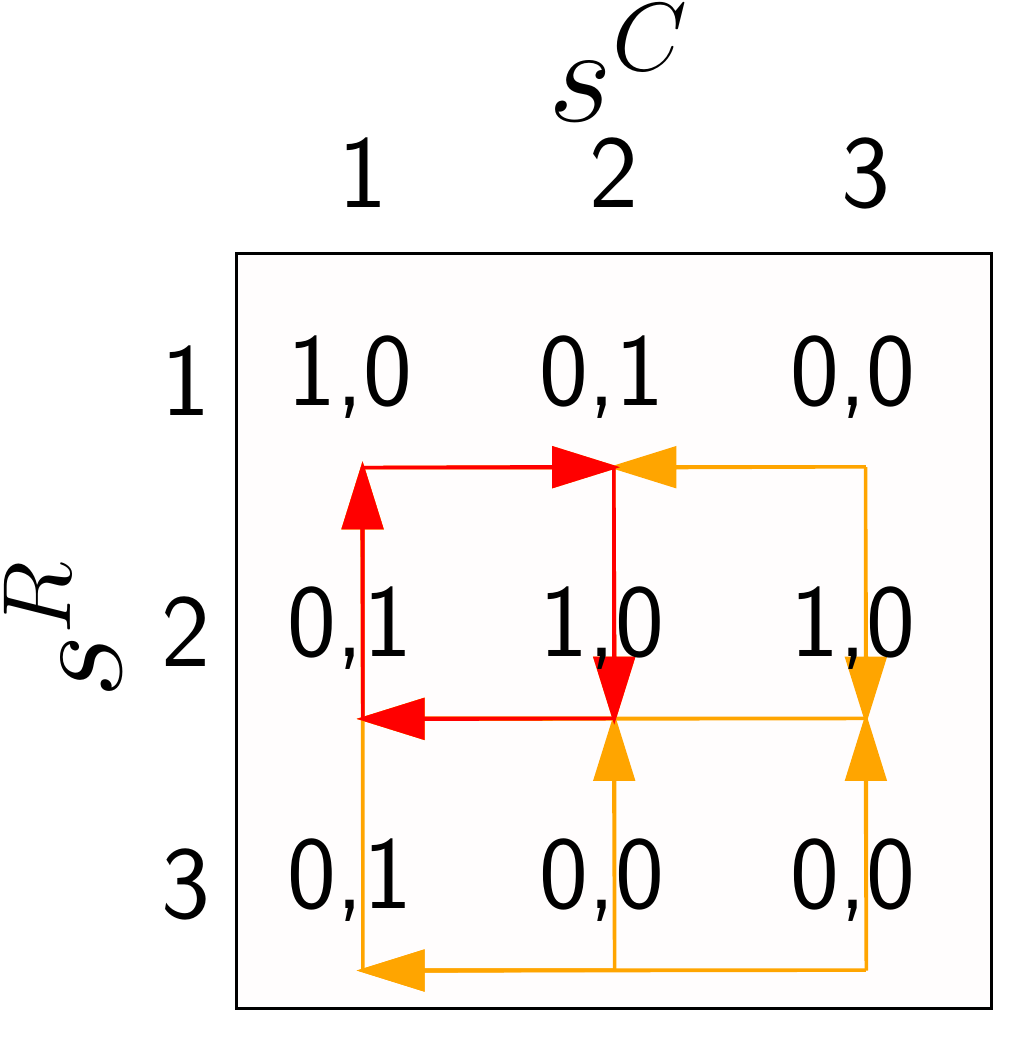}
                \label{fig:supp3h}
        \end{subfigure} 
				\caption{All possible $3^2-1 = 8$ ways to choose the two remaining best replies, so that they do not form a fixed point at $(3,3)$. The color code has been kept consistent with the main text. Using Eq. \eqref{eq:gsupp}, $g_3(1,0)=8$.}\label{fig:supp3}
\end{figure}

We finally calculate the ways to place the free best replies, which are not part of either cycles or fixed points. We begin again by example. In Fig. \ref{fig:supp3} we show payoff matrices with one free best reply per player. In the top left panel, the best reply of Row to Column playing $s^C=3$ is $s^R=2$; the best reply of Column to Row playing $s^R=3$ is $s^C=3$. The free best replies can be chosen freely, except for both of them to be move 3, in which case they would form another fixed point. In this example there are $3^2-1 = 8$ ways to choose free best replies so that they do not form other cycles or fixed points. 

In general,
\begin{equation}
g_N(n,d) = N^{2n} - \sum_{k=1}^n f(n,k)g_N(n-k,d+1)/(d+1)
\label{eq:gsupp}
\end{equation}
counts all possible ways to combine $n$ free best replies in a $N\times N$ payoff matrix, so that they do not form other cycles or fixed points. We provide a more complete example for Eq. \eqref{eq:gsupp} at the end of this section. Note that $N$ is a parameter and therefore is indicated as a subscript, while $n$ is a recursion variable: even when the number of available moves $n$ is smaller than $N$, the free best replies can be chosen out of all the $N$ moves (see Fig. \ref{fig:supp3}), in $N^{2n}$ ways.  The second term counts the ``forbidden'' combinations, i.e. the ones that form cycles or fixed points. This term has a recursive structure. It counts the number of ways to form each type of attractor, and then the number of ways not to have other attractors. $d$ denotes the recursion depth. The division by $d+1$ is needed to prevent double, triple, etc. counting of attractors. 

We now combine all the ways to have cycles, fixed points and free best replies to calculate the number of best reply configurations that correspond to a generic best reply vector $\boldsymbol{v}=(n_N, n_{N-1}, ...  n_k, ... n_2, n_1)$. We denote by $n_1$ the number of fixed points and by $n_k$, with $2 \leq k \leq N$, the number of $k$-cycles. Of course $\boldsymbol{v}$ has to obey the obvious constraint that fixed points and $k$-cycles do not take up more than $N$ moves:  $\sum_{k=1}^N n_k k \leq N$.  The frequency of the best reply vector $\boldsymbol{v}$ is 
\begin{equation}
\rho(\boldsymbol{v})=\left( \prod_{k=1}^N \prod_{j=1}^{n_k} \frac{f \left(N-\sum_{l=k+1}^N{n_l l} - (j-1)k , k\right)}{j} \right) g_N  \left. \left(N-\sum_{l=2}^N{n_l l}- n_1,0\right) \middle/ \left( N^{2N} \right)\right..
\label{eq:BRvectorsupp}
\end{equation}
 
Eq. \eqref{eq:BRvectorsupp} is Eq. (3) in the main paper. The first term with $f$ counts all the ways to have $k$-cycles, by multiplying the counts for all values of $k$ (first product) and for all $k$-cycles for a specific value of $k$ (second product).  Note that we progressively reduce the number of moves available to form $k$-cycles, as more and more moves become part of $k$-cycles (see below for an example that clarifies this point). If there are multiple $k$-cycles, $n_k>1$, we divide the count by $j=1,...n_k$ so to avoid double, triple, etc. counting. The case $k=1$ accounts for fixed points. The second term $g_N$ counts all the ways to choose the remaining $N-\sum_{l=2}^N{n_l l}-n_1$ free best replies. The product of the three terms gives the number of best reply configurations that correspond to the best reply vector $\boldsymbol{v}$. We divide this number by the possible configurations $N^{2N}$ and we obtain the frequency $\rho(\boldsymbol{v})$. 

\begin{figure}
\centering
\includegraphics[width=.5\textwidth]{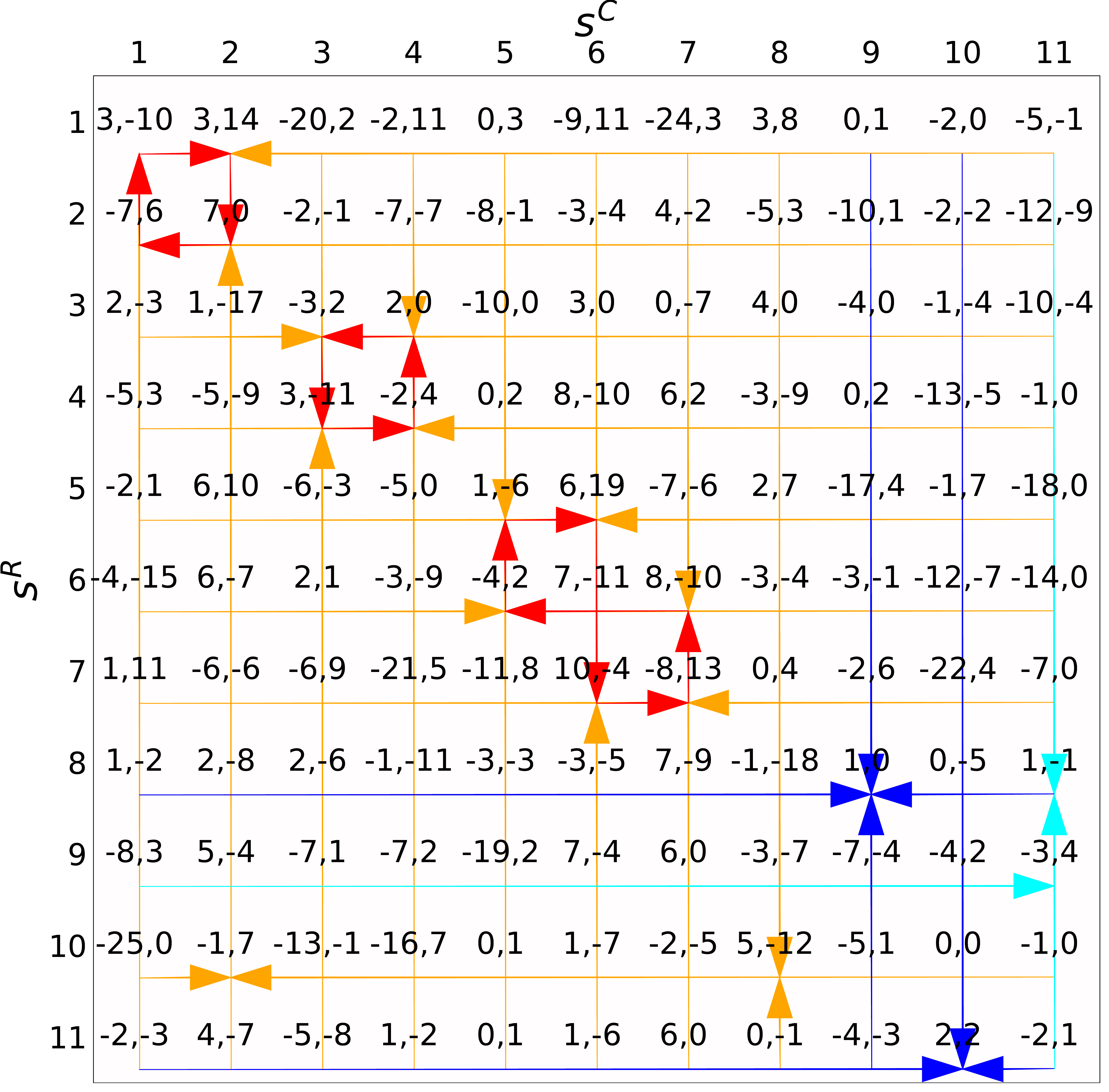}
\caption{Payoff matrix with $N=11$. The color code has been kept consistent with the main text. The set of attractors of best reply dynamics in the payoff matrix is $\boldsymbol{v}=(0,0,0,0,0,0,0,0,1,2,2)$, with $n_3=1$, $n_2=2$, $n_1=2$ and $n_k=0$, if $k>3$. It is $\sum_{k=1}^{11} n_k k = 9 < 11$.}
\label{fig:suppex1c}
\end{figure}

As an example, we calculate the number of best reply configurations with the same set of attractors as in Fig. \ref{fig:suppex1c}. We start counting the ways to form 3-cycles. We can choose any 3 moves out of 11 for both players to be part of a 3-cycle, meaning that there are $\binom{11}{3}^2$ possibilities. Once we have selected 3 moves per player, we can obtain 12 cycles for each choice by choosing $3!2! = 12$ sequences of moves. So the number of ways to form 3-cycles is $f(11,3)$.  The same reasoning applies to the two 2-cycles, except that there are only 8 and 6 moves per player still available and that the count of the ways to have 2-cycles needs to be divided by 2 in order to avoid double counting. So we multiply $f(11,3)$ by $f(8,2)f(6,2)/2$.  The number of best reply configurations with 2 fixed points in the remaining 4 moves can be calculated similarly: each player can choose the first fixed point out of 4 moves, and the second out of 3, but we have to consider double counting. So $f(4,1)f(3,1)/2$ gives the ways to form the two fixed points out of the 4 remaining moves. We are left with 2 moves per player that are not part of cycles or fixed points. There are $11^4$ ways to choose the free best replies, but we have to exclude the cases in which they would form another 2-cycle or one or more fixed points. There are 2 ways they could form a 2-cycle ($f(2,2)$), and 4 ways they could form 1 fixed point ($f(2,1)$). But for each of the latter we have to consider all compatible configurations, i.e. calculate $g_{11}(1,1)$: there are $11^2$ ways to choose the free best replies, minus the way in which this choice would form another fixed point (divided by 2, to account for the situation with two fixed points).  In summary, the number of best reply configurations is given by 
\begin{equation}
\rho(0,0,0,0,0,0,0,0,1,2,2)=f(11,3)f(8,2)\frac{f(6,2)}{2} f(4,1)\frac{f(3,1)}{2}g_{11}(2,0)/(11^{22}),
\end{equation}
with $f(11,3)=\binom{11}{3}^2 3\cdot 2 \cdot2$, $f(8,2)=\binom{8}{2}^2 2$, $f(6,2)=\binom{6}{2}^2 2$, $f(4,1)=\binom{4}{1}^2 1$, $f(3,1)=\binom{3}{1}^2 1$ and $g_{11}(2,0)=11^4-2-4 \cdot g_{11}(1,1)$, with $g_{11}(1,1)=11^2-1/2$. 

The explicit computation of the frequency gives $\rho(0,0,0,0,0,0,0,0,1,2,2)=1.44\cdot 10^{-6}$, so the best reply vector in Fig. \ref{fig:suppex1c} is very infrequent. For $N=11$, the most common best reply vectors are:

\begin{equation}
\begin{split}
& \rho(0,0,0,0,0,0,0,0,0,0,1)=0.17,\\
& \rho(0,0,0,0,0,0,0,0,0,0,2)=0.14,\\
& \rho(0,0,0,0,0,0,0,0,0,1,0)=0.14,\\
& \rho(0,0,0,0,0,0,0,0,0,1,1)=0.13,\\
& \rho(0,0,0,0,0,0,0,0,1,0,0)=0.09.\\
\end{split}
\end{equation}

For $N=20$, the most common best reply vectors are:

\begin{equation}
\begin{split}
& \rho(0, 0, 0, 0, 0, 0, 0, 0, 0, 0, 0, 0, 0, 0, 0, 0, 0, 0, 1, 1)=0.10,\\
& \rho(0, 0, 0, 0, 0, 0, 0, 0, 0, 0, 0, 0, 0, 0, 0, 0, 0, 0, 0, 1)=0.10,\\
& \rho(0, 0, 0, 0, 0, 0, 0, 0, 0, 0, 0, 0, 0, 0, 0, 0, 0, 0, 0, 2)=0.09,\\
& \rho(0, 0, 0, 0, 0, 0, 0, 0, 0, 0, 0, 0, 0, 0, 0, 0, 0, 0, 1, 0)=0.09,\\
& \rho(0, 0, 0, 0, 0, 0, 0, 0, 0, 0, 0, 0, 0, 0, 0, 0, 0, 1, 0, 0)=0.07.\\
\end{split}
\end{equation}

We observe that $k$-cycles with high values of $k$ are never really frequent; the frequency of any specific best reply vector decreases with $N$ (because there are many more best reply vectors with positive frequency); the best reply vectors with cycles become more frequent as $N$ increases, consistently with Fig. 4 of the main paper. Note that an accurate numerical estimate of the most common best reply vectors might be challenging due to the extremely high number of best reply configurations: the analytical result makes it possible to obtain exact estimates.

\subsection{Frequency of cycles and fixed points}
\label{sec:freqcfp}

So far we have provided an analytical expression to calculate the frequency of a specific best reply vector. In this section we obtain equations for the frequency of payoff matrices with at least one fixed point or one cycle of any specific length, and then for the frequency of payoff matrices with at least one cycle of any length. These expressions are useful because it is computationally very expensive to calculate the frequency of all best reply vectors and then consider the ensemble average. Indeed, in Fig. 4 of the main paper the analytical line with the frequency of non-convergence under best reply dynamics  (middle green line, $\mathcal{F}_N$) stops at $N=50$. On the contrary, the analytical lines for the fraction of payoff matrices with at least one cycle (top blue line, $\mathcal{F}(\boldsymbol{v})>0$) and with no fixed points (bottom red line, $\mathcal{F}(\boldsymbol{v})=1$) continue up to $N=400$. This is due to the fact that to compute the middle line we need to explicitly calculate the frequency of all best reply vectors, whereas to compute the top and bottom lines we use the expressions derived in this section.

Define 
\begin{equation}
h_N(n,k,d)= f(n,k)\left[ N^{2(n-k)} - \frac{h_N(n-k,k,d+1)}{d+2} \right].
\label{eq:hN}
\end{equation}

$h_N$ counts the number of configurations with at least one $k$-cycle in a $N\times N$ payoff matrix, with $n$ moves that are not already part of other $k$-cycles, at recursion depth $d$. The reasoning is similar to that in the previous section. Consider for instance the calculation of the number of 2-cycles in a $4\times 4$ payoff matrix: $N=n=4,k=2,d=0$. By using Eq. \eqref{eq:hN}, $h_4(4,2,0)=f(4,2) \left[ 4^{2\cdot 2} - h_4(2,2,1)/2 \right] $, where $h_4(2,2,1)=f(2,2)\left[ 4^0 \right] = 2$. There is a number $f(4,2)$ of 2-cycles, and for each of these there are $4^4$ ways to place the two remaining best replies of the players. But if those are combined so that they form another 2-cycle, we would count 2-cycles twice, so we need to remove one best reply configuration from the count.

We use the shorthand
\begin{equation}
\rho(N,k)=\frac{h_N(N,k,0)}{N^{2N}}
\label{eq:h}
\end{equation}
for the fraction of $N \times N$ payoff matrices with at least one $k$-cycle. Because a fixed point is a cycle of length one, Eq. \eqref{eq:h} can be used to calculate the number of payoff matrices with at least one fixed point, and
\begin{equation}
\rho_N(n_1=0) = 1-\frac{h(N,1)}{N^{2N}}
\label{eq:noNE}
\end{equation} 
is the fraction of payoff matrices with no fixed points. Eq. \eqref{eq:noNE} has been used for the bottom red analytical line in Fig. 4 of the main paper. Best reply dynamics never converges to a fixed point in these games, and other learning algorithms are very unlikely to converge as well (consider Fig. 2 in the main paper). Therefore, $\rho_N(n_1=0)$ is a lower bound for the frequency of non-convergence in generic games with $N$ moves.

Now define
\begin{equation}
h'_N(n,d)= \sum_{k=2}^n f(n,k)\left[ N^{2(n-k)} - \frac{h_N(n-k,k,d+1)}{d+2} \right].
\label{eq:hNprime}
\end{equation}
This expression is analogous to Eq. \eqref{eq:hN}, but it considers $k$-cycles of any length (except $k=1$), as opposed to $k$-cycles of a specific length. Indeed, we sum over all possible values of $k$, and the term with the double counting also considers cycles of any length. The fraction of configurations with at least one cycle is
\begin{equation}
\rho_N\left(\sum_{k=2}^N n_k>0\right) = \frac{h'_N(N,0)}{N^{2N}},
\label{eq:atleastonecycle}
\end{equation}
and this expression has been used for the top blue analytical line in Fig. 4 of the main paper. It represents an upper bound for the frequency of non-convergence in generic games with $N$ moves, because the lack of best reply cycles implies convergence in most cases.

Note that $\sum_{k=2}^{N} \rho(N,k)$ sums to more than $N^{2N}$, because several best reply configurations have multiple cycles of different length. On the contrary, $h'_N(N,0)$ is always less than $N^{2N}$, because some configurations have cycles but no fixed points. Had we started the summation in Eq. \eqref{eq:hNprime} from $k=1$, the count would sum exactly to $N^{2N}$, because all configurations have at least one cycle or one fixed point.

\subsection{Asymptotic frequency of attractors}
\label{sec:asympfreqattr}

Eq. \eqref{eq:hN} can be used, in the limit $N\rightarrow \infty$, to calculate analytically the absolute and relative frequencies of payoff matrices with at least one $k$-cycle or fixed point. Note that this calculation could potentially be related to the recently proposed concept of \textit{graphons} \cite{borgs2017graphons}, namely graphs of infinite size. We make the following ansatz:
\begin{equation}
\lim_{N\rightarrow \infty} \frac{h_N(N,k,0)}{N^{2N}} \approx \frac{h_N(N-k,k,1)}{(N-k)^{2(N-k)}}.
\label{eq:ansatz}
\end{equation}
We are making two approximations whose validity will be verified ex-post. First, the frequency of $k$-cycles reaches a fixed point as $N\rightarrow \infty$. Second, the functional form of $h_N(n,k,0)$ is very similar to that of $h_N(n,k,1)$. We know that this is not the case, as the term used to avoid multiple counting -- namely $h_N(n-k,k,d+1)$ -- is divided by 2 for $d=0$ and by 3 for $d=1$. The approximation becomes exact only for $d\rightarrow \infty$ (because $1/d$ and $1/(d+1)$ are very similar), but the quantity we are interested into has $d=0$. 

We can write
\begin{equation}
\frac{h_N(N,k,0)}{N^{2N}} = \frac{N^2 (N-1)^2 ... (N-k+1)^2}{\left(k!\right)^2} k! (k-1)! \frac{\left[ N^{2(N-k)} - \frac{h_N(N-k,k,1)}{2} \right]}{N^{2N}}.
\label{eq:kBRcycles}
\end{equation}
By applying the ansatz in Eq. \eqref{eq:ansatz} and after some algebra we obtain 
\begin{equation}
\lim_{N\rightarrow \infty} \frac{h_N(N,k,0)}{N^{2N}} := \rho(k) = \frac{1}{\left(k!\right)^2} k! (k-1)! \left( 1 - \rho(k)/2 \right),
\end{equation}
which can be solved self-consistently to yield 
\begin{equation}
\rho(k) = \frac{2}{2k+1}.
\label{eq:fk}
\end{equation}
So for $N\rightarrow \infty$, fixed points appear in 2/3 of the payoff matrices, 2-cycles appear in 2/5 of the payoff matrices, 3-cycles in 2/7, 4-cycles in 2/9, etc. Eq. \eqref{eq:fk} has been used to calculate the asymptotic frequency of configurations with no fixed points (1/3) in Fig. 4 of the main paper. We can easily obtain the relative frequencies (with respect to fixed points):
\begin{equation}
\frac{\rho(k)}{\rho(1)} = \frac{3}{2k+1},
\end{equation}
so 2-cycles appear 3/5 as often as fixed points, 3-cycles appear 3/7 as often, 4-cycles 3/9 as often, 5-cycles 3/11 as often, etc.

\begin{figure}
\centering
\includegraphics[width=.6\textwidth]{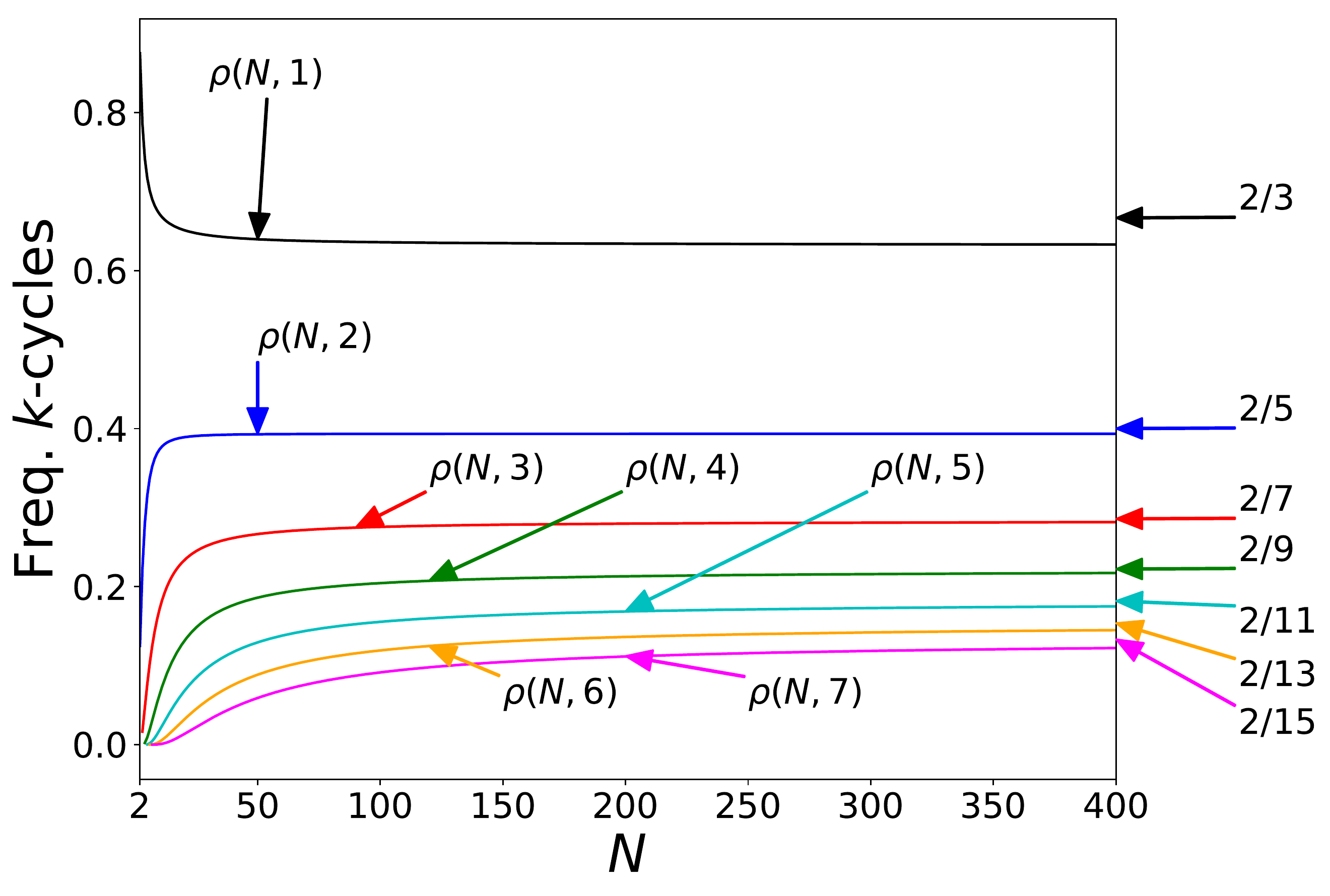}
\caption{Frequency of $k$-cycles, $\rho(N,k)$, as a function of the number of moves $N$. The numbers annotated on the right are the asymptotic frequencies of $k$-cycles, as calculated using Eq. \eqref{eq:fk}. The approximations tend to slightly overestimate the frequency, at least up to $N=400$, even more so for larger values of $k$. The exception is the fixed points, in which the approximation tends to be worse. }
\label{fig:supp4}
\end{figure}

In Fig. \ref{fig:supp4} we report the frequency of $k$-cycles, as calculated using Eq. \eqref{eq:kBRcycles}, as a function of the number of available moves $N$. There is a good correspondence between the asymptotic behavior in Eq. \eqref{eq:fk} and the explicit computation up to $N=400$, at least for the smallest values of $k$ (excluding the fixed points).

\FloatBarrier


\begin{thebibliography}{10}

\bibitem{myerson2013game}
R.~B. Myerson, {\it Game theory\/} (Harvard university press, 2013).

\bibitem{smith1982evolution}
J.~M. Smith, {\it Evolution and the Theory of Games\/} (Cambridge university
  press, 1982).

\bibitem{axelrod1981evolution}
R.~Axelrod, W.~D. Hamilton, The evolution of cooperation.
\newblock {\it Science\/} {\bf 211}, 1390--1396 (1981).

\bibitem{Nowak06071999}
M.~A. Nowak, D.~C. Krakauer, The evolution of language.
\newblock {\it Proceedings of the National Academy of Sciences\/} {\bf 96},
  8028-8033 (1999).

\bibitem{rosenthal1973class}
R.~W. Rosenthal, A class of games possessing pure-strategy nash equilibria.
\newblock {\it International Journal of Game Theory\/} {\bf 2}, 65--67 (1973).

\bibitem{kauffman1969metabolic}
S.~A. Kauffman, Metabolic stability and epigenesis in randomly constructed
  genetic nets.
\newblock {\it Journal of theoretical biology\/} {\bf 22}, 437--467 (1969).

\bibitem{may1973qualitative}
R.~M. May, Qualitative stability in model ecosystems.
\newblock {\it Ecology\/} {\bf 54}, 638--641 (1973).

\bibitem{fudenberg1998theory}
D.~Fudenberg, D.~K. Levine, {\it The theory of learning in games\/}, vol.~2
  (MIT press, 1998).

\bibitem{shapley1964some}
L.~S. Shapley, Some topics in two-person games.
\newblock {\it Advances in game theory, Annals of Mathematical Studies\/} {\bf
  52}, 1--29 (1964).

\bibitem{gintis2000game}
H.~Gintis, {\it Game theory evolving: A problem-centered introduction to
  modeling strategic behavior\/} (Princeton university press, 2000).

\bibitem{sato2002chaos}
Y.~Sato, E.~Akiyama, J.~D. Farmer, Chaos in learning a simple two-person game.
\newblock {\it Proceedings of the National Academy of Sciences\/} {\bf 99},
  4748--4751 (2002).

\bibitem{nachbar1990evolutionary}
J.~H. Nachbar, ``evolutionary'' selection dynamics in games: Convergence and
  limit properties.
\newblock {\it International journal of game theory\/} {\bf 19}, 59--89 (1990).

\bibitem{foster1998nonconvergence}
D.~P. Foster, H.~P. Young, On the nonconvergence of fictitious play in
  coordination games.
\newblock {\it Games and Economic Behavior\/} {\bf 25}, 79--96 (1998).

\bibitem{monderer1996fictitious}
D.~Monderer, L.~S. Shapley, Fictitious play property for games with identical
  interests.
\newblock {\it Journal of economic theory\/} {\bf 68}, 258--265 (1996).

\bibitem{milgrom1990rationalizability}
P.~Milgrom, J.~Roberts, Rationalizability, learning, and equilibrium in games
  with strategic complementarities.
\newblock {\it Econometrica: Journal of the Econometric Society\/} pp.
  1255--1277 (1990).

\bibitem{arieli2016stochastic}
I.~Arieli, H.~P. Young, Stochastic learning dynamics and speed of convergence
  in population games.
\newblock {\it Econometrica\/} {\bf 84}, 627--676 (2016).

\bibitem{may1972will}
R.~M. May, Will a large complex system be stable?
\newblock {\it Nature\/} {\bf 238}, 413--414 (1972).

\bibitem{blume1993statistical}
L.~E. Blume, The statistical mechanics of strategic interaction.
\newblock {\it Games and economic behavior\/} {\bf 5}, 387--424 (1993).

\bibitem{goldberg1968probability}
K.~Goldberg, A.~Goldman, M.~Newman, The probability of an equilibrium point.
\newblock {\it Journal of Research of the National Bureau of Standards\/} {\bf
  72}, 93--101 (1968).

\bibitem{dresher1970probability}
M.~Dresher, Probability of a pure equilibrium point in n-person games.
\newblock {\it Journal of Combinatorial Theory\/} {\bf 8}, 134--145 (1970).

\bibitem{powers1990limiting}
I.~Y. Powers, Limiting distributions of the number of pure strategy nash
  equilibria in n-person games.
\newblock {\it International Journal of Game Theory\/} {\bf 19}, 277--286
  (1990).

\bibitem{berg1998matrix}
J.~Berg, A.~Engel, Matrix games, mixed strategies, and statistical mechanics.
\newblock {\it Physical Review Letters\/} {\bf 81}, 4999-5002 (1998).

\bibitem{berg2000statistical}
J.~Berg, Statistical mechanics of random two-player games.
\newblock {\it Physical Review E\/} {\bf 61}, 2327-2339 (2000).

\bibitem{cohen1998cooperation}
J.~E. Cohen, Cooperation and self-interest: Pareto-inefficiency of nash
  equilibria in finite random games.
\newblock {\it Proceedings of the National Academy of Sciences\/} {\bf 95},
  9724--9731 (1998).

\bibitem{galla2013complex}
T.~Galla, J.~D. Farmer, Complex dynamics in learning complicated games.
\newblock {\it Proceedings of the National Academy of Sciences\/} {\bf 110},
  1232--1236 (2013).

\bibitem{skyrms1992chaos}
B.~Skyrms, Chaos in game dynamics.
\newblock {\it Journal of Logic, Language and Information\/} {\bf 1}, 111--130
  (1992).

\bibitem{blume1982learning}
L.~E. Blume, D.~Easley, Learning to be rational.
\newblock {\it Journal of Economic Theory\/} {\bf 26}, 340--351 (1982).

\bibitem{boldrin1986indeterminacy}
M.~Boldrin, L.~Montrucchio, On the indeterminacy of capital accumulation paths.
\newblock {\it Journal of Economic theory\/} {\bf 40}, 26--39 (1986).

\bibitem{hommes1998consistent}
C.~Hommes, G.~Sorger, Consistent expectations equilibria.
\newblock {\it Macroeconomic Dynamics\/} {\bf 2}, 287--321 (1998).

\bibitem{gigerenzer1999simple}
G.~Gigerenzer, P.~M. Todd, {\it Simple heuristics that make us smart\/} (Oxford
  University Press, 1999).

\bibitem{papadimitriou2016nash}
C.~Papadimitriou, G.~Piliouras, {\it Proceedings of the 2016 ACM Conference on
  Innovations in Theoretical Computer Science\/} (ACM, 2016), pp. 227--235.

\bibitem{goemans2005sink}
M.~Goemans, V.~Mirrokni, A.~Vetta, {\it Foundations of Computer Science, 2005.
  FOCS 2005. 46th Annual IEEE Symposium on\/} (IEEE, 2005), pp. 142--151.

\bibitem{erev1998predicting}
I.~Erev, A.~E. Roth, Predicting how people play games: Reinforcement learning
  in experimental games with unique, mixed strategy equilibria.
\newblock {\it American economic review\/} {\bf 88}, 848--881 (1998).

\bibitem{bush1955stochastic}
R.~R. Bush, F.~Mosteller, {\it Stochastic models for learning.\/} (John Wiley
  \& Sons, Inc., 1955).

\bibitem{robinson1951iterative}
J.~Robinson, An iterative method of solving a game.
\newblock {\it Annals of mathematics\/} pp. 296--301 (1951).

\bibitem{brown1951iterative}
G.~W. Brown, {\it Activity analysis of production and allocation\/},
  T.~Koopmans, ed. (Wiley, New York, 1951), pp. 374--376.

\bibitem{hofbauer1998evolutionary}
J.~Hofbauer, K.~Sigmund, {\it Evolutionary games and population dynamics\/}
  (Cambridge university press, 1998).

\bibitem{borgers1997learning}
T.~B{\"o}rgers, R.~Sarin, Learning through reinforcement and replicator
  dynamics.
\newblock {\it Journal of Economic Theory\/} {\bf 77}, 1--14 (1997).

\bibitem{camerer1999experience}
C.~Camerer, T.~Ho, Experience-weighted attraction learning in normal form
  games.
\newblock {\it Econometrica\/} {\bf 67}, 827--874 (1999).

\bibitem{nagel1995unraveling}
R.~Nagel, Unraveling in guessing games: An experimental study.
\newblock {\it The American Economic Review\/} {\bf 85}, 1313--1326 (1995).

\bibitem{selten1991anticipatory}
R.~Selten, {\it Game Equilibrium Models I\/}, R.~Selten, ed. (Springer-Verlag,
  Berlin-Heidelberg, 1991), pp. 98--154.

\bibitem{borrelli2015selection}
J.~J. Borrelli, Selection against instability: stable subgraphs are most
  frequent in empirical food webs.
\newblock {\it Oikos\/} {\bf 124}, 1583--1588 (2015).

\end{thebibliography}

\begin{thebibliography}{10}

\bibitem{crawford1974learning}
Vincent~P Crawford.
\newblock Learning the optimal strategy in a zero-sum game.
\newblock {\em Econometrica: Journal of the Econometric Society}, pages
  885--891, 1974.

\bibitem{conlisk1993adaptation}
John Conlisk.
\newblock Adaptation in games: Two solutions to the crawford puzzle.
\newblock {\em Journal of Economic Behavior \& Organization}, 22(1):25--50,
  1993.

\bibitem{bloomfield1994learning}
Robert Bloomfield.
\newblock Learning a mixed strategy equilibrium in the laboratory.
\newblock {\em Journal of Economic Behavior \& Organization}, 25(3):411--436,
  1994.

\bibitem{robinson1951iterativesupp}
Julia Robinson.
\newblock An iterative method of solving a game.
\newblock {\em Annals of mathematics}, pages 296--301, 1951.

\bibitem{gintis2000gamesupp}
Herbert Gintis.
\newblock {\em Game theory evolving: A problem-centered introduction to
  modeling strategic behavior}.
\newblock Princeton university press, 2000.

\bibitem{pangallo2017taxonomy}
Marco Pangallo, James~BT Sanders, Tobias Galla, and J~Doyne Farmer.
\newblock A taxonomy of learning dynamics in 2 x 2 games.
\newblock Preprint available at https://arxiv.org/abs/1701.09043, 2017.

\bibitem{bush1955stochasticsupp}
Robert~R Bush and Frederick Mosteller.
\newblock {\em Stochastic models for learning.}
\newblock John Wiley \& Sons, Inc., 1955.

\bibitem{macy2002learning}
Michael~W Macy and Andreas Flache.
\newblock Learning dynamics in social dilemmas.
\newblock {\em Proceedings of the National Academy of Sciences}, 99(suppl
  3):7229--7236, 2002.

\bibitem{galla2013complexsupp}
Tobias Galla and J~Doyne Farmer.
\newblock Complex dynamics in learning complicated games.
\newblock {\em Proceedings of the National Academy of Sciences},
  110(4):1232--1236, 2013.

\bibitem{erev1998predictingsupp}
Ido Erev and Alvin~E Roth.
\newblock Predicting how people play games: Reinforcement learning in
  experimental games with unique, mixed strategy equilibria.
\newblock {\em American economic review}, 88:848--881, 1998.

\bibitem{brown1951iterativesupp}
G.~W. Brown.
\newblock Iterative solution of games by fictitious play.
\newblock In T.C. Koopmans, editor, {\em Activity analysis of production and
  allocation}, pages 374--376. Wiley, New York, 1951.

\bibitem{fudenberg1998theorysupp}
Drew Fudenberg and David~K Levine.
\newblock {\em The theory of learning in games}, volume~2.
\newblock MIT press, 1998.

\bibitem{smith1982evolutionsupp}
John~Maynard Smith.
\newblock {\em Evolution and the Theory of Games}.
\newblock Cambridge university press, 1982.

\bibitem{hofbauer1998evolutionarysupp}
Josef Hofbauer and Karl Sigmund.
\newblock {\em Evolutionary games and population dynamics}.
\newblock Cambridge university press, 1998.

\bibitem{borgers1997learningsupp}
Tilman B{\"o}rgers and Rajiv Sarin.
\newblock Learning through reinforcement and replicator dynamics.
\newblock {\em Journal of Economic Theory}, 77(1):1--14, 1997.

\bibitem{camerer1999experiencesupp}
Colin Camerer and Teck Ho.
\newblock Experience-weighted attraction learning in normal form games.
\newblock {\em Econometrica}, 67(4):827--874, 1999.

\bibitem{sato2005stability}
Yuzuru Sato, Eizo Akiyama, and James~P Crutchfield.
\newblock Stability and diversity in collective adaptation.
\newblock {\em Physica D: Nonlinear Phenomena}, 210(1):21--57, 2005.

\bibitem{selten1991anticipatorysupp}
R.~Selten.
\newblock Anticipatory learning in two-person games.
\newblock In R.~Selten, editor, {\em Game Equilibrium Models I}, pages 98--154.
  Springer-Verlag, Berlin-Heidelberg, 1991.

\bibitem{lecutier2013stochastic}
David Lecutier.
\newblock Stochastic dynamics of game learning.
\newblock Master's thesis, University of Manchester, 2013.

\bibitem{evans2013klevel}
Theodore Evans.
\newblock k-level reasoning: A dynamic model of game learning.
\newblock Master's thesis, University of Manchester, 2013.

\bibitem{nagel1995unravelingsupp}
Rosemarie Nagel.
\newblock Unraveling in guessing games: An experimental study.
\newblock {\em The American Economic Review}, 85(5):1313--1326, 1995.

\bibitem{crawford2013structural}
Vincent~P Crawford, Miguel~A Costa-Gomes, and Nagore Iriberri.
\newblock Structural models of nonequilibrium strategic thinking: Theory,
  evidence, and applications.
\newblock {\em Journal of Economic Literature}, 51(1):5--62, 2013.

\bibitem{borgs2017graphons}
Christian Borgs and Jennifer~T Chayes.
\newblock Graphons: A nonparametric method to model, estimate, and design
  algorithms for massive networks.
\newblock {\em arXiv preprint arXiv:1706.01143}, 2017.

\end{thebibliography}
\end{document}